\documentclass[11pt]{article}
\usepackage{slashed}
\usepackage{amsfonts}
\usepackage{amsmath}
\usepackage{amssymb}
\usepackage{hyperref}
\usepackage{epsfig}%pstricks,pst-node}
\usepackage{graphicx}
\usepackage{mathrsfs}

\usepackage{pgf,tikz}
\usetikzlibrary{arrows}
\usetikzlibrary{patterns}

\usepackage{braket}
\usepackage{setspace}
\definecolor{r}{rgb}{1,0,0}
\definecolor{b}{rgb}{0,0,1}
 \newcommand{\tr}[1]{}

\newcommand{\sep}{\mathord{:}}
\newcommand{\fra}[2]{\textstyle{\frac{#1}{#2}}}
\newcommand{\beqn}{\begin{eqnarray}\begin{aligned}}
\newcommand{\eqn}{\end{aligned}\end{eqnarray}}

\newcommand{\unit}{{\rm 1\hspace*{-0.4ex}%
\rule{0.1ex}{1.52ex}\hspace*{0.2ex}}}

\newcommand{\st}{\sqrt{2}}

\newcommand{\ovw}{\overline{\omega}}

\newcommand{\uli}{\underline{i}}

\newcommand{\ulk}{\underline{k}}
\newcommand{\ull}{\underline{\ell}}

\newcommand{\ulr}{\underline{r}}

\newcommand{\who}{\widehat{1}}
\newcommand{\wht}{\widehat{2}}
\newcommand{\whth}{\widehat{3}}
\newcommand{\whf}{\widehat{4}}
\newcommand{\nrm}{\scalebox{1.2}{$|\!|$}}

\def\scirc{\scalebox{.8}{$\circ$} }

\begin{document}

\title{\protect {\Large \textsc{Systematics and symmetry in molecular phylogenetic modelling: 
%and inference: computational group theory and insights from physics}}}
%and inference: perspectives from physics.}}}
perspectives from physics.}}}

%\author{Bertfried Fauser$^1$\footnote{Present address:}, Peter D Jarvis$^2$ and Ronald C King$^3$}
\author{Peter D Jarvis and Jeremy G Sumner
\footnote{School of Natural Sciences (Mathematics and Physics), College of Science and Engineering, University of Tasmania, Tas 7001 Australia 
\texttt{peter.jarvis@utas.edu.au,jeremy.sumner@utas.edu.au}} }

%\eads{\mailto{bfauser@gmail.com},  \mailto{peter.jarvis@utas.edu.au}, \mailto{r.c.king@soton.ac.uk}}

%\begin{abstract}
%\end{abstract}
%\pacs{}
%\submitto{}
%\keywords

\maketitle

\abstract
The aim of this review is to present and analyze the probabilistic models of mathematical phylogenetics which have been intensively used in recent years in biology as the cornerstone of attempts to infer and reconstruct the ancestral relationships between species. We outline the development of theoretical phylogenetics, from the earliest studies based on morphological characters, through to the use of molecular data in a wide variety of forms. We bring the lens of mathematical physics to bear on the formulation of theoretical models, focussing on the applicability of many methods from the toolkit of that tradition -- techniques of groups and representations to guide model specification and to exploit the multilinear setting of the models in the presence of underlying symmetries; extensions to coalgebraic properties of the generators associated to rate matrices underlying the models, and possibilities to marry these with the graphical structures (trees and networks) which form the search space for inferring evolutionary trees.

Particular aspects which we wish to present to a readership accustomed to thinking from physics, include relating model classes to
structural data on relevant matrix Lie algebras, as well as using manipulations with group characters (especially the operation of plethysm, for computing tensor powers) to enumerate various natural polynomial invariants, which can be enormously helpful in tying down robust, low-parameter quantities for use in inference (some of which have only come to light through our perspective). Above all, we wish to emphasize the many features of multipartite entanglement which are shared between descriptions of quantum states on the physics side, and the multi-way tensor probability arrays arising in phylogenetics. In some instances, well-known objects such as the Cayley hyperdeterminant (the `tangle') can be directly imported into the formalism -- in this case, for models with binary character traits, and for providing information about triplets of taxa. In other cases new objects appear, such as the remarkable `squangle' invariants for quartet tree discrimination, which for DNA data are of quintic degree, with their own unique interpretation in the phylogenetic modelling context. All this hints strongly at the natural and universal presence of entanglement as a phenomenon which reaches across disciplines. We hope that this broad perspective may in turn furnish new insights of use in physics. 

\mbox{}\\
\vfill
\pagebreak
\tableofcontents
\pagebreak

	%%%%%%%%%%%%%%%%%%%%%%%%%%%%%%%%%%%%%%%%%%%%%%%%%%%%%%%%%%%%%%%%%%
	%\input{secs/IntroOverview.tex}
	%\section{Introduction to phylogenetics: the challenge of ancestry reconstruction}
	%\label{sec:IntroOverview}
%%%%%%%%%%%%%%%%%%%%%%%%%%%%%%%%%%%%%%%%%%%%%%%%%%%%%%%
\section{Introduction to phylogenetics: the challenge of recovering evolutionary history}
\label{sec:IntroOverview}
The intellectual road to modern phylogenetics has been a long and gradual journey, with progress along it intimately bound to our developing understanding of the biological world itself. Its current evolutionary focus subsumes the older study of the taxonomy of living systems, epitomized by the classic work of Linnaeus, and also owes much to the development of an ecological awareness of nature, furnished by groundbreaking insights from earlier founding figures such as von Humboldt, as well as Darwin himself. Mendel's seminal 
work on inheritance and variability, done around the same period as that of Darwin and Wallace but ignored for four decades before its belated rediscovery, provided the crucial underpinnings for the synthesis and universal acceptance of the evolutionary paradigm into biology. 

The first modern `phylogenetic tree' drawn within an evolutionary framework was famously Darwin's notebook sketch, \\
% SEE http://phylonetworks.blogspot.com.au/2012/06/charles-darwins-unpublished-tree.html
\begin{center}
\includegraphics[width=3cm]{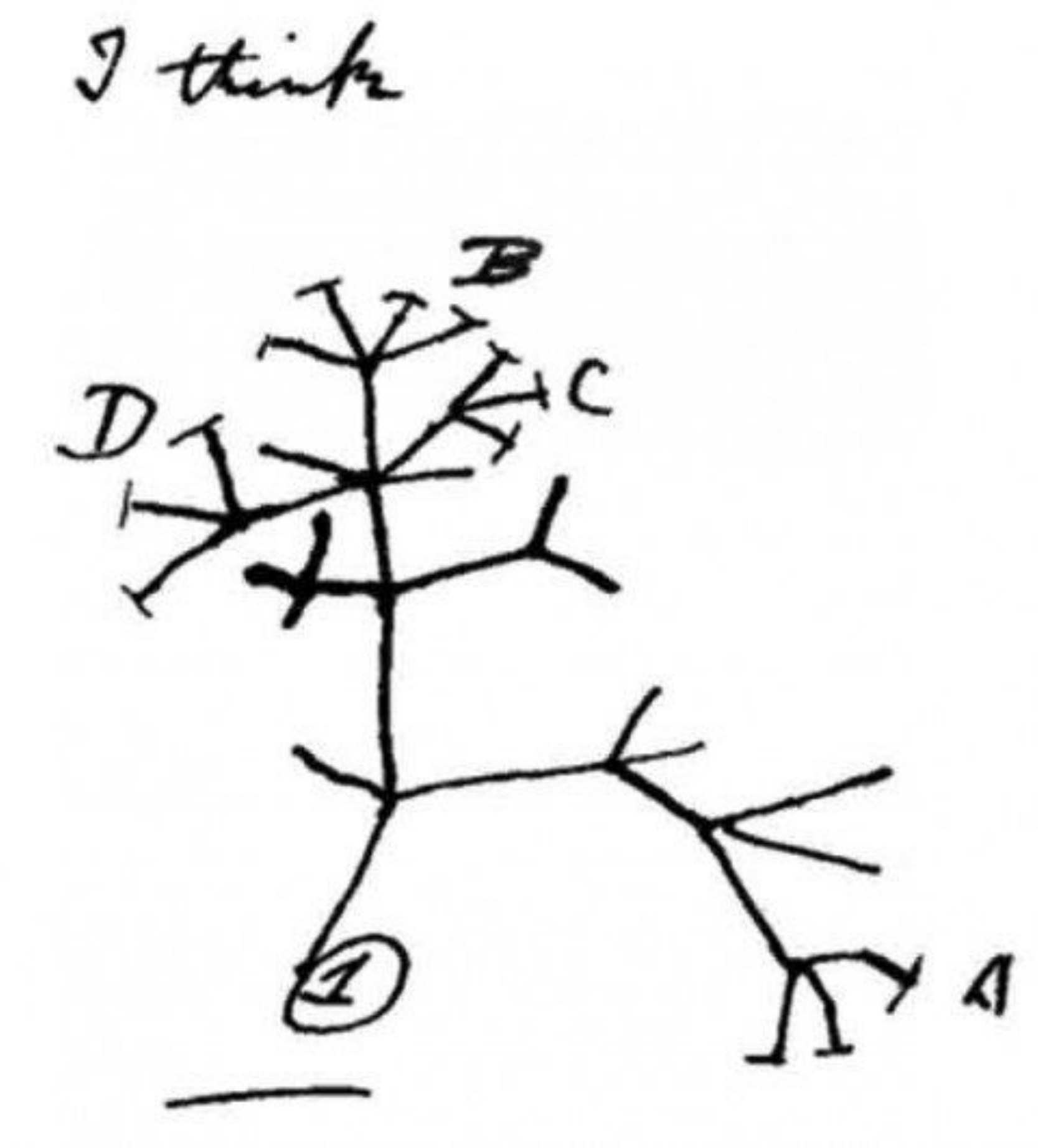}
\end{center}
setting out his ideas of species evolution, which he later elaborated into his grand thesis, \emph{On the Origin of Species}. Judging from the accompanying text \cite{Darwin1837}, this astonishing feat of insight already entails much of the `fine print' of modern, quantitative phylogenetics: the mathematical structure of a tree, with an attributed (assumed, but unknown) `common ancestor' as root, and a direction away from the root; a major branch or evolutionary line, with side off-shoots, not all of which appear to be extant, or available to observation; several nodes with more than 2 out edges, suggesting either the need for further observations, or perhaps points of rapid radiation; as well as variable edge lengths, suggesting a relative time scale for events -- all within a species-level description.
It is only in the past three decades or so, especially with the advent of molecular data, that what might be called a `standard model' of phylogenetics has emerged. From this perspective the depth of the above insights is reinforced --
as shall be seen in more detail below, all of the implied qualifications, such as missing branches, unresolved vertices, and variable-length edges, turn out to be challenging questions beyond the standard model.

The aim of this review is to present in some detail, the formulation and systematic structure of the above-mentioned standard model of theoretical molecular phylogenetics, to an audience familiar with physics, and not necessarily schooled in biological fundamentals. The reader who absorbs the main points should come to appreciate the significance of some of the above details, in the context of the setting and problematics of the modelling. Here however we return to some further broad-brush remarks, in order to develop the subject in context (a summary of the detailed contents of our review material appears at the end of this section).

Attempts at systematic analysis of Darwin's intuition without exploiting the molecular context, concern the description of observed species\footnote{A term applicable to phylogenetic methods in more generic contexts is `taxonomic units' (see below).} by certain attributed `characters' or `traits'.
These provided a quantitative score based on morphological, or possibly also heritable, attributes, typically of a coarse-grained nature, for example binary (present/absent, large/small), or perhaps ternary (positive, negative, absent), and so on.
The assembly of a phylogenetic tree can then proceed under the assumption of `maximum parsimony', representing an assumption of evolutionary conservatism, in that the ancestral
features of common species' progenitors should manifest as few as possible character changes between themselves and their descendants.

In the last half century or more however, the era of molecular biology has engendered sweeping changes to 
the study of the taxonomy and phylogeny of living systems. Both in terms of information storage and retrieval
(genetic data), and in terms of structural features including morphology and behaviour (phenotypic data), the fabric of 
taxonomy and phylogeny has now become the quantitative parametrization of the molecular sequences
involved -- from genes, or generally DNA or RNA (that is, strings whose individual subunits are nucleotides\footnote{The monomer units comprising heterocylic carbon-nitrogen rings with attached sugar and phosphate groups, linked via phosphodiester bonds to form the nucleic acid polymers. Named for the two types of purine, $\texttt{R}$, rings (adenosine \texttt{A}, guanine \texttt{G}), and the two types of
pyrimidine, \texttt{Y}, rings (thymine \texttt{T}, cytosine \texttt{C}, with uracil \texttt{U} instead of \texttt{T} occurring in RNA).} ), and from protein structure (that is, strings comprised of amino acids\footnote{Amino-carboxyl molecules of the form $\mathrm{H}_2\mathrm{N\,CHR\,COOH}$
polymerized by linking peptide bonds. The 20 amino acids occurring in proteins are named for the residue molecule $\mathrm{R}$, and referred to by letter, for example 
arginine $R$, histidine $H$, and further $K,D,E,S,T,N,Q,C,G,P,A,V,I,L,M,F,Y,W$.}). What allows molecular phylogenetic inference to become a rigorous discipline, however, is the coupling of the sequence-level description to the so-called Kimura neutral theory of evolution \cite{kimura1983neutral} which asserts that variability in the molecular sequences themselves
occurs as a result of population-level fixing of random mutations, which in first approximation, are selectively neutral and have no effect on their biological function.  This is indeed consistent with the famous insight of Schr\"{o}dinger \cite{schrodinger1944lifeDE} that the informational content of the macromolecules of life is aperiodically distributed, and hence robust to such mutations. 

A presentation of the standard dogma of molecular biology which underpins the above assertions is
beyond the scope of this review. Although not needed in the sequel, it is helpful if the reader has some basic acquaintance with the structure and machinery of information storage, transcription and translation at the molecular level, and we refer
to standard texts for further details \cite{krebs2017lewin}.
Rather, we focus here on the narrower setting of the phylogenetic modelling itself.
We assume therefore that the molecular sequences, used as inputs to phylogenetic analysis, and representing
various `taxonomic units' -- be they species, genes, geographically defined cohorts, or some other aggregations --
albeit taken from extant individuals, validly sample distributions from the whole population. Further, pre-processing 
of this data is taken to have been infallibly carried out, to produce a so-called molecular alignment, so that like can be compared with like in terms of molecular changes.  All of these steps are contestible in practice, and indeed inform some of the core questions about the limitations of phylogenetic analysis, so presciently glimpsed by Darwin himself.

We comment briefly on just two of these aspects. Firstly, it will be evident that the manipulation of data 
leading to the sequence alignment, to be used as input into phylogenetic analysis, is itself subject to both theoretical and 
experimental error. Only species or taxonomic entities can be used, where functional and other considerations 
can identify sequences which are comparable -- for example, coding for orthologous genes
or control regions in DNA and RNA, or for identifiably analogous structural regions, in the case of protein structure and amino acid sequence data. The very nature of alignment, and the necessity to arrive at equal length sequences for comparison purposes, forces insertions or deletions of subunits within the input strings to be overlooked or removed, thus omitting a whole class of
random changes which themselves could be argued to be valid parts of the neutral evolution hypothesis \cite{saurabh2012gaps}, and restricting focus to single subunit (nucleotide or amino acid) substitutions. Whether the observed features in each aligned string are indeed substitutions valid for the entire taxonomic entity, or are merely mutations, whose appearance is an artefact of incomplete 
population sampling, is yet a further point of contention\footnote{Indeed, in recent years the study of single nucleotide polymorphisms 
\emph{within} the human genome has had a profound influence on medicine, not to mention forensic science.}. Furthermore, it will become clear from 
the assumptions underlying the probabilistic modelling that the data also should ideally derive from sampling sequences of infinite length -- again, a limit which is manifestly never realized in practice. Beyond all of the foregoing is a central question for
evolution itself: is the process correctly thought of as a bifurcating tree, or are biological mechanisms at play whereby `speciation' can be accompanied occasionally by reticulation such that the overall flow is along a network rather than a tree?

The otherwise sound theoretical framework notwithstanding, the above serious caveats form the backdrop of modern molecular phylogenetics methodologies. These were first advocated in the works of Felsenstein and others (references given below), and have since 
provided the framework for immense progress, perhaps best symbolised in the global `Tree of Life' project -- a modern-day
effort of Manhattan proportions, to build an open-access, interacting and searchable representation of all 10 million or so species of life on earth.

\begin{figure}[tbp]
\label{fig:YangsDataSet}
%\hskip1ex\includegraphics[width=18.5cm]{pics/Yang96T1.pdf}\\
\hskip1ex\includegraphics[width=18.5cm]{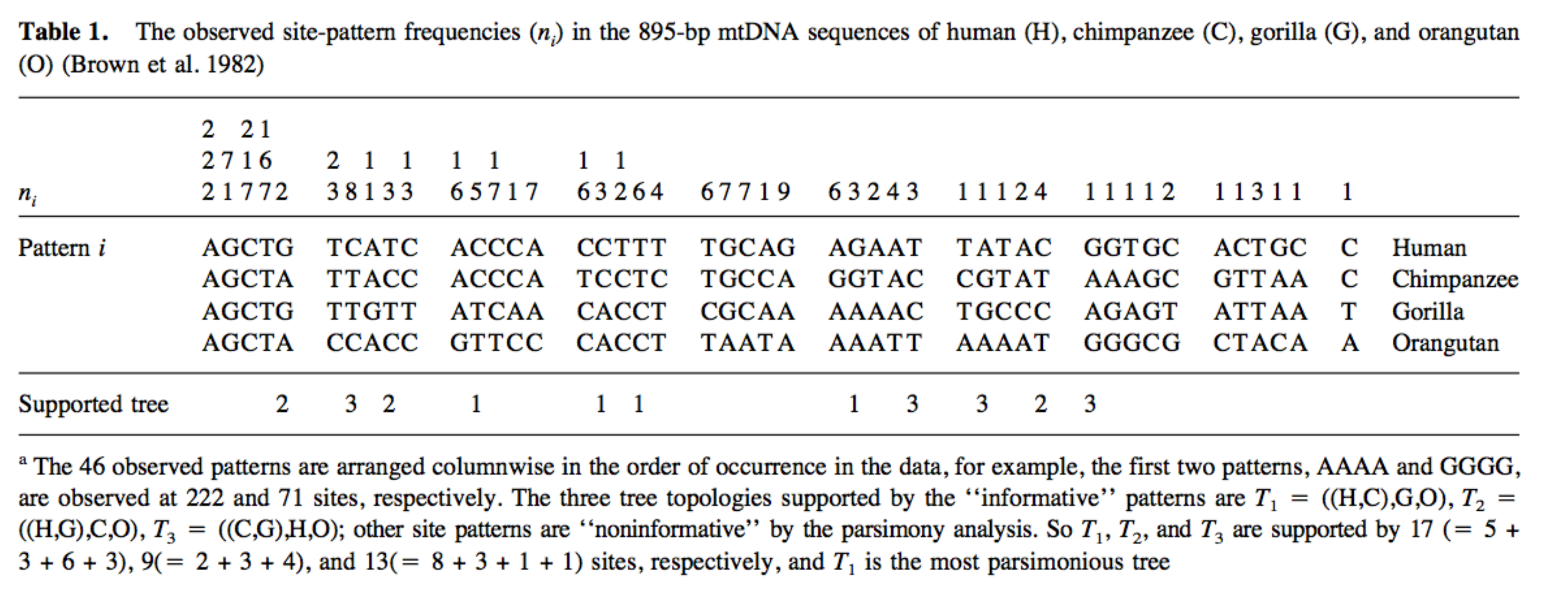}\\\mbox{}
%\hskip0ex\includegraphics[width=18.5cm]{pics/Yang96T2.pdf}
\hskip0ex\includegraphics[width=18.5cm]{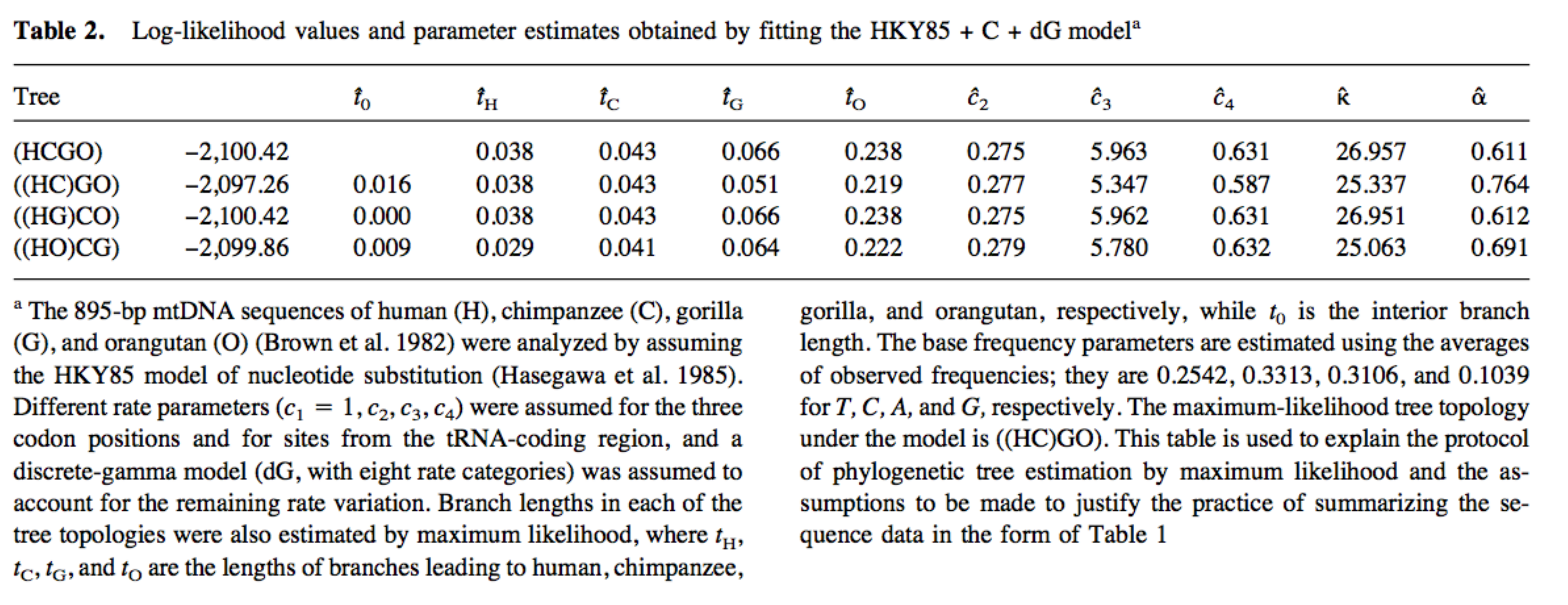}
\caption{A phylogenetic dataset from Yang, 1996 \cite{Yang1996} on human evolutionary origins (reproduced by permission of the author).
Table 1, site frequency patterns and parsimony; table 2, likelihood analysis.}
\end{figure}

To end this introduction, we provide an experimental dataset, to which the methods 
to be presented in the remainder of this review can apply (figure \ref{fig:YangsDataSet}, tables 1 \& 2 from Yang, 1996 \cite{Yang1996}). 
The paper is an analysis of a 
phylogenetic alignment of a 895-nucleotide gene (coding for a mitochondrial tRNA) in 4 primate (great ape) species, and serves to illustrate selection and pre-processing of input sequence data. In this case, necessarily, such a basic functional gene must be present across all species so there is no ambiguity in the selection of homologues. Nor, given billions of years of evolution and stabilisation in ancestral lineages far more basal than the primates, is there need for significant adjustment for alignment purposes. Population sampling issues should furthermore be negligible -- allelic variations 
can be assumed to have been fixed in favour of dominant types within each species. Finally, while the relatively modest sequence length certainly departs from the ideal of infinite length referred to above\footnote{Alignments of thousands of base pairs are not uncommon.}, but at the level of inference, merely introduces issues of finite size sampling variance 
which are statistically controllable.

This dataset provides the alignment of the four 895 nucleotide sequence for the particular mitochondrial tRNA
gene in question for each species\footnote{Differently from functional genes for enzymes for example, which encode proteins,
this gene encodes RNA, so that the translation process is absent. Indicative perhaps of the deep origins of the associated molecular mechanisms in the evolution of life itself, this detail does not however affect the phylogenetic analysis, which is not concerned with the fate of the sequences `downstream'.}. The sequences are not given explicitly, however, but are rather summarized in a table of the numbers of repetitions in each possible 4-way pattern  -- irrespective of order along the sequence in which they occur, these list the number of all $4^4 = 256$ possible combinations of
4 base letters at different sites (clearly, the entries should sum to 895). Normalized to this total, the data thus represents
a multi-way table of pattern frequencies, which under the statistical assumptions of the model, are taken to be a sample from
some multinomial distribution with allocated probabilities for each pattern's occurrence. 

It is the task of phylogenetic modelling to provide theoretical proposal
distributions, against which the data can be assessed. Phylogenetic inference, on the other hand, in all its forms, is concerned with the statistical question of the `goodness of fit', and deducing the `best' underlying theoretical parameters used in the model. As is evident from our overview given above, such inputs will include the combinatorial description of an underlying presumed ancestral tree, thus returning the logic to Darwin's fundamental insight. 

In relation to data handling and computational aspects, modern phylogenetic analyses may bear on broad questions of ecology, phylogeography, conservation management, disease control or even climate change, and typically require species-wide or higher level
aggregations of sequence data, and hence typically entail many dozens or thousands of taxa. While the demands of data processing in such studies are enormous, nonetheless the underlying theoretical modelling is still of the kind that we wish to present here. In this context we believe it all the more important and useful to display and expose in a formal and systematic manner, the basic models and some limitations thereof.
The above-chosen dataset is of course much more modest, as its context is rather
an analysis of competing inference methods, but here serves as an example for illustrative purposes.

As mentioned, this review is solely concerned with theoretical and structural aspects of the modelling, and seeks to explain in detail the model building in the so-called the `general Markov model on trees' which plays the role of a standard formalism for molecular phylogenetics. It will become apparent that the theoretical inputs to which inference must be 
subjected, include not only the discrete data of the tree, but many additional parameters setting rates, time intervals and so on.
These details will be explained incrementally in the subsections to follow. Suffice it to re-emphasize that we are not directly concerned here with statistical questions of inference, which are manifold, and of crucial importance in their own right\footnote{Our chosen sample phylogenetic dataset \cite{Yang1996} in fact forms the subject of a comparative analysis of inference methods, in relation to some questions of deep phylogeny: a favoured tree, under the traditional assumption of `parsimony', is compared with a computation of `maximum likelihood' (see Fig. \ref{fig:YangsDataSet}, Table 2) under a specific parametric Markov model belonging to one of the classes which we describe in the subsequent sections.}. That said, we are nonetheless guided by statistical considerations, particularly the bias-variance tradeoff in any statistical estimation. Phylogenetics modelling is replete with choices adopted to respect the need to limit unnecessary parameter explosion\footnote{An obvious case in point is that in many instances, the position of a root (common ancestor) is not identifiable, or is fixed by inclusion of an `outgroup'
for analysis purposes, thus effecting a drastic reduction in the search space of admissible trees. In the sequel, for theoretical purposes we disregard this convention, and simply work with rooted trees.}. For details of inference methods used in phylogenetics, we refer to standard texts (see for example \cite{felsenstein2004,semple2003phylogenetics24,steel2016phylogeny}). 

We now lay out the contents for the remainder of this review. Having set out in this brief overview the scope of the subject of
molecular phylogenetics, in \S \ref{sec:ModellingIntro} below we provide a mathematical scaffolding for 
the theoretical context of phylogenetic model building, as used to generate parametrized, probabilistic descriptions of 
pattern frequencies for molecular sequence data for the purpose of inference. Namely, we adapt the language of multilinear tensor algebra to the stochastic context, and moreover for the purpose of analysis, we allow our objects to take complex values (the justification of this choice will be elaborated in detail in subsequent sections). The building of a `phylogenetic tree', \S \ref{subsec:ModellingGMM}, then becomes an algorithm for constructing a certain `phylogenetic tensor'
by starting with a tree represented as a graph, and interpreting the labels with which it is decorated. A hierarchical framework for identifying Markov substitution models via the imposition of Lie algebras on the rate matrices, called the `Lie-Markov models', is 
analyzed (\S\S \ref{subsubsec:Catalogue}, \ref{subsubsec:LieMarkovClass}).This leads on (in the case of nucleotide data, with four states) to a number of examples 
of popular substitution models, and these in turn (\S\S \ref{subsubsec:HadamardFourier},\ref{subsubsec:Diagonalization}) yield illustrative case studies of how basis transformations (including discrete Fourier
transforms) can in principle `invert' the model (that is, give the direct means to read off the tree and branch lengths from the data or simple transformations on it). These methods also illustrate the difficulty of using network descriptions instead of trees
(\S\S \ref{subsubsec:Coproducts}, \ref{subsubsec:StarLemma}).
In all of these considerations, model closure is vital, and in \S \ref{subsubsec:ModelClosure} we are able to relate this concept directly to matrix Lie algebras. 
Our use of multilinear algebra and the transformation properties of phylogenetic tensors lends itself to the methods of classical invariant theory, and in \S \ref{sec:Entanglement} we review the construction of various useful relative invariants for the action of the group of invertible stochastic transformations. In the context of tensor analysis, these so-called Markov invariants are none other than types of entanglement measures whose unitary equivalents have long been of interest in quantum information. The appendices, \S\S \ref{subsec:CyclicBasis}, 
\ref{subsec:BCHapproximants}, \ref{subsec:GroupCharacters},  provide technical details on permutation groups,
matrix products, and group characters and representations.

The review concludes in \S \ref{sec:Conclusions} with further discussion from our work, and beyond this, also gives a lightning overview of  a bewildering variety of alternative characterizations of phylogenetic trees and their mathematical foundations available in the literature,  
whose richness itself speaks to the depth and current interest in the subject at hand.
%The review concludes in \S \ref{sec:Conclusions} with further discussion from our work, and beyond this, also gives a lightning introduction to  sources in the literature of a bewildering variety of alternative characterizations of phylogenetic trees and their mathematical foundations 
%-- from abstractions such as Frobenius algebras and the language of quantum information, conformal field theory and operads, to toric algebraic geometry, through to advocacy for a fundational shift to the language and formalism of `stochastic dynamics', and its structure, symmetries and conservation laws. 
%\vfill

	%%%%%%%%%%%%%%%%%%%%%%%%%%%%%%%%%%%%%%%%%%%%%%%%%%%%%%%
	%\input{secs/ModellingIntro.tex}
	%\section{Molecular phylogenetic modelling and multilinear tensor analysis}
	%\label{sec:ModellingIntro}
\section{Molecular phylogenetic modelling and tensor analysis}
\label{sec:ModellingIntro}
In the following we present an approach to formulating theoretical models for the multi-way alignment scores which are the raw material of phylogenetic analysis, as outlined in \S \ref{sec:IntroOverview} above. Given the presumed stochastic nature of the biological events underlying the alignment data, the rigorous theoretical setting is in terms of generalizations of Markov chains, to Markov processes on trees, and possibly networks. In such graphical models, the vertices (or nodes) represent extant or ancestral taxa, to which are assigned random variables, evaluated in the appropriate character sample space. Edges represent transition probabilities, and the usual Markov assumption of dependence only the prior state, becomes independence across nodes, except for dependence on the prior state of a common source node. In order to arrive at a joint probability distribution for pendant leaf nodes (possibly even including a root), which is the outcome of the stochastic model to be compared with the alignment data, a sum must be taken over all intermediate states at the unobserved, internal nodes, conferring on the overall model the status of a hidden Markov process.

In order to bring the language and algebraic tools of multilinear tensor calculus to bear on the phylogenetic models, we adopt the language of finite dimensional complex linear spaces to handle
probability vectors, whose components will be proxies for distributions, and we assume these to be subject to complex linear transformations, implemented by certain nonsingular matrices with constraints which ensure probability conservation for parameters in the correct domain. 

In the next subsection we present, in algebraic form, the most general probabilistic model which is assumed to underly the stochastic data represented by molecular phylogenetic alignment arrays -- the so-called general Markov model. In subsequent sections, we take up specific details of how various evolutionary change models are classified and parametrized, from the point of view of symmetry aspects (either motivated from biological and statistical considerations, or simply for completeness).
This leads on to an investigation of the role of a certain class of Lie algebras underlying the models, and to a discusson of subalgebras and embeddings. At the same time, specific cases 
avail themselves of transform techniques (related to discrete Fourier transforms, referred to as Hadamard transforms in the phylogenetics literature). These have the property of being compatible with the algebraic structure on the whole tree, leading to the possibility of subjecting the entire alignment dataset to a coordinate change from which all theoretical parameters (that is, the tree adjacency array and its edge lengths, as well as the parameters describing evolution on the edges) can be read off directly. A final subsection is devoted to generalizations of (and obstructions to) the transform idea, for more general models on trees and networks. 

As a disclaimer to the technical details, it should be borne in mind that the following material is also covered in standard texts on molecular phylogenetics (see for example \cite{felsenstein2004,semple2003phylogenetics24,steel2016phylogeny}). As mentioned above, our presentation here serves to establish the notation for the perspective of tensor algebra and group representations which we adopt and apply in subsequent sections of this review.

	%%%%%%%%%%%%%%%%%%%%%%%%%%%%%%%%%%%%%%%%%%%%%%%%%%%%%%%
	%\input{secs/ModellingGMM.tex}
	%\subsection{The general Markov model on trees}
	%\label{subsec:ModellingGMM}
\subsection{The general Markov model on trees}
\label{subsec:ModellingGMM}
For characters that can assume $K$ different types, we consider probabilities as vectors (states) in ${\mathbb C}^K$ with basis\footnote{In the probabilistic language, the $e_i$ can be considered as point measures $\delta_i$ on the discrete sample space.} of unit vectors $\{e_i\,, i=1,2,\cdots, K\}$\,, of the form $p=  \left.\sum\right._{i=1}^K p^i e_i\,$\, with nonnegative coefficients. They undergo stochastic transformations via $K \times K$ matrices $M^i{}_j$\,, which satisfy $0\le M^i{}_j \le 1$ and
\begin{align*}
\left.\sum\right._{i=1}^K  M^i{}_j = 1\,, \quad j = 1,2,\cdots, K\,,
\end{align*}
such that with $p$ transforming as 
\begin{align}
\label{eq:VectorStateTransform}
p \rightarrow p' = &\, M \cdot p\,, \qquad p'{}^i= \left.\sum\right._{j=1}^K M^i{}_j p^j\,, \qquad 
\end{align}
we have probability conservation,
\[
\left.\sum\right._{i=1}^K p^i = 1 = \left.\sum\right._{i=1}^K p'{}^i\,.
\]
Concretely, we can picture the $e_i$ as unit column vectors, and $p$ as general column vectors, respectively. 
%Thus the $K \!\times \!K$ matrices $M^i{}_j$ are constrained to have unit column sums, which of course ensures that 
%\[
%\left.\sum\right._{i=1}^K p^i = 1 = \left.\sum\right._{i=1}^K p'{}^i\,.
%\]
Correspondingly, we will consider \emph{multi-way} arrays $P$ to be elements of the appropriate tensor product spaces 
${\mathbb C}^K\otimes {\mathbb C}^K$, ${\mathbb C}^K\otimes {\mathbb C}^K\otimes {\mathbb C}^K$, $\cdots$, of the form
\[
P = \left.\sum\right._{i,j=1}^K P^{ij} e_i\otimes e_j\,,\qquad P = \left.\sum\right._{i,j,k=1}^K P^{ijk} e_i\otimes e_j\otimes e_k\,,\qquad \cdots\,, \quad \mbox{and so on.}
\]
Such tensors, in turn, are subject to stochastic transformations
\begin{align}
\label{eq:TensorStateTransform}
P \rightarrow P' = \big( M_1\otimes M_2 \otimes \cdots \big) \cdot P
\end{align}
with the distinct stochastic matrices $M_1$, $M_2$, $\cdots$, specified in number by the arity of the tensor array $P$. At this stage, a vital aspect of the modelling is immediately apparent: such arrays are to describe a number of subsystems subject to \emph{independent}
stochastic variation\footnote{A formal description would invoke random variables which are conditionally independent (see below for 
further details).}, precisely reflecting the presence of the independent stochastic matrices within the tensor product.

The analogue of the unit probability condition for an $L$-way tensor array is the total marginalization condition
\[
\sum_{i_1 i_2 \cdots i_L} P^{i_1 i_2 \cdots i_L} =1
\]
summing over all components. Naturally, there exist also marginalizations of intermediate arity: in the absence of extra conditions, there are for example $L$ different $L\!-\!1$-way sub-tensors, corresponding to different choices of single marginalization, of the form
\[
P^{i_1 i_2 \cdots i_{r\!-\!1} i_{r\!+\!1}\cdots i_L}_{(r)}= \sum_i P^{i_1 i_2 \cdots i_{r\!-\!1} i i_{r\!+\!1} \cdots i_L}\,,
\]
together with $L\!-\!2$-way tensors of $\textstyle{\frac 12}L(L\!-\!1)$ different types, and so on.

The device of working over ${\mathbb C}$ rather than ${\mathbb R}$ is of course a standard step enabling 
various intended algebraic operations to be fully implemented (see below); the parametrizations required for applications can always be recovered by specialization. This applies both to probability vectors (non-negative real numbers), and to the linear transformations which they undergo, corresponding to the effect of stochastic changes. % (in the standard basis, non-negative real off-diagonal entries and unit column sums). 
Below we shall discuss in some detail the technical restriction to nonsingular matrices, and also establish the parametric
restrictions (within ${\mathbb R}$) which obtain for the corresponding matrix elements. It suffices here to note \cite{johnson1985,mourad2004} that in the complex regime, the set of $K \!\times \!K$ nonsingular, unit column-sum matrices (which we denote $GL_1(K)$) forms a matrix Lie group equivalent to the $K\!-\!1$ dimensional \emph{complex affine group}, $GL_1(K)\cong A\!f\!f(K\!-\!1)$\,. For simplicity, in the sequel  we shall refer to the corresponding complex matrix group as the Markov, or sometimes as the stochastic, group\footnote{An equivalent characterization of $GL_1(K)$\,, is as nonsingular linear transformations (of ${\mathbb C}^K$) which fix a nontrivial linear form. Note that, the stochastic matrices as strictly defined above (see equation (\ref{eq:VectorStateTransform})) in fact form a matrix \emph{semigroup}.}.

While the linear setting technically admits the operation on probability vectors and tensors of simple addition (rather than convex combination), which has no direct meaning in the biological sense, its great utility will be seen in the key role played by the use of certain changes of basis for these objects. These typically will arise from systematic features of the interrelationships between the character values themselves, specific to the application at hand. By contrast, the unified manner in which the action of the requisite linear transformations on them can be handled, confers a natural representation-theoretical setting on the formalism, where the structural group is the complex affine group. The consequences of obtaining basis-independent characterizations of such transformations yields a complementary perspective, leading to powerful insights into phylogenetic inference coming from classical invariant theory, as we shall show below. 

Before proceeding, we give some additional notation and conventions. Firstly, we note a default set of choices of bases which are always available, and have the virtue of exposing the nature of the transformation group. Alongside the standard or \emph{natural} basis $\{ e_i\,, i=1,2,\cdots, K \}$\,, we introduce an \emph{affine} basis
$\{ \widetilde{e}_i\,, i=1,2,\cdots, K \}$\, as follows. This is generated by any linearly independent set of new basis elements, whose \emph{dual}  must include the mandatory uniform sum over all elements in the standard basis. Concretely, this is any choice of new row vectors, which must include the all-ones row vector from the standard basis. If the standard unit row vectors\footnote{As row vectors, the $f^i$ are the transposes $f^i = (e_i){}^\top$\,.}  form the set $\{ f^i, i=1,2,\cdots, K\}$, then we define $\widetilde{f}^K = f^1 + f^2 + \cdots + f^K$ and extend to the full set  $\{ \widetilde{f}^i, i=1,2,\cdots, K\}$ as follows. For some parameters $x^a{}_b$\,, $\eta_a$\,, $a,b = 1,2,\cdots, K\!-\!1$\,, define
\begin{align*}
\widetilde{f}^a = &\, \left.\sum\right._{b=1}^{K\!-\!1}x^a{}_b f^b + \eta^a f^K\,; \qquad
\widetilde{f}^K =  \left.\sum\right._{a=1}^{K\!-\!1} f^a\,+ f^K\,. 
%\widetilde{e}_a = &\, \left.\sum\right._{b=1}^{K\!-\!1}e_b y^b{}_a  +  e_K \xi_a \,; \quad a= 1,2,\cdots,K\!-\!1\,; \qquad
%\widetilde{e}_K =  \lambda e_K + \left.\sum\right._{a=1}^{K\!-\!1} e_a \zeta^a\,.
\end{align*}
Conversely, for some parameters $y^a{}_b$\, $\xi^a$\,, $\zeta^a$\,, $a,b = 1,2,\cdots, K\!-\!1$\,, $\lambda$\,, consider the general basis transform
\begin{align*}
%\widetilde{f}^a = &\, \left.\sum\right._{b=1}^{K\!-\!1}x^a{}_b f^b + \eta^a f^K\,; \quad a= 1,2,\cdots,K\!-\!1\,; \qquad
%\widetilde{f}^K =  \left.\sum\right._{a=1}^{K\!-\!1} f^a\,+ f^K\,; \\
\widetilde{e}_a = &\, \left.\sum\right._{b=1}^{K\!-\!1}e_b y^b{}_a  +  e_K \xi_a \,, \quad a= 1,2,\cdots,K\!-\!1\,; \qquad
\widetilde{e}_K =  \lambda e_K + \left.\sum\right._{a=1}^{K\!-\!1} e_a \zeta^a\,,
\end{align*}
such that the dual elements have the above form including the crucial $\widetilde{f}^K$ as defined.
In concrete terms, the above redefinitions simply amount to a particular similarity transform, to which probability vectors, and in general tensor arrays, are subjected. We summarize the change of coordinates as follows.\\

\noindent
\begin{quotation}
\noindent
\textbf{Coordinates for probability tensors and Markov matrices in an affine\footnote{So named because the stochastic transformations have the form of affine maps (see below). The \emph{natural} basis might alternatively be referred to as the \emph{biological} basis,
in contrast to the \emph{affine} or (because of the evident simplification of the stochastic transformations) \emph{computational} basis.} basis:}\\
Write
$e_i = \sum_{j=1}^K \widetilde{e}_j X^j{}_i$, $\widetilde{f}^i = \sum_{j=1}^K(X^{-1})^i{}_j f^k$, and $p = \sum_{i=1}^K p^ie_i = \sum_{i=1}^K \widetilde{p}{}^i\widetilde{e}_i$, and again display the components in the matrix form (ordering rows and columns $i,j = 1,2,\cdots,K$
with $a,b = 1,2,\cdots, K\!-\!1$\,) as
\begin{align*}
M = &\,  \left[\begin{array}{cc} m^a{}_b & \ell^a \\ \lambda_b &  \mu \end{array}\right]\,,\qquad 
X = \left[\begin{array}{cc} y^a{}_b & \zeta^a\\ \xi_b & c  \end{array}\right]\,,\qquad 
X^{-1} = \left[\begin{array}{cc}  x^a{}_b & \eta^a \\ 1 \cdots 1 & 1 \end{array}\right]\,. 
\end{align*}
Components in a generic affine basis have a simplified block form 
\begin{align*}
\widetilde{p} =&\, X^{-1}p\,,\qquad \left[\begin{array}{c} \widetilde{p}^a \\ \widetilde{p}^K\end{array}\right] \equiv \left[\begin{array}{c} \widetilde{p}^a \\ 1 \end{array}\right]\,;\qquad \\
\widetilde{M} = &\, X^{-1}MX =  \left[\begin{array}{cc} \widetilde{m}^a{}_b & \widetilde{\ell}^a \\ 0 \cdots 0 &  1 \end{array}\right]\,.
\end{align*}
In these coordinates the transformation law under stochastic changes is
\begin{align*}
\widetilde{p}\rightarrow &\,  \widetilde{p}\,' = \widetilde{M}\,\widetilde{p} \equiv 
\left[\begin{array}{c} \sum_b\widetilde{m}^a{}_b\,\widetilde{p}^b + \widetilde{\ell}^a \\[.1cm] 1\end{array}\right]\,.
%of $p$, $M$, $X$ and $X^{-1}$ 
\end{align*}
%\mbox{}\hfill $\blacksquare$\\
\mbox{}\\[-.5cm]
\mbox{}\hfill $\Box$\\
\noindent
\textbf{Corollary: doubly stochastic Markov matrices in the affine basis:}\\
For $M$ doubly stochastic, choose  $X^{-1}$ such that the off-diagonal components are $\eta^a:= -\sum_i x^a{}_i$\,.
Then $\widetilde{M}$ has block diagonal form,
\[
\widetilde{M} = \left[\begin{array}{cc} \widetilde{m}^a{}_b & 0 \\ 0 \cdots 0 &  1 \end{array}\right]\,.
\]
%\mbox{}\hfill $\blacksquare$\\
\mbox{}\\[-.5cm]\mbox{}\hfill $\Box$
\end{quotation}

Clearly, in an affine basis, we achieve the desired simplification that the probability vectors possess an invariant component
$\widetilde{p}{}^K$, and correspondingly the complementary subspace undergoes (inhomogeneous) general  linear transformations, characterizing the $K\!-\!1$-dimensional affine group (see below). For later use, we find it convenient to re-label this component as $0$
rather than $K$, and adopt the index convention $a,b,c = 1,2, \cdots, K\!-\!1$, while in concrete matrix calculations retaining the index ordering $1,2, \cdots, K\!-\!1,0 $\,. By the same token, for a general $L$-way phylogenetic tensor with components 
$P^{i_1 i_2 \cdots i_L}$ in the natural basis, the marginalizations referred to above, correspond to the choice of components in an affine basis with one or more indices set as `$0$', giving, in principle, $L$ different $L\!-\!1$-way subtensors of the form
\[
\widetilde{P}_{(r)}^{a_1 a_2 \cdots a_{L\!-\!1}}:= P^{a_1 a_2 \cdots a_{r\!-\!1} 0 a_{r\!+\!1} \cdots a_{L\!-\!1}}\,,
\]
together with $\textstyle{\frac 12}L(L\!-\!1)$ different $L\!-\!2$-way tensors, and so on, leaving finally
\[
\widetilde{P}^{0 0 \cdots 0} \equiv 1
\]
reflecting the unit probability mass for the distribution being described\footnote{Formally, the probability mass for a single vector, and the various partial marginalizations in the tensor case, signify the presence of invariant subspaces under the stochastic transformation group.}.

As a variation on this theme, note the above coordinate transform for the class of \emph{doubly stochastic} transformations, taken up in the above as the statement of the corollary (and needed subsequently). These are matrices $M$ for which both row and column sums are unity. At one level, this may be seen as a specialization within the hierarchy of available models (about which we shall have much more to say in the sequel); more abstractly, such a restriction may arise if there is further structure on state space, such as an inner product allowing a canonical identification with its dual. Given the extra constraints on $M$, however, we can exploit the rather arbitrary nature of the affine basis\footnote{A simple choice is for example $x = {\mathbb I}_{K\!-\!1}$, and then $\eta^a:= -1$.} whereby the choice of the off-diagonal components $\eta^a:= -\sum_i x^a{}_i$ within $X^{-1}$ above, results in $\widetilde{\ell}^a\equiv 0$ in $\widetilde{M}$\,. Clearly, the affine transformations in the new coordinates are no longer inhomogeneous, but instead, equivalent to general linear transformations in dimension $K\!-\!1$.

The second technical matter is the introduction of linear operators between tensor spaces which will be a crucial underpinning of our formulation of the theoretical models. The basic object on the model space is the \emph{splitting operator}\footnote{In the algebraic context to be developed (see \S \ref{subsubsec:Coproducts}), $\delta$ is identified as a \emph{comultiplication}.}  $\delta$, and its extensions to higher tensor products gives the hierarchy $\delta^{(r)}{}_{(\ell)}$\,.\\

\noindent
\begin{quotation}
\noindent
\textbf{Phylogenetic splitting operators $\delta$, $\delta^{(r)}_{(\ell)}$:}\\
Given the model space $V \cong {\mathbb C}^K$, we simply define the linear operator
$\delta: {\mathbb C}^K \rightarrow {\mathbb C}^K\otimes {\mathbb C}^K$ \emph{with respect to the natural basis} by its action
\[
\delta(e_i) = e_i \otimes e_i\,, \qquad i=1,2,\cdots, K\,.
\] 
and extending to all of ${\mathbb C}^K\otimes {\mathbb C}^K$ by linearity. More generally
given $\ell$ copies $\otimes^\ell V$, and position
$1\le r\le \ell$, we define the operator $\delta^{(r)}_{(\ell)}: \otimes^\ell V \rightarrow \otimes^{(\ell\!+\!1)}V$\,, by the insertion of $\delta$ at position $r$ in the tensor product:
\begin{align*}
%\delta^{(r)} = &\, \stackrel{r\!-\!1}{\overbrace{{\sf Id} \otimes  {\sf Id} \otimes \cdots \otimes  {\sf Id}}} \otimes\, \delta \, \otimes 
%\stackrel{\ell\!-\!r\!+\!1}{\overbrace{{\sf Id}\otimes  {\sf Id}\otimes \cdots \otimes  {\sf Id}}}\,,\\
\delta^{(r)}_{(\ell)}  =\,  \stackrel{r\!-\!1}{\overbrace{{\sf Id} \otimes  {\sf Id} \otimes \cdots \otimes  {\sf Id}}} \otimes\, &\, \delta \, \otimes 
\stackrel{\ell\!-\!r\!}{\overbrace{{\sf Id}\otimes  {\sf Id}\otimes \cdots \otimes  {\sf Id}}}\,,\\
\delta^{(r)}_{(\ell)}\big(e_{i_1}\otimes e_{i_2}\otimes \cdots \otimes  e_{i_r} \otimes \cdots e_{i_\ell}\big)
=&\, e_{i_1}\otimes e_{i_2}\otimes \cdots \otimes  e_{i_r} \otimes  e_{i_r} \otimes \cdots e_{i_\ell}\,.
\end{align*}
\mbox{}\\[-.5cm]
\mbox{}\hfill $\Box$\\
\end{quotation}

\noindent
We are now in a position to develop in this formalism the structure of the main theoretical model underlying molecular phylogenetics. 
Its aim is to provide a mathematical scenario whereby a given set of phylogenetic array data, corresponding to some molecular alignment (as explained in the preliminary remarks above, \S \ref{sec:ModellingIntro}), can be fitted to a parametrized model for the purpose of 
statistical inference and recovery of parameters of biological interest (see also the introductory discussion, \S \ref{sec:IntroOverview}). The statistical inference methods themselves can of course vary, from 
deterministic ways of directly inverting or transforming the data, through to optimization routines such as Bayesian or maximum likelihood algorithms. Which methods and statistical analyses are applicable depends on the specific context, including the nature of the biological setting and the quality of the data, and a detailed discussion is beyond the purpose of this review. However, one of our main aims is to 
discuss model classes where a deeper systematic analysis reveals various robust theoretical tools which 
can help to overcome statistical issues.

Returning to the key parameters, these primarily include the pattern of ancestral relationships 
and speciation events which best explain the data at hand: namely, an evolutionary history in the form of an ancestral tree, assumed to start at some common ancestor or root, and to proceed by branching (ideally, bifurcation) down to the extant species set represented by the leaves of the tree. In terms of the multiway phylogenetic array, it will be apparent that species (leaf nodes) from more recently separated edges should contain more correlations in their corresponding content than more distantly related ones, and it is the task of the theoretical model to quantify this. Numerical parameters which provide further explanatory detail are `lengths', or `distances', equivalent to evolutionary divergence times from the root, to the observed species.

Consider an $L$-part alignment describing a phylogenetic dataset for $L$ species or `taxonomic units'. To this experimentally sampled entity we associate a theoretical phylogenetic tensor belonging to the tensor product of $L$ copies of the
fundamental model space $V \cong {\mathbb C}^K$, constructed as follows.
We hypothesize an underlying, planted, binary tree ${\mathcal T}$, with an assigned root node and its edge (the common ancestor), and $2L\!-\!1$ edges: $L\!+\!1$ external (including the root edge, and the $L$ pendant leaf edges terminated by leaves), and $L\!-\!2$ internal edges. Without loss of generality we adopt an equivalent description whereby the root is the unique, degree 2 node (effectively, the corresponding edge has vanishing length), so that ${\mathcal T}$ has $2(L\!-\!1)$ edges, including $L$ leaf edges and $L\!-\!2$ internal edges. 

\begin{figure}[htbp]
   \centering
%\hskip-2ex \rotatebox{-1.5}{\includegraphics[height=5cm]{pics/6LeafTree.jpg}}\hskip-2ex
%\rotatebox{-1}{\includegraphics[height=5cm]{pics/6LeafPells.jpg}}   
%%% FULLTREE -- NEW
\mbox{}\\
\scalebox{.9}{\begin{tikzpicture} 
%1. \draw [help lines] (0,0) grid (7,8); 
3. \shade[fill, top color=gray!20, bottom color=gray!0] (-.3,0) rectangle (7.3,1);
1. \draw[thick] (-.3,0) -- (7.3,0); 
1. \node[right] at (7.3,0) {$6\quad P^{(6)}={\mathbb I}^3\!\otimes\! M_5\!\otimes\! M_5'\!\otimes\! {\mathbb I}\!\circ\! {\sf Id}^3\!\otimes\!\delta\!\otimes\! {\sf Id} \!\circ\! P^{(5)}$};
3. \shade[fill, top color=gray!20, bottom color=gray!0] (-.3,1) rectangle (7.3,2);
1. \draw[thick] (-.3,1) -- (7.3,1); 
1. \node[left] at (-.3,1) {$r_5=4$};
1. \node[right] at (7.3,1) {$5\quad P^{(5)}={\mathbb I}^3\!\otimes\! M_4\!\otimes\! M_4'\!\circ\! {\sf Id}^3\!\otimes\!\delta \!\circ\! P^{(4)}$};

3. \shade[fill, top color=gray!20, bottom color=gray!0] (-.3,2) rectangle (7.3,3);
1. \draw[thick] (-.3,2) -- (7.3,2); 
1. \node[left] at (-.3,2) {$r_4=4$};
1. \node[right] at (7.3,2) {$4\quad P^{(4)}={\mathbb I}\!\otimes\! M_3\!\otimes\! M_3'\!\otimes\! {\mathbb I}\!\circ\! {\sf Id}\!\otimes\!\delta\!\otimes\! {\sf Id} \!\circ\! P^{(3)}$};
3. \shade[fill, top color=gray!20, bottom color=gray!0] (-.3,3) rectangle (7.3,4);
1. \draw[thick] (-.3,3) -- (7.3,3); 
1. \node[left] at (-.3,3) {$r_3=2$};
1. \node[right] at (7.3,3) {$3\quad P^{(3)}=M_2\!\otimes\! M_2'\!\otimes\! {\mathbb I}\!\circ\! \delta\!\otimes\! {\sf Id} \!\circ\! P^{(2)}$};
3. \shade[fill, top color=gray!20, bottom color=gray!0] (-.3,4) rectangle (7.3,7);
1. \draw[thick] (-.3,4) -- (7.3,4); 
1. \node[left] at (-.3,4) {$r_2=1$};
1. \node[right] at (7.3,4) {$2\quad P^{(2)}=M_1\!\otimes\! M_1'\!\circ\! \delta\!\circ\! P^{(1)}$};
3. \shade[fill, top color=gray!20, bottom color=gray!0] (-.3,7) rectangle (7.3,8);
1. \draw[thick] (-.3,7) -- (7.3,7); 
1. \node[left] at (-.3,7) {$r_1=1$};
1. \node[right] at (7.3,7) {$1\quad P^{(1)}=\pi$};
3. \shade[fill, top color=gray!20, bottom color=gray!0] (-.3,0) rectangle (7.3,1);
%1. \draw[thick] (-.3,8) -- (7.3,8); 
1. \node[right] at (7.3,8) {$0$};
2. \draw[dashed,ultra thick] (3.5,7) to  (3.5,8); 
2. \draw[ultra thick] (0,0) --  (3.5,7) -- (7,0);
4. \draw[fill, color=white] (0,0) circle (.1);
4. \draw (0,0) circle (.1);
4. \draw[fill, color=white] (7,0) circle (.1);
4. \draw (7,0) circle (.1);
2. \draw[ultra thick] (2,4) --  (4,0);
4. \draw[fill, color=white] (4,0) circle (.1);
4. \draw (4,0) circle (.1);
4. \draw[fill, color=white] (2,4) circle (.1);
4. \draw (2,4) circle (.1);
 2. \draw[ultra thick] (2.5,3) --  (1,0);
4. \draw[fill, color=white] (2.5,3) circle (.1);
4. \draw (2.5,3) circle (.1);
4. \draw[fill, color=white] (1,0) circle (.1);
4. \draw (1,0) circle (.1);
2. \draw[ultra thick] (6,2) --  (5,0); 
4. \draw[fill, color=white] (6,2) circle (.1);
4. \draw (6,2) circle (.1);
4. \draw[fill, color=white] (5,0) circle (.1);
4. \draw (5,0) circle (.1);
2. \draw[ultra thick] (5.5,1) --  (6,0);
4. \draw[fill, color=white] (6,0) circle (.1);
4. \draw (6,0) circle (.1);
4. \draw[fill, color=white] (5.5,1) circle (.1);
4. \draw (5.5,1) circle (.1);
 %
%4. \draw[fill, color=white] (3.5,7) circle (.1);
4. \draw[fill, color=white] (3.5,7) circle (.1);
4. \draw (3.5,7) circle (.1);
\end{tikzpicture}}\mbox{}\\
\caption{\small{Illustrating the construction of the phylogenetic tensor $P_{\mathcal T}$, for the case of a 6-leaf tree ($L=6$),
the tree presentation with internal node positions $(r_1,r_2,r_3, r_4,r_5)= (1,1,2,4,4)$. The
$2(L\!-\!1) = 10$ edges ($L=6$ leaf edges and $L\!-\!2=4$ internal) are decorated with $(L\!-\!1)=5$ pairs of stochastic matrices
$\{M_1\, M_1'\,; M_1\, M_1'\,;\cdots; M_5, M_5'\}$. Without loss of generality, a root node (at level 0) is replaced by an equivalent, unique valence 2 root node (at level 1), assigned an (initial) phylogenetic vector $\pi \in V$. The nodes at level 6 are the leaves. See the text for the recursive construction of $P_{\mathcal T}$.}}
   \label{fig:6LeafTree}
\end{figure}
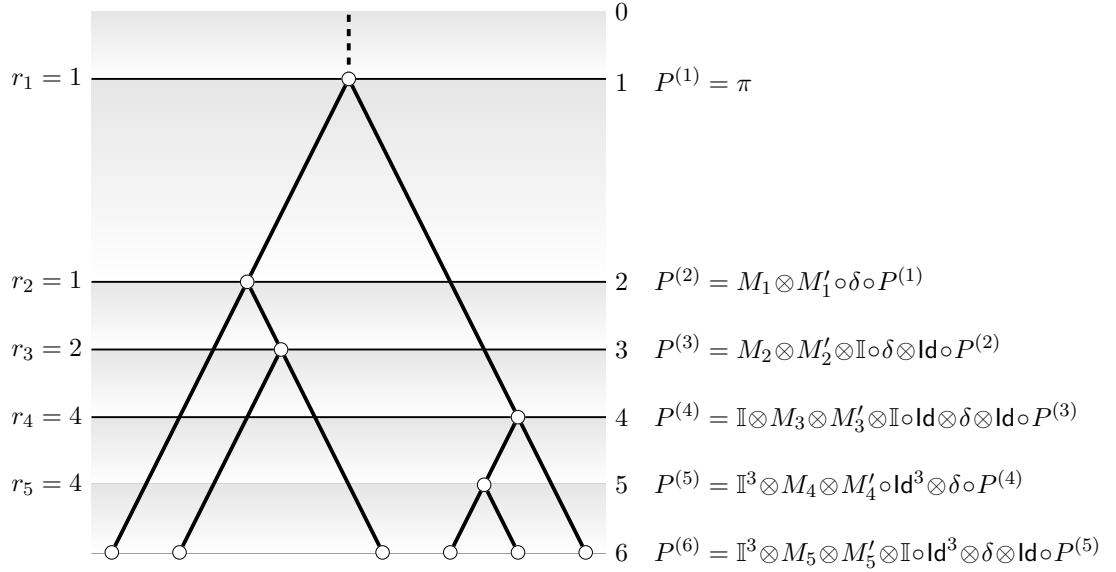

The following abstract description is illustrated in a concrete case by figure \ref{fig:6LeafTree}\,, where specific values
are taken for the chosen of leaves and related enumerative data. See the caption for further details. We shall return to this example below in making some further comments on the notation and formalism.
Let there be given a graphical presentation of the tree ${\mathcal T}$ whereby, at each depth $\ell = 1,2, \cdots, L\!-\!1$, there is only one internal node (of valence 3). Depth $L$ corresponds of course to the $L$ valence 1 leaf nodes.  Draw $L$ horizontals such that the $\ell$'th
horizontal strip corresponding to the half-open vertical interval $[\ell, \ell+1)$ -- occupying depths $h$ with $ \ell \le h < \ell+1 $ -- includes the aforesaid $\ell$'th horizontal line, and the part of the graph occupied by the edges up to (but not intersecting) the horizontal at depth $\ell \!+\!1$. Given this presentation, the structure of the tree is encoded by the sequence
$\{ r_\ell\,, 1\le r_\ell \le \ell, \ell = 1,2,\cdots, L\!-\!1 \}$\, of depths at which horizontal $\ell$ intersects the unique internal branching node at depth $\ell$. The tree data for the model is completed by assigning $2(L\!-\!1)$ Markov matrices $\{ M_\ell\,, M'_\ell\,\,,  \ell =1,2,\cdots, (L\!-\!1)\}$. For the theoretical model of the associated probability tensor based on this tree, $P_{\mathcal T} \in \otimes^L V$, 
we proceed as follows:\\

%\noindent
\begin{quotation}
\noindent
\textbf{Phylogenetic tree tensor construction:}\\
Construct  $P_{\mathcal T} \in \otimes^L V$ recursively as the composition of strip operators $S_{\mathcal T}^\ell\,, \ell = 1,2,\cdots, L\!-\!1$\,, $S^\ell: \otimes^\ell V \rightarrow \otimes^{(\ell\!+\!1)}V$\,.
In turn, each $S_{\mathcal T}^\ell$ is a composition $S_{\mathcal T}^\ell = \Delta_{\mathcal T}^\ell \circ E_{\mathcal T}^\ell$ of a branching part, 
$\Delta_{\mathcal T}^\ell: \otimes^\ell V \rightarrow \otimes^{(\ell\!+\!1)}V$\, comprising the $\delta$ splitting operator insertion at the correct position, together with an edge operator $E_{\mathcal T}^\ell: \otimes^{(\ell\!+\!1)}V \rightarrow \otimes^{(\ell\!+\!1)}V$\, involving the correct insertion of the pair of stochastic transformations $\{ M_\ell\,, M'_\ell\,\}$ on the out edges of the just-split internal node. Explicitly we have the construction
\begin{align}
\label{eq:AbstractTree}
P_{\mathcal T} = &\, \Big(\overleftarrow{\prod}_{\ell=1}^{(L\!-\!1)} S_{\mathcal T}^\ell \Big)\circ \pi\,,\qquad S_{\mathcal T}^\ell := E_{\mathcal T}^\ell  \circ\Delta_{\mathcal T}^\ell 
\,,   \\
\mbox{where}\qquad  \Delta_{\mathcal T}^{(\ell)} = \delta^{(r_\ell)}_{(\ell)}=&\,  \stackrel{r\!-\!1}{\overbrace{{\sf Id} \otimes  {\sf Id} \otimes \cdots \otimes  {\sf Id}}} \otimes\,  \delta \, \otimes 
\stackrel{\ell\!-\!r\!}{\overbrace{{\sf Id}\otimes  {\sf Id}\otimes \cdots \otimes  {\sf Id}}}\,, \\
\mbox{and}\qquad E_{\mathcal T}^{(\ell)} = &\, \stackrel{r\!-\!1}{\overbrace{{\mathbb I} \otimes  {\mathbb I} \otimes \cdots \otimes  {\mathbb I}}} \otimes\,  
 M_\ell \otimes M_{\ell'} \otimes \, 
 \otimes \stackrel{\ell\!-\!r\!}{\overbrace{{\mathbb I}\otimes  {\mathbb I}\otimes \cdots \otimes  {\mathbb I}}}\,.
\end{align}
In recursive form, a sequence of tensors $P_{\mathcal T}^{(\ell)} \in \otimes^\ell V$\,, $\ell = 1,2,\cdots, L$ is built up where  
\begin{align*}
P_{\mathcal T}^{(1)} \equiv \,  \pi\,, \qquad 
P_{\mathcal T}^{(2)} =  &\,  S_{\mathcal T}^{(1)} \circ \pi\,, \quad  
{\cdots} \quad P_{\mathcal T}^{(\ell\!+\!1)} = S_{\mathcal T}^{(\ell)} \circ P_{\mathcal T}^{(\ell)}\,; \quad  \ell = 1, 2,\cdots, L\!-\!1\,,
%\mbox{with} \qquad P_{\mathcal T} \equiv &\, P_{\mathcal T}^{(L)}\,.
\end{align*}
yielding the final phylogenetic tensor for this tree and model setting choice, $P_{\mathcal T} \equiv  P_{\mathcal T}^{(L)}$\,.\\
\mbox{}\hfill $\Box$
\end{quotation}
%\mbox{}\hfill $\Box$

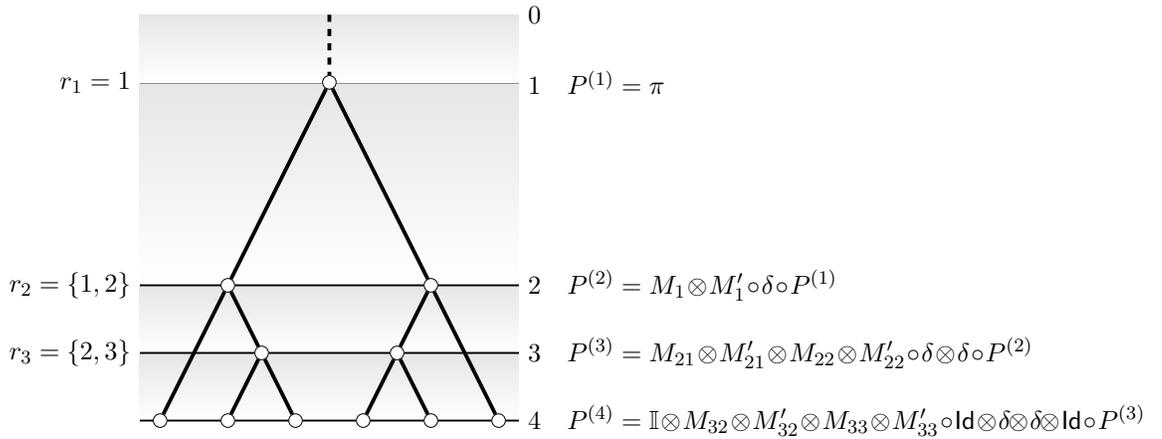
\begin{figure}[htbp]
   \centering
%{\includegraphics[height=7cm]{pics/6LeafTreeAlt.jpg}}   
\mbox{}\\
\scalebox{.9}{\begin{tikzpicture} 
%1. \draw [help lines] (0,0) grid (7,8); 
3. \shade[fill, top color=gray!20, bottom color=gray!0] (-.3,0) rectangle (5.3,1);
1. \draw[thick] (-.3,0) -- (5.3,0);
\node[right] at (5.3,0) 
{$4\quad P^{(4)}={\mathbb I}\!\otimes\! M_{32}\!\otimes\! M_{32}'\!\otimes\!M_{33}\!\otimes\! M_{33}' 
\!\circ\! {\sf Id}\!\otimes\!\delta\!\!\otimes\!\delta\!\!\otimes\!{\sf Id}\!\circ\!P^{(3)}
$}; 
%1. \node[right] at (7.3,0) {$6\quad P^{(6)}={\mathbb I}^3\!\otimes\! M_5\!\otimes\! M_5'\!\otimes\! {\mathbb I}^3\!\circ\! {\sf Id}\!\otimes\!\delta\!\otimes\! {\sf Id} \!\circ\! P^{(5)}$};
%
3. \shade[fill, top color=gray!20, bottom color=gray!0] (-.3,1) rectangle (5.3,2);
1. \draw[thick] (-.3,1) -- (5.3,1); 
\node[left] at (-.3,1) {$r_3=\{2,3\}$};
\node[right] at (5.3,1) {$3\quad P^{(3)}=
M_{21}\!\otimes\! M_{21}'\!\otimes\!M_{22}\!\otimes\! M_{22}'\!\circ\!\delta\!\otimes\!\delta\!\circ\!P^{(2)}$};
3. \shade[fill, top color=gray!20, bottom color=gray!0] (-.3,2) rectangle (5.3,5);
1. \draw[thick] (-.3,2) -- (5.3,2); 
1. \draw[thick] (-.3,5) -- (5.3,5); 
\node[left] at (-.3,2) {$r_2=\{1,2\}$};
\node[right] at (5.3,2) {$2\quad P^{(2)}=M_{1}\!\otimes\! M_{1}'\!\circ\!\delta\!\circ\!P^{(1)}$};
3. \shade[fill, top color=gray!20, bottom color=gray!0] (-.3,5) rectangle (5.3,6);
%1. \draw[thick] (-.3,6) -- (5.3,6); 
1. \node[left] at (-.3,5) {$r_1=1$};
1. \node[right] at (5.3,5) {$1\quad {P}{}^{(1)}=\pi$};
1. \node[right] at (5.3,6) {$0$};
2. \draw[dashed,ultra thick] (2.5,6) to  (2.5,5); 
2. \draw[ultra thick] (0,0) --  (2.5,5) -- (5,0);
4. \draw[fill, color=white] (0,0) circle (.1);
4. \draw (0,0) circle (.1);
4. \draw[fill, color=white] (5,0) circle (.1);
4. \draw (5,0) circle (.1);
2. \draw[ultra thick] (1,2) --  (2,0);
4. \draw[fill, color=white] (2,0) circle (.1);
4. \draw (2,0) circle (.1);
4. \draw[fill, color=white] (1,2) circle (.1);
4. \draw (1,2) circle (.1);
 2. \draw[ultra thick] (4,2) --  (3,0);
4. \draw[fill, color=white] (4,2) circle (.1);
4. \draw (4,2) circle (.1);
4. \draw[fill, color=white] (3,0) circle (.1);
4. \draw (3,0) circle (.1);
2. \draw[ultra thick] (1.5,1) --  (1,0); 
4. \draw[fill, color=white] (1.5,1) circle (.1);
4. \draw (1.5,1) circle (.1);
4. \draw[fill, color=white] (1,0) circle (.1);
4. \draw (1,0) circle (.1);
2. \draw[ultra thick] (3.5,1) --  (4,0);
4. \draw[fill, color=white] (4,0) circle (.1);
4. \draw (4,0) circle (.1);
4. \draw[fill, color=white] (3.5,1) circle (.1);
4. \draw (3.5,1) circle (.1);
4. \draw[fill, color=white] (2.5,5) circle (.1);
4. \draw (2.5,5) circle (.1);
\hskip10ex \end{tikzpicture} 
}
\mbox{}\\
\caption{\small{An tree equivalent to that of figure \ref{fig:6LeafTree}, allowing more than one internal node at each depth. In this case the depth is 4, and the coding is $(r_1=1,r_2=\{1,2\},r_3=\{2,3\})$\,. At depths 2 and 3, the two required splitting operators $\delta$ and out-edge evolution insertions $M \otimes M'$ can be combined (see text).}}
   \label{fig:Depth3Tree}
\end{figure}

We close this development with some further information about the tree tensor construction. 
We keep the discussion informal by simply stating various important properties and illustrating our remarks 
with reference to specific examples (figures \ref{fig:6LeafTree}, \ref{fig:Depth3Tree} and \ref{fig:TruncTree} ).

Firstly, it is clear that the same tree (regarded formally via its vertex and edge sets and their incidence relations, with labelled leaves)
can be associated with a multitude of different `height' functions (assignments of depth levels to internal nodes from the root down to the leaves). For the canoncial construction we have insisted on only one node per level. A given tree may, depending on its structure,
admit several such assignments (an exception being a `caterpillar tree', with one leaf edge emanating from the root, and all other leaf edges branching from the partner of this edge from the root). For example, for the 6-leaf tree illustrated, the depth function
$(r_1,r_2,r_3, r_4,r_5)= (1,2,1,3,2)$\, is also possible. Degenerate assignments are also admissible, such as that illustrated 
in figure \ref{fig:Depth3Tree}, where only 4 levels (depths) appear, whose branching node coordinates now become sets,
$(r_1,r_2,r_3) = ( 1,\{1,2\},\{2,3\})$\,. The associated splitting and edge operators at each level $\ell$ now consist of products of
the identity operator at each crossing entering the $\ell$'th strip which does not mark an internal node, together at the marked nodes, with insertion of the appropriately located splitting operator, or alternatively the pair of assigned substitution matrices $M\otimes M'$\,. 
These operators can themselves be regarded as the product of corresponding non-degenerate cases obtained by a slight tilting of the tree relative to the depth rulings. Indeed, an anticlockwise rotation of the tree of figure \ref{fig:Depth3Tree} to remove the degeneracy, will recover the crossings $(r_1,r_2,r_3, r_4,r_5)= (1,2,1,3,2)$\, of the alternative, non-degenerate presentation. The point is that in the degenerate case, the components of the edge and splitting contributions in each strip commute as operators -- they are always separated across the tensor product, such that they act on different subspaces therein\footnote{A formal discussion of these points (which is beyond the scope of the present review) would of course require adeqate notions of the equivalence of phylogenetic trees, and establishing the correct behaviour of the phylogenetic tensor under appropriate graph isomorphisms. Note finally that for modelling purposes, the non-identifiability of the root entrains a further set of equivalences between various phylogenetic tensors.}. We shall have more to say about such patterns of edge and splitting operators in relation to the possibility of phylogenetic `network' models, in \S \ref{subsec:TreesNetworks} below.

An important variant on the phylogenetic tensor $P_{\mathcal T}$ which we shall require in the sequel, is the 
associated \emph{truncated} or \emph{clipped} tensor $\overline{P}_{\mathcal T}$\,, figure \ref{fig:TruncTree}. Informally, this is the tensor of rank $L$ built using the standard construction, but \emph{as if there had been no evolution on pendant edges}\footnote{Relative to the central problem of recovering the evolutionary tree,
modelling stochastic events on the pendant edges interposes `nuisance'
parameters which only obscure the underlying common ancestry. In the
methods of invariant theory which we develop in \S \ref{sec:Entanglement} below,
it is the clipped tensor which will play a pivotal role.}\,. Given that the substitution matrices are all invertible, this tensor $\overline{P}_{\mathcal T}$ could thus be obtained from the `fully evolved' ${P}_{\mathcal T}$ simply by acting with the tensor product of the respective \emph{inverse} leaf edge substitution matrices $M^{-1}$\,. Alternatively, 
$\overline{P}_{\mathcal T}$ is the tensor constructed by `holding' all edge evolution undergone by non-cherry leaves (that is, leaves whose edges arise higher in the tree) until a final joint substitution process on all leaves including cherries recovers the full tree\footnote{The `left-to-right' ordering of substitution matrices
$1,2,\cdots, L$ is by abuse of notation a (relabelled) subset of the full set $\{ M_1\, M_1'\,; M_2\, M_2'\,;\cdots; M_{L\!-\!1}, M_{L\!-\!1}'\}$ including internal edges used in the iterative construction. For example it is clear that $M_4'$ (figure \ref{fig:6LeafTree}) is re-labelled as $M_6$ in the construction of equation (\ref{eq:PfromClipP}) involving the clipped tensor (see figure \ref{fig:TruncTree} and caption).}:
\begin{align}
\label{eq:PfromClipP}
{P}_{\mathcal T}=&\, M_1 \otimes M_2 \otimes \cdots \otimes M_L \cdot \overline{P}_{\mathcal T}\,.
\end{align}
Formally, $\overline{P}_{\mathcal T}$ is built using the iterative algorithm described above, by taking the 
full list of $2L-2$ of edge substitution matrices $M_{\ell}, M_{\ell}'$ at each level, and replacing the $M$'s corresponding to leaf edges by the identity matrix ${\mathbb I}$\,. The effect of this on the clipped tensor $\overline{P}_{\mathcal T}$ so obtained, is naturally that a `final' epoch of evolution across all leaf edges is necessary in order to recover the full ${P}_{\mathcal T}$\,. In figure \ref{fig:TruncTree} this is indicated by the 
thinner lines used for these terminal (non-evolved) edges. 
\begin{figure}[htbp]
   \centering
%{\includegraphics[height=7cm]{pics/6LeafTreeTrunc.jpg}} 
\mbox{}\\
%%% TRUNCTREE -- NEW WITH THIN LINES FOR LEAF EDGES
\scalebox{.9}{\begin{tikzpicture} 
%1. \draw [help lines] (0,0) grid (7,8); 
3. \shade[fill, top color=gray!20, bottom color=gray!0] (-.3,0) rectangle (7.3,1);
1. \draw[thick] (-.3,0) -- (7.3,0); 
%1. \node[right] at (7.3,0) {$6\quad\overline{P}{}^{(6)}={\mathbb I}^3\!\otimes\! {\mathbb I}\!\otimes\!{\mathbb I}\!\otimes\! {\mathbb I}\!\circ\! {\sf Id}\!\otimes\!\delta\!\otimes\! {\sf Id} \!\circ\! \overline{P}{}^{(5)}$};
1. \node[right] at (7.3,0) {$6\quad\overline{P}{}^{(6)}={\mathbb I}^6\!\circ\! {\sf Id}^3\!\otimes\!\delta\!\otimes\! {\sf Id} \!\circ\! \overline{P}{}^{(5)}$};
3. \shade[fill, top color=gray!20, bottom color=gray!0] (-.3,1) rectangle (7.3,2);
1. \draw[thick] (-.3,1) -- (7.3,1); 
1. \node[left] at (-.3,1) {$r_5=4$};
1. \node[right] at (7.3,1) {$5\quad\overline{P}{}^{(5)}={\mathbb I}^3\!\otimes\! M_4\!\otimes\! {\mathbb I}\!\circ\! {\sf Id}^3\!\otimes\!\delta \!\circ\! \overline{P}{}^{(4)}$};

3. \shade[fill, top color=gray!20, bottom color=gray!0] (-.3,2) rectangle (7.3,3);
1. \draw[thick] (-.3,2) -- (7.3,2); 
1. \node[left] at (-.3,2) {$r_4=4$};
1. \node[right] at (7.3,2) {$4\quad\overline{P}{}^{(4)}={\mathbb I}\!\otimes\! {\mathbb I}\!\otimes\! {\mathbb I}\!\otimes\! {\mathbb I}\!\circ\! {\sf Id}\!\otimes\!\delta\!\otimes\! {\sf Id} \!\circ\! \overline{P}{}^{(3)}$};
3. \shade[fill, top color=gray!20, bottom color=gray!0] (-.3,3) rectangle (7.3,4);
1. \draw[thick] (-.3,3) -- (7.3,3); 
1. \node[left] at (-.3,3) {$r_3=2$};
1. \node[right] at (7.3,3) {$3\quad\overline{P}{}^{(3)}={\mathbb I}\!\otimes\! M_2'\!\otimes\! {\mathbb I}\!\circ\! \delta\!\otimes\! {\sf Id} \!\circ\! \overline{P}{}^{(2)}$};
3. \shade[fill, top color=gray!20, bottom color=gray!0] (-.3,4) rectangle (7.3,7);
1. \draw[thick] (-.3,4) -- (7.3,4); 
%\draw[thick] (2,4) --  (2.65,2.7);
1. \node[left] at (-.3,4) {$r_2=1$};
1. \node[right] at (7.3,4) {$2\quad\overline{P}{}^{(2)}=M_1\!\otimes\! M_1'\!\circ\! \delta\!\circ\! \overline{P}{}^{(1)}$};
3. \shade[fill, top color=gray!20, bottom color=gray!0] (-.3,7) rectangle (7.3,8);
1. \draw[thick] (-.3,7) -- (7.3,7); 
1. \node[left] at (-.3,7) {$r_1=1$};
1. \node[right] at (7.3,7) {$1\quad\overline{P}{}^{(1)}=\pi$};
3. \shade[fill, top color=gray!20, bottom color=gray!0] (-.3,0) rectangle (7.3,1);
1. \node[right] at (7.3,8) {$0$};
2. \draw[dashed,ultra thick] (3.5,7) to  (3.5,8); 
2. \draw[ultra thick] (1.85,3.7) --  (3.5,7) -- (6.15,1.7);
2. \draw[ultra thick] (2,4) --  (2.65,2.7);
 2. \draw[thick] (1,0) --  (2.35,2.7); 
4. \draw[fill, color=white] (2,4) circle (.1);
4. \draw (2,4) circle (.1);
 2. \draw[ultra thick] (2.5,3) --  (2.35,2.7);
 %%%
 2. \draw[thick] (1,0) --  (2.35,2.7); 
 2. \draw[thick] (0,0) --  (1.85,3.7);
 2. \draw[thick] (4,0) --  (2.65,2.7);
2. \draw[thick] (5,0) --  (5.35,0.7);
2. \draw[thick] (6,0) --  (5.65,0.7);
2. \draw[thick] (7,0) --  (6.15,1.7);
%%%
4. \draw[fill, color=white] (2.5,3) circle (.1);
4. \draw (2.5,3) circle (.1);
2. \draw[ultra thick] (6,2) --  (5.35,0.7); 
4. \draw[fill, color=white] (6,2) circle (.1);
4. \draw (6,2) circle (.1);
2. \draw[ultra thick] (5.5,1) --  (5.65,0.7);
4. \draw[fill, color=white] (5.5,1) circle (.1);
4. \draw (5.5,1) circle (.1);
 %
%4. \draw[fill, color=white] (3.5,7) circle (.1);
4. \draw[fill, color=white] (3.5,7) circle (.1);
4. \draw (3.5,7) circle (.1);
%1. \draw [help lines] (0,0) grid (7,8); 
\draw[fill, color=white] (0,0) circle (.1);
4. \draw (0,0) circle (.1);
\draw[fill, color=white] (1,0) circle (.1);
4. \draw (1,0) circle (.1);
\draw[fill, color=white] (4,0) circle (.1);
4. \draw (4,0) circle (.1);
\draw[fill, color=white] (5,0) circle (.1);
4. \draw (5,0) circle (.1);
\draw[fill, color=white] (6,0) circle (.1);
4. \draw (6,0) circle (.1);
\draw[fill, color=white] (7,0) circle (.1);
4. \draw (7,0) circle (.1);
\end{tikzpicture}
}
\mbox{}\\  
\caption{\small{Illustrating the construction of the truncated phylogenetic tensor $\overline{P}_{\mathcal T}$, corresponding to the case of the 6-leaf tree of figure \ref{fig:6LeafTree}. In the notation used in that example, we have
${P}_{\mathcal T}=M_2 \!\otimes\! M_3 \!\otimes\! M_3'\!\otimes\! M_5 \!\otimes\! M_5'\!\otimes\! M_4'\!\circ \!\overline{P}_{\mathcal T}$\,.}}
   \label{fig:TruncTree}
\end{figure}

	%\input{secs/LieMarkov.tex}
	%\subsection{Continuous time models: Lie-Markov classification.}
	%\label{subsec:LieMarkov}
\subsection{Continuous time models and the Lie-Markov classification.}
\label{subsec:LieMarkov}

Until this point, we have assumed that the evolutionary changes leading to candidate theoretical phylogenetic frequency arrays
are attributed to Markov matrices $M$, acting on probability vectors and tensors. Recall that these are generically complex,
and, for the purposes of our analysis, have been declared to be nonsingular matrices (of size $K\!\times\!K$\,, in the case of $K$ character types). In the biological context, of course, the matrices must be stochastic with non-negative real entries (in the standard basis), so that they correctly describe changes to probabilities attained by some assumed random variable; the column sum condition of course takes care of conservation of probability. If it also be assumed that the random variable belongs to a continuous time process, and moreover that at a particular time the appropriate Kolmogorov smoothness equations are satisfied, then there is an affiliated matrix generator $Q$,
called the rate matrix, for the continuous time Markov process, such that $M=\exp Q$\,.
  
In our general analysis, we adopt a continuous time picture,
but we choose to work in the ambient complex group. As already discussed, the complex matrix Lie group, $GL_1(K)$ is equivalent to $A\!f\!f(K\!-\!1)$\,, the affine group in dimension $K\!-\!1$\,.  
We work in the connected component of the identity, covered by the exponential map, so that there is by default a natural rate generator candidate. Moreover, the requisite conditions on such
$Q$, such that $M=\exp Q$ is a bona fide real stochastic matrix with positive entries -- which of course is mandatory for applications -- are that $Q$ should have nonnegative real off-diagonal entries, and negative diagonal entries with overall zero column sum (note that $\exp Q$, defined by the exponential series with powers of $Q$, is well defined)\footnote{Indeed, the zero column sum condition comes from the identity $M =\exp Q = {\mathbb I} + Q\big(\big({\exp Q - {\mathbb I}}\big)/{Q}\big)$\,
(where the fraction stands for the residual terms in the power series expansion, with $Q$ power reduced by 1). Evidently the first term,
${\mathbb I}$\,, ensures that $M$ has the correct unit column sum, given that the $Q$ prefactor confers column sum zero on the second term. On the other hand, if $-q$ is the smallest (largest negative) diagonal matrix entry of $Q$, we have $M = \exp Q = \exp{\big(-q{\mathbb I} + \big(Q+ q{\mathbb I}\big)\big)} = e^{-q}\exp{\big(Q+ q{\mathbb I}\big)}$\,,
the result being (up to a positive scale factor) the exponential of a matrix with all-nonnegative entries.}.

Further examination of the interrelationship between the rate generator and the group matrices of the affine group, and their parametrizations, is given in \cite{johnson1985,mourad2004}\,. For a complete discussion in the specific case of $K=2$\,, see \cite{sumner:2013:lg2x2}\,. The inverse, or so-called embedding problem -- of determining, from given stochastic data, whether a continuous time Markov process is a valid description, and inferring the existence of a rate generator -- is a difficult question in general, and beyond the scope of this review (see for example \cite{davies2010embeddable,Lencastre:Raischel:Rogers:Lind:2016a} and references therein), except insofar as 
explicit computations in tractable cases of some of the models which we treat (in the $K=4$, nucleotide substitution case) can lead to complete answers, along the lines of \cite{sumner:2013:lg2x2}\ for $K=2$. We discuss the question of model closure in more detail in \S
\ref{subsubsec:ModelClosure} below.

%%%%%%%%%%%%%%%%%%
\subsubsection{A catalogue of nucleotide rate models.}
\label{subsubsec:Catalogue}

In molecular phylogenetics within the continuous time formalism, a wide variety of rate models -- that is, specific choices of the parametrization of the matrix $Q$ -- are in common useage, for example as input in standard software packages for phylogenetic reconstruction\footnote{The remainder of the discussion will be centred mainly on the nucleotide case $K=4$, although some constructions will be general.}. These range from the simplest, with as few parameters as possible (in view of the bias-variance tradeoff, useful when data is likely to be noisy, or derives from sequences where substitution processes tend to act on all characters in the same way), to having specific patterns based on biological or biochemical assumptions about how characters interchange differentially under structural molecular constraints; through to intermediate forms chosen for generality, but compromising on computational overheads; through to the most general parametrizations. As we will demonstrate, in the nucleotide case $K=4$,  it is possible to give a powerful model classification, on the basis of very natural symmetry assumptions relating to how chosen rate parameters respond to certain types of permutations which act on the state labels (and hence on the underlying linear space $V \cong {\mathbb C}^4$). Together with the demand of multiplicative closure for algebraic consistency, and with a view to exploiting group actions, this classification gives rise to the so-called Lie-Markov models \cite{sumner2011BF,fernandez2015lie,woodhams:fernandez-sanchez:sumner:2015a} (for a condensed overview, see figure \ref{fig:LieMarkovModelsFlowChart} and its accompanying caption).
Before presenting a formal analysis and a summary of the results, we present and discuss several illustrative examples of 
particular rate models, both known and newly identified, as an introduction to the topic and to serve as a vehicle for 
framing the theory\footnote{In the following selection the rate matrices $Q$, exemplifying different rate model classes
(the sets denoted $\mathfrak Q$ in \S \ref{subsubsec:LieMarkovClass} below), are tagged if possible by explanatory acronyms, as well as by the specific label identifiers from our papers. These are given here merely for completeness; their systematic meaning will be addressed subsequently. For typographical convenience a `$*$' symbol is sometimes used on rate matrix diagonals, 
to stand for the negative sum of entries from the respective column; we avoid this convention where possible. Note finally that in our conventions  a specific matrix element, say $Q_{AG}$, denotes the rate $A \leftarrow G$.}. 

 The \emph{Jukes-Cantor} model \cite{jukes1969} assumes evolution via a constant rate of substitution between different nucleotide bases.
Ordering these\footnote{In this context there should be no confusion with amino acid symbols $C$ (cysteine) and $G$ (glycine).}  as ${A,G,C,T}$ we have the one parameter model
\[
Q^{JC} = Q^{(1.1)}= \left[\begin{array}{cccc} -3\alpha & \alpha &\alpha& \alpha \\ \alpha & -3\alpha &\alpha &\alpha\\
			 \alpha &\alpha & -3\alpha &\alpha \\ \alpha & \alpha&\alpha &-3\alpha \end{array}\right]\,,
\]
useful for its analytical simplicity and avoidance of over-parametrization. 
%An elaboration is the \emph{Kimura three-parameter} model
%\cite{kimura1981estimation},
%\[
%Q^{K3ST} =Q^{3.3a}= \left[\begin{array}{cccc} \!-\!\alpha\!-\!\beta\!-\!\gamma & \alpha &\beta & \gamma \\ 
%\alpha & \!-\!\alpha\!-\!\beta\!-\!\gamma &\gamma& \beta\\
%			 \beta& \gamma & \!-\!\alpha\!-\!\beta\!-\!\gamma &\alpha \\ \gamma& \beta&\alpha & \!-\!\alpha\!-\!\beta\!-\!\gamma\end{array}\right]\,,
%\]
%in which the base substitution rates for transitions, $A \leftrightarrow G$ and $C\leftrightarrow T$ are assigned to value
%$\alpha$, in contrast to the transversions $A \leftrightarrow C$\,, $G \leftrightarrow T$ (rate parameter $\beta$),
%and transversions $A \leftrightarrow T$\,, $G \leftrightarrow C$  (rate parameter $\gamma$)\,.
%(Variations on this theme include, for example, the special case of setting the two transversion rate parameters equal (the \emph{Kimura two-parameter model} \cite{Kimura1980}); or, to distinguish the different pair
%transition rates, but to have a third parameter covering all transversion\footnote{A non-symmetric four-paramater extension is also admitted, where rates $C,T \rightarrow A,G$ are different from $C,T \leftarrow A,G$.} rates $C,T \leftrightarrow A,G$).

A generalization is the \emph{Kimura two-parameter} model (2.2b)
\cite{Kimura1980},
\[
Q^{K2ST} =Q^{(2.2b)}= \left[\begin{array}{cccc} \!-\!\alpha\!-\!2\beta & \alpha &\beta & \beta \\ 
\alpha & \!-\!\alpha\!-\!2\beta &\beta& \beta\\
			 \beta& \beta & \!-\!\alpha\!-\!2\beta &\alpha \\ \beta& \beta&\alpha & \!-\!\alpha\!-\!2\beta\end{array}\right]\,,
\]
in which the base substitution rates for \emph{transitions}, $A \leftrightarrow G$ and $C\leftrightarrow T$ are assigned to value
$\alpha$, in contrast to the \emph{transversions} $\texttt{R} \leftrightarrow \texttt{Y}$ (that is, the substitutions
$A \leftrightarrow C$\,, $G \leftrightarrow T$, and $A \leftrightarrow T$\,, $G \leftrightarrow C$), with rate parameter $\beta$\,.

Elaborations on this theme include, for example, the  \emph{Kimura three-parameter model} (3.3a) \cite{kimura1981estimation}, in which the base substitution rates for transitions, $A \leftrightarrow G$ and $C\leftrightarrow T$ are assigned to value
$\alpha$, with the transversions $A \leftrightarrow C$\,, $G \leftrightarrow T$  assigned rate parameter $\beta$, while the 
transversions $A \leftrightarrow T$\,, $G \leftrightarrow C$  are modelled with a different rate parameter $\gamma$\,.
Conversely, in the \emph{Tamura-Nei equal frequency model} (3.3c) \cite{tamura1993estimation}, the distinct pairs of transition rates are be modelled differentially, but with a third parameter covering all transversion rates $A,G \leftrightarrow C,T$\,. A non-symmetric, four-parameter extension, model (4.4b) is also admitted, where rates $\texttt{R}\rightarrow \texttt{Y}$\,, $A,G \rightarrow C,T$\,, are different from $\texttt{R}\leftarrow \texttt{Y}$\,, $A,G \leftarrow C,T$. We shall return to the specific form of the rate matrices for all these variants in \S \ref{subsubsec:Diagonalization} below.

The above rate models (with the exception of the last, four-parameter variant) have symmetric rate matrices. Consequently, the all-ones left null eigenvector (responsible for probability conservation of the Markov transition matrix) is also a right eigenvector (so that they are doubly stochastic); moreover, these models have the uniform distribution as stationary state -- that is, with each character having probability $\textstyle{\frac 14}$. A commonly used model which allows for data with non-uniform base frequencies, in the most parsimonious parametrization, is the \emph{Felsenstein (1981)} model\footnote{Also known as the equal-input model in the $K$-state case.} 
\cite{felsenstein1981},
\[
Q^{F81}=Q^{(4.4a)} = \left[\begin{array}{cccc} \!-\!\beta\!-\!\gamma\!-\!\delta & \alpha &\alpha& \alpha \\ \beta & \!-\!\alpha\!-\!\gamma\!-\!\delta &\beta &\beta\\
			 \gamma &\gamma & \!-\!\alpha\!-\!\beta\!-\!\delta &\gamma \\ \delta & \delta&\delta &\!-\!\alpha\!-\!\beta\!-\!\gamma \end{array}\right]\,,
\]
which implements four different substitution rates, one for any base to evolve into each of $A, G, C,T$ in turn. The (unique) stationary distribution is constructed from these rates by scaling, as the probability vector
\[
\pi =  \frac{1}{\lambda}\big( {\alpha}{}, {\beta}{}, {\gamma}{}, {\delta}{}\big)^\top\,,
\]
where $\lambda = \alpha + \beta+\gamma+\delta$.

We mention one further specific choice of rate model which is of intermediate complexity, but serves to illustrate the
full power of the Lie-Markov classification: model (5.6b) \cite{fernandez2015lie},
\[
Q^{(5.6b)} = \left[\begin{array}{cccc} \!-\!\alpha\!-\!2\beta\!+\!x & \alpha\!+\!x &\beta\!+\!x& \beta\!+\!x \\ \alpha\!+\!y & \!-\!\alpha\!-\!2\beta\!+\!y & \beta\!+\!y& \beta\!+\!y \\
			 \beta\!+\!z& \beta\!+\!z &\!-\!\alpha\!-\!2\beta\!+\!z &\alpha\!+\!z \\ \beta\!+\!t& \beta\!+\!t&\alpha\!+\!t & \!-\!\alpha\!-\!2\beta\!+\!t \end{array}\right]\,.
\]
This model has 5 parameters, due to the constraint $x+y+z+t=0$\,; the reason for the presentation given in this way will be explained below.

At the other extreme, the so-called \emph{general Markov} model
 \cite{rodriguez1990} assumes \emph{no} such interrelationships between different rates, giving the full, 12 parameter model,
\[
Q^{GM} = \left[\begin{array}{cccc} \alpha_{11} & \alpha_{12} &\alpha_{13}& \alpha_{14} \\ 
\alpha_{21} & \alpha_{22} &\alpha_{23}& \alpha_{24}\\
\alpha_{31} & \alpha_{32} &\alpha_{33}& \alpha_{34}\\ 
\alpha_{41} & \alpha_{42} &\alpha_{43}& \alpha_{44} 
 \end{array}\right]\,,
\]
wherein the diagonal entries
 ensure the vanishing of column sums, for example
\[
\alpha_{22} = -\big( \alpha_{12}+\alpha_{32}+\alpha_{42}\big)\,.
\]

%%%%%%%%%%%%%%%%%%
\subsubsection{The Lie-Markov hierarchy.}
\label{subsubsec:LieMarkovClass}

In order to bring some systematics to bear on the types of rate models exemplified by the above selection, we 
introduce some useful notation. Recall that we assume that our stochastic matrices are appropriate elements of the
complex Markov matrix group. We present a definite basis for the corresponding complex rate matrices, for the Lie algebra\footnote{A succinct notation is $L(GL(K))=gl(K)$\,, $L(GL_1(K)) = gl_1(K)$.} $L(GL_1(K))$ of the group $GL_1(K)$ (the presentation of the affine group which appears in the standard character basis), as follows. To each of the off-diagonal $K\!\times \!K$ elementary matrices $e_i{}^j$, $i \ne j$, generators of $GL(K)$, we define an associated $GL_1(K)$ generator, ${\textsl l}_i{}^j$, by simply subtracting the unit diagonal $e_j{}^j$ which makes the column sum vanish; the diagonals $e_i{}^i$with $i=j$  must obviously be excluded as they cannot be so adjusted. These matrices are of course closed under multiplication\footnote{Each column sum of $M_1 M_2$ reduces to the corresponding column sum of $M_2$\,, after imposing the condition for $M_1$.}, but as we wish to use the notion of Lie algebras, we require the corresponding Lie bracket (as a matrix commutator). Using the standard product of elementary matrices, 
we have:
\begin{align*}
e_i{}^j e_k{}^\ell = &\,\delta^j{}_k e_i{}^\ell\,,\\
{[}e_i{}^j, e_k{}^\ell{]}=&\,\delta^j{}_k{e}_i{}^\ell  -\delta^\ell{}_i{e}_k{}^j\,,\\
{\textsl l}_i{}^j := &\, e_i{}^j - e_j{}^j\,; \\
{\textsl l}_i{}^j {\textsl l}_k{}^\ell 
=  &\, \delta^j{}_k\big({\textsl l}_i{}^\ell - {\textsl l}_k{}^\ell \big) - \delta^{j\ell} {\textsl l}_i{}^j ;\\
{[}{\textsl l}_i{}^j, {\textsl l}_k{}^\ell{]}=&\,
\big(\delta^j{}_k{\textsl l}_i{}^\ell  -\delta^\ell{}_i{\textsl l}_k{}^j\big) - (\delta^j{}_k{\textsl l}_k{}^\ell-\delta^\ell{}_i{\textsl l}_i{}^j \big)
- \delta^{j\ell}\big( {\textsl l}_i{}^j -{\textsl l}_k{}^\ell\big)\,.
\end{align*}
It is very convenient to work with the overdetermined set of ${\textsl l}_i{}^j$ for all index choices $i,j = 1,2, \cdots, K$ -- not only does each ${\textsl l}_i{}^j$ have zero column sum, by construction,
%\[
%\sum_m  \big({\textsl l}_i{}^j\big)^m{}_n = \sum_m \big(e_i{}^j\big)^m{}_n - \sum_m\big(e_j{}^j\big)^m{}_n
%=   \delta^j{}_n \big(\sum_m ( \delta_i{}^m - \delta_j{}^m)\big)  \equiv 0\,,
%\]
but in addition, there are the $K$ relations  ${\textsl l}_i{}^i \equiv 0$\,, $i =1,2,\cdots, K$\,, which trivially means that the dimension of the Lie algebra is $K(K\!-\!1)$ $= (K\!-\!1)^2\!+\!(K\!-\!1)$ as it should be. Based in the above, we adopt the definitons 
\begin{align*}
{[}E_i{}^j, E_k{}^\ell{]}=&\,\delta^j{}_k{E}_i{}^\ell  -\delta_i{}^\ell{E}_k{}^j\,,\\
{[}{\textsl L}_i{}^j, {\textsl L}_k{}^\ell{]}=&\,
\big(\delta^j{}_k{\textsl L}_i{}^\ell  -\delta_i{}^\ell{\textsl L}_k{}^j\big) - (\delta^j{}_k{\textsl L}_k{}^\ell-\delta_i{}^\ell{\textsl L}_i{}^j \big)
- \delta^{j\ell}\big( {\textsl L}_i{}^j -{\textsl L}_k{}^\ell\big)\,
\end{align*}
to be the defining relations of the Lie algebras $L\big(GL(K)\big)$, for $K^2$ abstract generators $E_i^j$, and $L\big(GL_1(K)\big)$, for $K(K\!-\!1)$ generators ${\textsl L}_i{}^j$, respectively.  

To simplify notation let ${\mathfrak L}_{GM}:=L\big(GL_1(K)\big)$ denote the Lie algebra of the general Markov rate model. Consider the convex cone 
\begin{equation}
\label{eq:GMdef}
{\mathfrak L}_{GM}^+ = \left\{ \sum_{i,j=1}^K \alpha_i{}^j {\textsl l}_j{}^i\,,\quad  \alpha_i{}^j \ge 0 \right\}\,.
\end{equation}
We define a \emph{stochastic Markov model} ${\mathfrak M}$ to be some fixed subset of the set of all stochastic Markov matrices, and a \emph{stochastic rate model} ${\mathfrak Q}$ to be a distinguished subset of the Lie algebra ${\mathfrak L}_{GM}$ each of whose elements is a linear
combination of the $\{ {\textsl l}_i{}^j\} $\, with positive real coefficients. Corresponding to a given stochastic rate model is the corresponding stochastic Markov model, the set of matrix exponentials $\exp{\mathfrak Q}$\,.
A \emph{Lie-Markov model} \cite{sumner2011BF,fernandez2015lie,woodhams:fernandez-sanchez:sumner:2015a} is a (complex) Lie subalgebra ${\mathfrak L}$
of ${\mathfrak L}_{GM}$ (see below). We define the \emph{stochastic cone} of a Lie-Markov model 
${\mathfrak L}^+$ of ${\mathfrak L}$ to be the intersection ${\mathfrak L}^+:= {\mathfrak L}\cap {\mathfrak L}_{GM}^+$\,.
If the Lie algebra ${\mathfrak L}$ of a Lie-Markov model has a \emph{stochastic basis} $\{L_1,L_2,\cdots, L_d\}$\,, that is, a basis each of whose elements is a linear combination of the $\{ {\textsl l}_i{}^j\}$\,, with positive real coefficients, then these basis elements can be taken as the extremal rays of the stochastic cone; otherwise the extremal elements (which are necessary for the specification of the model parametrization) must be specified separately from the generators themselves\footnote{An instance of this common situation amongst the Lie-Markov models is model (5.6b) mentioned already: there are 6 extremal rays corresponding to the presence of 
the parameters $\alpha, \beta, x,y,z,t$ in the rate matrix, but the Lie algebra has dimension 5, in view of the constraint $x+y+z+t=0$\,.}.

It is evident that the identification of Lie-Markov models ${\mathfrak L}$\,, including the characterization of their associated stochastic cones ${\mathfrak L}^+$\,, provides a significant selection criterion for models whose substitution matrices
(in this case the set ${\mathfrak M}= \exp{{\mathfrak L}^+}$\,) have the potential for multiplicative closure -- that is, allowing inhomogeneous processes such as successive substitutions $M_1M_2$ to be modelled by an effective average homogeneous
substitution $\overline{M}$ from the same rate class (for further exposition of the nature and importance of model closure see \S 
\ref{subsubsec:ModelClosure} below). 
As already foreshadowed however, the algebraic criterion for the existence of Lie-Markov models alone, is insufficiently powerful to guide model selection without further refinement. In particular, it is straightforward to construct examples of a (parametrized) infinite family of Lie-Markov models \cite{sumner2011BF}. From the statistical, model fitting point of view, such a situation is unsatisfactory, as one is then required to use the data to decide which model parameter to choose. As we presently outline, the resolution of this dilemma is to apply model permutation symmetries, to identify a distinguished model from amongst such a family.

To this end it is necessary to view the nature of model parametrization in relation to the goals of statistical inference and parameter recovery. From this point of view the rather surprising realization is that, in the absence of errors, a standard computational technique such as maximum likelihood, should return parameter estimates which are quite independent of the order of the trial parameters used as input. Concretely, for the part of the modelling of concern here, if from a theoretical model a list of numerical   
rate parameters $\{\alpha_1,\alpha_2, \cdots, \alpha_d \}$ is input to an analysis, the output will be the corresponding estimates
$\{\widehat{\alpha}_1 ,\widehat{\alpha}_2, \cdots, \widehat{\alpha}_d \}$\,. If the trial parameters
are however subject to some permutation, and the list remains the same, $\{\alpha'_1,\alpha'_2, \cdots, \alpha'_d \} \equiv \{\alpha_1,\alpha_2, \cdots, \alpha_d \}$ (the same parameters, input in a different ordering), \emph{the maximum likelihood estimates should be unchanged}. Another way of formulating this is with regard to the graph of the rate model in question (for an irreducible process, this is the complete graph on $K$ nodes, with directed edges labelled by the appropriate substitution rate parameters). Should the nodes be subject to some permutation belonging to the group ${\mathfrak S}_K$, yielding a graph with shuffled edge rates, there should be some compensating permutation of the edge parameters (which may be $d$ in number) which restores the original labelled graph. As we shall see presently, this property is trivially true for both the Jukes-Cantor model and the general Markov model, but a simple analysis shows that it also holds for the Kimura models as well as the equal-input models (see \S \ref{subsubsec:Catalogue} above for definitions). The point of the classification is to identify models where the implied homomorphism ${\mathfrak S}_K \rightarrow {\mathfrak S}_d$ is associated with the action of some subgroup $G < {\mathfrak S}_K$\,.

For the following analysis we drop the convention of using contra- and co-variant indices in matrix elements and generator labels and for typographical clarity, we adopt a capitalized notation $\{L_{ij}\,, i,j = 1,2,\cdots, K\,; L_{ii}=0 \}$ for the generators, even in the fundamental
$K\!\times \!K$ representation. The formal description appeals to the action of ${\mathfrak S}_K$ on ${\mathfrak L}_{GMM}$ induced by the permutation group acting on $V={\mathbb C}^K$. For $\sigma\in{\mathfrak S}_K$ we have
the $K\!\times K$ permutation matrix $K_\sigma$\, 
\begin{align*}
% \sigma\big(L_{ij}\big) = &\, L_{\sigma(i)\sigma(j)} \equiv K_\sigma L_{ij}K_\sigma^{-1}\,, \\
%\mbox{where} \qquad 
\sigma\big(e_{i}\big) =&\, \left.\sum\right._{j=1}^K e_j \big(K_\sigma\big){}_{ji}\,,\\
\mbox{and the action}\qquad 
 \rho(\sigma)\cdot L_{ij} := &\, L_{\sigma(i)\sigma(j)} \equiv K_\sigma L_{ij}K_\sigma^{-1}\,, \qquad\qquad
%\mbox{where} \qquad 
%\sigma\big(e_{i}\big) =&\, \left.\sum\right._{j=1}^K e_j \big(K_\sigma\big){}_{ji}
\end{align*}
which transforms rate matrices as 
\begin{align*}
Q= \sum \alpha_{ij}L_{ij} &\,\rightarrow  \sigma\cdot Q= K_\sigma Q K_\sigma^{-1}
= \sum  \alpha_{ij} L_{\sigma(i)\sigma(j)} %\equiv K_\sigma L_{ij}K_\sigma^{-1}\,. 
%\mbox{where} \qquad 
%\sigma\big(e_{i}\big) =&\, \left.\sum\right._{j=1}^K e_j \big(K_\sigma\big){}_{ji}
\end{align*}
We say that a Lie-Markov model $\mathfrak{L}$ has the symmetry of a group $G\leq \mathfrak{S}_K$ if $\mathfrak{L}$ is invariant under the natural action of $G$, that is:
\[
\sigma\cdot Q:=K_{\sigma }QK_{\sigma}^{-1}\in \mathfrak{L} \text{ for all }Q\in \mathfrak{L}.
\]
To further reduce the number of possibilities, we also have recourse to consider the stronger property that $\mathfrak{L}$ has a basis 
${B}_{\mathfrak L} = \{ L_1, L_2, L_3, \cdots, L_d \}$ of ${\mathfrak L}$ which is invariant under $G$: for all $\sigma \in G$, we have
\begin{align*}
\sigma \cdot {B}_{\mathfrak L} := \{ K_\sigma L_1K_\sigma^{-1}, K_\sigma L_2K_\sigma^{-1}, K_\sigma L_3K_\sigma^{-1}, \cdots, K_\sigma L_dK_\sigma^{-1} \} \equiv \{ L_1, L_2, L_3, \cdots, L_d \}
\end{align*}
Although complicating the definition, demanding the stronger basis property further assists in identifying Lie-Markov models. The method  entails a constructive analysis ordered by dimension, based on the decomposition of ${\mathfrak L}_{GM}$ with respect to the symmetry group $G$, and examination of all possible partitions of $\mbox{dim}({\mathfrak L})$ into $G$-modules, for which the Lie bracket closes .

Figure \ref{fig:LieMarkovModelsFlowChart} below gives a comprehensive tabulation of Lie-Markov models admitting the symmetry of the 8-dimensional wreath product group, the dihedral group ${\mathfrak S}_2\wr{\mathfrak S}_2$ \cite{fernandez2015lie} (see \cite{sumner2011BF} for the case $G={\mathfrak S}_4$\,). This is the biologically important symmetry on nucleotide bases which recognises the purine-pyrimidine pairings $A$, $G$ as well as $C$, $T$, and also the switch of these pairs, and thus is generated by the permutations $(AG)$\,, $(CT)$\,, and $(AC)(GT)$\,.  As can be seen from figure \ref{fig:LieMarkovModelsFlowChart} however, there is a large hierarchy of Lie-Markov models with this 
${\mathfrak S}_2\! \wr\! {\mathfrak S}_2$ pairing symmetry -- in fact, except for degenerate cases, there is a threefold multiplicity of such pairing groups -- one for purine-pyrimidine $\texttt{R}=\{A,G\}\,, \texttt{Y}=\{C,T\}$, one for weak-strong $\texttt{W}=\{A,T\}\,,
\texttt{S}=\{C,G\}$, and one for amino-keto
$\texttt{M}=\{A,C\}\,,\texttt{K}=\{T,G\}$ pairings (see \cite{woodhams:fernandez-sanchez:sumner:2015a}). In the so-called \emph{strand symmetric model} \cite{casanellas2005strand,jarvis:sumner:2016missm} 
the symmetry is with respect to the strong-weak pairings $(CG)$\,, $(AT)$\,, and $(CG)(AT)$\,, and \emph{only} these permutations are used in restoring the action of arbitrary permutations of the rate matrix\footnote{We comment further on the strand symmetric model in \S 
\ref{sec:Entanglement} below.}.

An obvious instance of Lie-Markov models admitting a discrete symmetry, occurs when \emph{each} of the basis elements
is \emph{invariant}, that is,  $\sigma\cdot L_a = K_\sigma L_a K_\sigma^{-1} = L_a$\,, $a=1,\cdots, d$\,. For nucleotides, $K=4$, this is the case for the model class of \emph{equivariant models} \cite{draisma2008} which are characterized by the subgroup $G \le {\mathfrak S}_4$ under which the rate matrices are invariant. These cases of course necessarily reappear, in the tabulation of models with 
${\mathfrak S}_2\! \wr\! {\mathfrak S}_2$ symmetry. An interesting case in point is the doubly-stochastic model, which is
clearly $\mathfrak{S}_4$-\emph{symmetric}, while not having an $\mathfrak{S}_4$-permutation basis. It does however appear as a Lie-Markov model, with a permutation basis, under the action of the subgroup $\mathfrak{S}_2\wr \mathfrak{S}_2$ \cite{fernandez2015lie}.

An important case beyond just $K=4$ is that of the so-called \emph{group-based models}. Consider the action of a group $\Gamma$ (not necessarily abelian) of order $K$, with characters labelled by group elements, wherein the components of the rate generator are of Toeplitz type, assumed to depend only on the connecting group element, $\big(L_\sigma \big){}_{ab} = \delta_{\sigma,ab^{-1}}-\delta_{e, ab^{-1}}$\,. The first term in this expression is nothing but the permutation representing the element $\sigma$ on this space\footnote{
Because $\sigma \cdot e_a = e_{\sigma^{-1} \cdot a} = \sum_b e_b \big(K_\sigma \big){}_{ba}$\,.}. The commutator of two such elements $L_\sigma = -\unit + K_\sigma$ is \
\begin{align}
{[}L_\sigma, L_{\sigma'}{]} =&\, {[}-\unit + K_\sigma, -\unit + K_{\sigma'}{]}
= K_\sigma  K_{\sigma'} - K_{\sigma'}K_\sigma \\
 = &\, K_{\sigma \sigma'} -K_{\sigma' \sigma} \equiv L_{\sigma \sigma'} -L_{\sigma' \sigma}\,,
\label{eq:GroupBased}
\end{align}
and we conclude that the generators $\{ L_\sigma\,, \sigma \in \Gamma \} $ of group-based models indeed form a Lie algebra, and are hence are also instances of Lie-Markov models\footnote{Since closure under the Lie bracket here does not use the existence of the inverses $\sigma^{-1}$, it is natural to assume that this model class generalizes to taking $\Gamma$ to be a (finite) semigroup. This idea is explored in detail in \cite{SumnerWoodhams2018}.}.

\begin{figure}[htbp]
   \centering
%\hskip-2ex \rotatebox{90}{\includegraphics[height=18cm]{pics/fig1_colour.pdf}}  
\hskip-2ex \rotatebox{90}{\includegraphics[height=18cm]{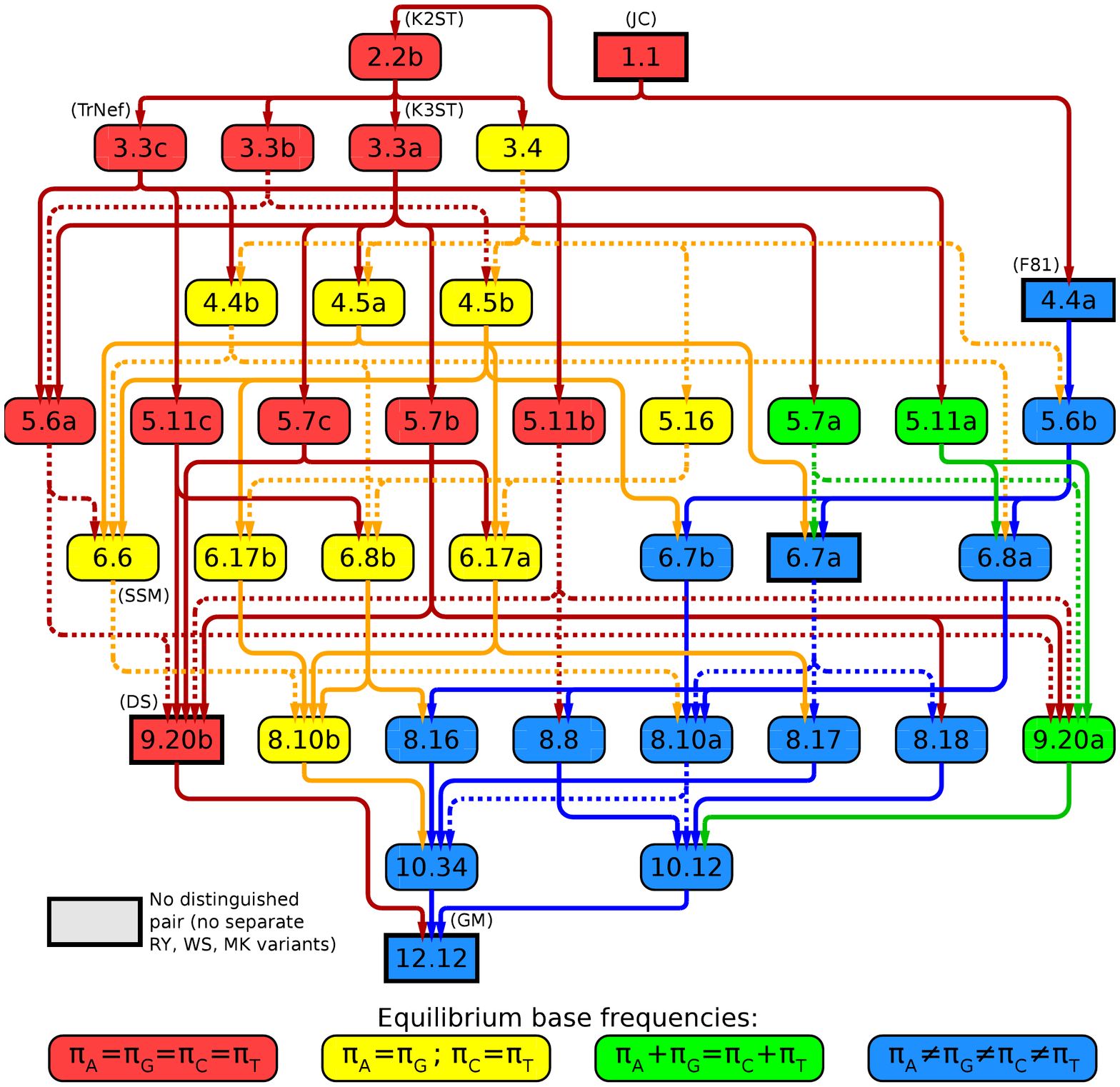}}  
\caption{\protect\small{Diagram depicting the hierarchy of Lie-Markov models with symmetry ${\mathfrak S}_2\! \wr \!{\mathfrak S}_2$ and their interrelationships. For specific forms of rate matrices, see \S\S \ref{subsubsec:Catalogue}, \ref{subsubsec:Diagonalization}.
Note that there are three hierarchies of Lie-Markov models, depending on the underlying nucleotide pairing; the entries labelled by common names in the chart generally occur within the $\texttt{RY}$ variant. Thus for example, in addition to the standard Tamura-Nei 
(equal frequency) $Tr\!N\!e\!f_{{\texttt{RY}}}$ rate model, 
there are two alternatives $Tr\!N\!e\!f_{{\texttt{WS}}}$ and 
$Tr\!N\!e\!f_{{\texttt{MK}}}$\,, all denoted $(3.3c)$. The strand symmetric model (6.6) on the other hand, entails Watson-Crick or strong-weak pairing, so that in addition to $SSM_{{\texttt{WS}}}$ there are the $SSM_{{\texttt{RY}}}$ and $SSM_{{\texttt{MK}}}$ variants. 
In some cases, the rate matrix is sufficiently degenerate that these variants are not distinguished; this is the case for the the one-parameter Jukes-Cantor $JC$ model $(1.1)$; the standard Felsenstein $F81$ rate model $(4.4a)$; model $(6.7)$; the doubly stochastic model $(9.20b)$, and of course the general Markov model $GM$\,, model $(12.12)$. In some cases, notably the doubly stochastic model, a permutation basis does not exist. The \emph{omission} of some expected common models -- with labels such as $HKY$ and $GTR$ 
(see \S\S \ref{subsubsec:ModelClosure}) -- from the chart illustrates the very point of the classification of Lie Markov models, that in such rate models, the generators do \emph{not} form  a closed Lie algebra, and thus are expected to perform badly in inhomogeneous settings. See the text for definitions and further elaboration of the arguments.
}
}
   \label{fig:LieMarkovModelsFlowChart}
\end{figure}
\mbox{}
\vfill
\pagebreak
%%%%%%%%%%%%%%%%%%
\subsubsection{Model closure.}
\label{subsubsec:ModelClosure}
%	\subsubsection{Model closure.}
%	\label{subsubsec:ModelClosure}
From the point of view of model selection and fitting, and statistical consistency, and as is evident from our category of nucleotide rate models, there is great biological interest in identifying different phylogenetic model types. Working within the class of continuous time Markov
models, and under the aegis of `multiplicative closure', we have developed the Lie-Markov hierarchy 
(\S\S \ref{subsec:LieMarkov}, \ref{subsubsec:LieMarkovClass}) wherein, in addition, the rate matrices form a Lie algebra. 

In this subsection we briefly elaborate on this point in view of its practical significance, as well as its theoretical importance in model classification. Model closure is simply the demand that the product $\overline{M}=M_1M_2$ of two substitution matrices can itself be parametrized as an element of the same model -- or at the level of the corresponding rate matrices, that an element $\overline{Q}$ belonging to the same model type exists such that $\exp \overline{Q} = \exp Q_1 \exp Q_2$\,. This is mandatory in the non-homogeneous case, whenever edges are allowed to be parametrized by rate matrices which are not merely related by scaling\footnote{If $Q_1 = t_1 Q$ and $Q_2 = t_2 Q$\, then of course $\overline{Q} = (t_1+t_2)Q$ and the model type is unaffected.}.

Recall from \S \ref{subsec:LieMarkov}, that the rate generators of a Lie-Markov model were identified as 
the intersection of the underlying Lie subalgebra, with the stochastic cone of the general Markov model (with the relevant Lie subalgebras being identified via invariance under a set of symmetry conditions). For the moment we adopt a slightly different starting point. Let us define a ``general rate model'' to be a set ${\mathfrak R}^+ := {\mathfrak R}\cap {\mathfrak L}^+ $ of stochastic rate generators, where ${\mathfrak R}$ comprises real parameters, which are the solution set of a finite collection of homogeneous polynomial constraints on the coordinates of the affine group Lie algebra (that is, the general Markov rate model). Additionally, by homogeneity, we have 
${\mathbb R}{\mathfrak R}= {\mathfrak R}$\,, and observe also that ${\mathfrak R}^+$ is a cone, ${\mathbb R}^+ {\mathfrak R}^+ = {\mathfrak R}^+$\, as required for biological utility. Calling ${\mathfrak M}_S$ the \emph{semigroup generated}\footnote{In \S \ref{subsubsec:LieMarkovClass} above, only the set of matrix exponentials was considered.} by matrix exponentials $\exp{{\mathfrak R}^+}$\,, a reasonable definition of such a model being multiplicatively closed \cite{Sumner_2017mult}, is that the corresponding matrix logarithms 
$\log(\exp Q_1 \exp Q_2)$\,, wherever these are defined by the standard series expansion, are themselves elements of ${\mathfrak R}$\, (but not necessarily of ${\mathfrak R}^+$). Imposing model multiplicative closure in this weak sense, by an argument based on the famous Baker Campbell Hausdorff (BCH) series formula expressing the product of matrix exponentials as the exponential of a series involving nested commutator brackets \cite{Baker1902,Campbell1897,Hausdorff1906}, it is then possible to show under reasonable conditions, that ${\mathfrak R}$ is in fact a vector space, and moreover, that it also contains commutators of its elements, and so forms a Lie algebra. Any such ``general rate model'' must therefore be of Lie-Markov type, and the polynomial constraints can only be linear. As this chain of reasoning is central to our characterization of model classes, in appendix \S \ref{subsec:BCHapproximants} we provide further justification, based on error estimates in the BCH series, which we explicitly derive following \cite{BlanesCasas2004}.

As an example we consider the Hasegawa Kishino Yano \cite{hasegawa1985} model class $HKY$, an instance of the so-called
`general time-reversible' type $GTR$ (for the formal definition see \cite{tavare1986}). The $GTR$ model is not multiplicatively closed\footnote{They are therefore not represented in the Lie-Markov model hierarchy.}, and used inappropriately, have been shown \cite{sumner2012a} to lead to errors in phylogenetic inference and modelling. The rate matrix is:
\[
Q^{HKY}=\left[
\begin{matrix}
-\kappa \alpha_G -\alpha_C -\alpha_T & \kappa\alpha_A & \alpha_A & \alpha_A \\
\kappa\alpha_G & -\kappa \alpha_A -\alpha_C -\alpha_T  & \alpha_G & \alpha_G \\
\alpha_C & \alpha_C & -\alpha_A - \alpha_G - \kappa \alpha_T &  \kappa\alpha_C \\
\alpha_T & \alpha_T & \kappa\alpha_T  &  -\alpha_A - \alpha_G - \kappa \alpha_C
\end{matrix}
\right]\,.
\]
In contrast to the Kimura $K3ST$ model, this model class accommodates non-uniform stationary nucleotide frequencies of the Markov chain (proportional to the by the nonnegative parameters $\alpha_A,\alpha_G,\alpha_C,\alpha_T$\,), while the weight $\kappa\geq 0$ is included to capture the `transition/transversion' rate ratio.
Equivalently, we may express the $HKY$ model as the subset of stochastic rate matrices of the general Markov model (\ref{eq:GMdef})
obeying the homogeneous linear and quadratic constraints
\[
\mathcal{R}^+_{\text{HKY}}=
\left\{ Q \in \mathcal{L}^+: \begin{matrix}\alpha_{13}\!=\!\alpha_{14},\alpha_{23}\!=\!\alpha_{24},\alpha_{31}\!=\!\alpha_{32},\alpha_{41}\!=\!\alpha_{42}\\\alpha_{12}\alpha_{23}\!=\!\alpha_{21}\alpha_{13},\alpha_{34}\alpha_{13}\!=\!\alpha_{12}\alpha_{31},\alpha_{43}\alpha_{13}\!=\!\alpha_{12}\alpha_{41}\end{matrix} \right\}\,.
\]
Since the defining constraints are not linear however, we can infer from the foregoing that $\mathcal{R}^+_{\text{HKY}}$ is not multiplicatively closed. 

This is borne out by a concrete, numerical example \cite{Sumner_2017mult}. For the product of the following  $Q_1,Q_2\in \mathcal{R}^+_{\text{HKY}}$\,,
%We chose $Q_1,Q_2\in \mathcal{R}^+_{\text{HKY}}$ via $(\alpha_A,\alpha_G,\alpha_C,\alpha_T;\kappa)=(.02,.01,.005,.009; 1.5)$ and $(.03,.01,.006,.008;1.4)$ respectively, and computed (using Mathematica):
\begin{align*}
Q_1 = &\,
\left[
\begin{matrix}
\!-\!.014 \!-\!1.5 \!\times\! .01  & 1.5 \!\times\! .02 & 0.02 & 0.02 \\
1.5 \!\times\! .01&  \!-\!0.014 \!-\!1.5 \!\times\! .02 & 0.01 & 0.01 \\
0.005 &  0.005 & \!-\!0.03 \!-\!1.5 \!\times\! .009  & 1.5\!\times\! .005 \\
0.009 & 0.009 & 1.5 \!\times\! 0.009 & \!-\!0.03 \!-\!1.5 \!\times\! .005 
\end{matrix}
\right]\,, \\
Q_2 = &\,
\left[
\begin{matrix}
\!-\!0.014 \!-\!1.4 \!\times\! .01 & 1.4 \!\times\! .03 &0.03&0.03  \\
 1.4 \!\times\!.01 &\!-\!0.014 \!-\!1.4 \!\times\! .03 & 0.01 & 0.01 \\
 0.006 & 0.006 & \!-\!.04 \!-\!1.4 \!\times\! .008 & 1.4 \!\times\! .006 \\ 
0.008 & 0.008 & 1.4 \!\times\! .008 & \!-\!0.04 \!-\!1.4 \!\times\! .006
\end{matrix}
\right]\,,
\end{align*}
a standard numerical package delivers
\[
\ln(e^{Q_1}e^{Q_2})=
\left[
\begin{array}{rrrr}
\! -0.0571752 & 0.0718248& 0.0498348& 
  0.0498348\\ 0.0291051& \!-0.0998949& 0.0200951& 
  0.0200951\\ 0.0109967& 0.0109967& \!-0.0947047& 
  0.0158953\\ 0.0170734& 0.0170734& 0.0247748&\! -0.0858252
\end{array}
\right].
\]
Matching this to a putative $HKY$ rate matrix $\overline{Q}$, we are immediately led to 
\[
(\overline{\alpha}_A,\overline{\alpha}_G,\overline{\alpha}_C,\overline{\alpha}_T)\!=\!(0.0498348,0.0200951,0.0109967, 0.0170734)\,,
\] 
but \emph{no} consistent solution for $\overline{\kappa}$ is obtainable (in fact \emph{four} different values are required).
Therefore,  $\ln(e^{Q_1}e^{Q_2})$ is \emph{not} a member of ${\mathcal R}^{{HKY}}$\,.
However, the discussion above suggests that the correct generalization to recover multiplicative closure is 
simply the \emph{linear} span
\[
Q^{({8.8})}=\left[
\begin{matrix}
-\kappa_2 -\gamma-\delta & \kappa_1  & \alpha &  \alpha\\ 
\kappa_2 & -\kappa_1 -\gamma-\delta & \beta & \beta\\ 
 \gamma & \gamma & -\alpha-\beta-\kappa_4 & \kappa_3 \\ 
 \delta & \delta & \kappa_4 &  -\alpha-\beta-\kappa_3\end{matrix}
\right]\,
\]
under the correct stochastic conditions; this rate model indeed appears in the Lie-Markov hierarchy of models with
${\mathfrak S}_2 \wr {\mathfrak S}_2$ symmetry, as model (8.8) (see figure \ref{fig:LieMarkovModelsFlowChart} ).
%
%
%\vfill
%\pagebreak
%%%%%%%%%%%%
%%%%%%%%%%%%
%%%%%%%%%%%%
\subsubsection{Rate matrix diagonalization.}
\label{subsubsec:Diagonalization}
The introductory presentation of the modelling framework in molecular phylogenetics, \S \ref{subsec:ModellingGMM}, 
emphasized that in addition to the natural basis for phylogenetic tensors, which is distinguished because of the
biological meaning of its tensor components, other coordinate systems such as the 
the affine basis are available for specific analysis. In light of the plethora of Lie-Markov models, it is of considerable
importance to identify basis transformations and alternative coordinate systems which can lead to better understanding
of the nature of specific models and the meaning of their parameters. To exemplify this we here examine the rate
matrices of selected models. In some special cases, the simplifications gained will indeed be able to be extended to the model on the entire tree. As we shall see presently, in these cases there are tremendous theoretical simplifications available, in that parameters for the underlying tree can be extracted in principle directly from the coordinate transformation applied to the complete phylogenetic tensor coming from the alignment. This will be developed in detail below, along with some accompanying formalism which 
arises in these cases and which can be seen as a deep algebraic generalization of the underlying graphical models.

Consider firstly the matter of rate matrix diagonalization\footnote{We emphasize that the involved similarity transformations
are to act on the \emph{entire set of rate matrices} belonging to the model class in question.}. We limit our discussion here to selected illustrative cases (see \cite{woodhams:fernandez-sanchez:sumner:2015a}, figure \ref{fig:LieMarkovModelsFlowChart} above). Consider for example, the three-dimensional models displayed there.
Model (3.3a), the Kimura three parameter model \cite{kimura1981estimation}, listed above in our introductory survey in \S \ref{subsec:LieMarkov} above, with
\[
Q^{K3ST} = Q^{(3.3a)}=\left[\begin{array}{cccc} \!-\!\alpha\!-\!\beta\!-\!\gamma & \alpha &\beta & \gamma \\ 
\alpha & \!-\!\alpha\!-\!\beta\!-\!\gamma &\gamma& \beta\\
			 \beta& \gamma & \!-\!\alpha\!-\!\beta\!-\!\gamma &\alpha \\ \gamma& \beta&\alpha & \!-\!\alpha\!-\!\beta\!-\!\gamma\end{array}\right]\,,
\]
is accompanied within the Lie-Markov hierarchy by model (3.3c), the Tamura-Nei (1993) model, with equal base frequencies 
\cite{tamura1993estimation},
\[
Q^{T\!N93ef} = Q^{(3.3c)}=\left[\begin{array}{cccc} \!-\!\alpha\!- \!2\beta  & \alpha &\beta & \beta \\ 
\alpha & \!-\!\alpha\!-\!2\beta &\beta & \beta\\
			 \beta& \beta & \!-\!\gamma\!-\!2\beta  &\gamma \\ \beta& \beta&\gamma &  \!-\!\gamma\!-\!2\beta\end{array}\right]\,,
\]
which provides an interesting variant on the Kimura model, in providing rate parameters capable of distinguishing between the transition rates, while lumping the transversion rates together as one average. Model (3.3b), with rate generator 
\[
Q^{\widetilde{K3ST}} =Q^{(3.3b)}= \left[\begin{array}{cccc} \!-\!\alpha\!-\!\beta\!-\!\gamma & \alpha &\beta& \gamma \\ \alpha & \!-\!\alpha\!-\!\beta\!-\!\gamma &\gamma &\beta\\
			 \gamma &\beta & \!-\!\alpha\!-\!\beta\!-\!\gamma &\alpha \\ \beta & \gamma&\alpha &\!-\!\alpha\!-\!\beta\!-\!\gamma \end{array}\right]\,,
\]
is a further variant, wherein the two parameters dedicated to transversions, interchange roles depending on the direction of substitution. Finally there is model (3.4), a three-parameter sub-model of (4.4b),
\[
Q^{(4.4b)} = \left[\begin{array}{cccc} \!-\!\alpha\!- \!2\gamma & \alpha &\beta & \beta \\ 
\alpha & \!-\!\alpha\!-\!2\gamma &\beta & \beta\\
			 \gamma& \gamma & \!-\!\delta\!-\!2\beta  &\delta \\ \gamma& \gamma&\delta &  \!-\!\delta\!-\!2\beta \end{array}\right]\,,
\]
subject to positive coefficients obeying in addition the condition $\alpha + \gamma = \beta +\delta$.

As noted previously, in the first two of these models, $Q^{K3ST}$ and $Q^{T\!N93ef}$, the parameters are such that the rate matrices are \emph{symmetric}; specifically, they belong to the class of doubly stochastic models mentioned in \S \ref{subsec:ModellingGMM} above. Evidently, their Lie algebra generators can be assigned in the appropriate affine basis, to the (purely block-diagonal) $gl(3)$ subalgebra. We have explicitly\footnote{The $K3ST$ model is of equivariant type, with invariance group the distinguished Klein four subgroup of ${\mathfrak S}_4$, whereas the ${T\!N93ef}$ model is built from a reducible representation of the the dihedral group ${\mathfrak S}_2\! \wr\! {\mathfrak S}_2$ (see \ref{subsec:LieMarkov} above  and \cite{sumner2011BF,fernandez2015lie} for further details).}
\begin{align*}
Q^{K3ST} = &\, \alpha L_{(12)(34)} + \beta L_{(13)(24)}  + \gamma L_{(14)(23)} \, ,\\
Q^{T\!N93ef} = &\, \alpha L_{(12)} + \beta L_{(34)}  + \gamma \big(L_{(1324)}+L_{(1423)}\big)\,,
\end{align*}
and the Lie algebras are given as
\begin{align*}
{\mathfrak L}^{K3ST} =&\, \langle L_{(12)(34)}, L_{(13)(24)} ,L_{(14)(23)} \rangle\,,\\
 \mbox{and}\qquad {\mathfrak L}^{T\!N93ef} =&\, \langle L_{(12)},L_{(34)}, L_{(1324)}+L_{(1423)}\rangle\,,
\end{align*}
respectively.
In the case of $Q^{T\!N93ef}$, we find
\[
%L_{(12)} = \left[\begin{array}{cccc} \oo &1&0&0\\ 1&\oo&0&0\\0&0&0&0\\0&0&0&0\end{array}\right]\,,
%L_{(34)}=\left[\begin{array}{cccc} 0&0&0&0\\0&0&0&0\\ 0&0&\oo &1\\ 0&0&1&\oo \end{array}\right]\,,\
%L_{(1324)}+L_{(1423)}= \left[\begin{array}{cccc} \ot &0&1&1\\ 0&\ot&1&1\\1&1&\ot&0\\1&1&0&\ot \end{array}\right]\,,
L_{(12)} = \left[\begin{array}{rrrr} -1 &1&0&0\\ 1&-1&0&0\\0&0&0&0\\0&0&0&0\end{array}\right]\,,
L_{(34)}=\left[\begin{array}{rrrr} 0&0&0&0\\0&0&0&0\\ 0&0&-1 &1\\ 0&0&1&-1 \end{array}\right]\,,\
L_{(1324)}+L_{(1423)}= \left[\begin{array}{rrrr} -2 &0&1&1\\ 0&-2&1&1\\1&1&-2&0\\1&1&0&-2 \end{array}\right]\,,
\]
and introducing a similarity transformation $\widehat{L}= X^{-1}{L}X$ by the orthogonal matrix:
\begin{align}
\label{eq:33c44aXdef}
%X = \textstyle{\frac 12}\left[\begin{array}{cccc} 1 &\st&0&1\\ 1&\ost &0&1\\1& 0&\st&\oo\\1& 0&\ost&\oo\end{array}\right]\, = \big(X^{-1}\big)^\top\,,
X = \textstyle{\frac 12}\left[\begin{array}{rrrr} 1 &\st&0&1\\ 1&-\st &0&1\\1& 0&\st&-1\\1& 0&-\st&-1\end{array}\right]\, = \big(X^{-1}\big)^\top\,,
%,\qquad 
%X^{-1} = \textstyle{\frac 12}\left[\begin{array}{cccc} 1 &1&1&1\\ \st&\ost &0&0\\0& 0&\st&\ost\\1& 1&\oo&\oo\end{array}\right]
\end{align}
produces the diagonal forms
\begin{align}
\label{eq:33c44aDiagForms}
%\widehat{L}_{(12)}= \left[\begin{array}{cccc} 0 &0&0&0\\ 0 &\ot&0&0\\0&0&0&0\\0&0&0&0\end{array}\right]\,,\quad
%\widehat{L}_{(34)} = \left[\begin{array}{cccc} 0 &0&0&0\\ 0 &0&0&0\\0&0&\ot&0\\0&0&0&0\end{array}\right]\,\quad
%\widehat{L}_{(1324)}+ 
%\widehat{L}_{(1423)}  = \left[\begin{array}{cccc} 0 &0&0&0\\ 0 &\ot&0&0\\0&0&\ot&0\\0&0&0&\of\end{array}\right]\,.
\widehat{L}_{(12)}= \left[\begin{array}{rrrr} 0 & 0&0&0\\ 0 &\hskip-1ex -2&0&0\\0&0&0&0\\0&0&0&0\end{array}\right]\,,\quad
\widehat{L}_{(34)} = \left[\begin{array}{rrrr} 0 &0&0&0\\ 0 &0&0&0\\0&0&\hskip-1ex -2&0\\0&0&0&0\end{array}\right]\,\quad
\widehat{L}_{(1324)}+ 
\widehat{L}_{(1423)}  = \left[\begin{array}{rrrr} 0 &0&0&0\\ 0 &\hskip-1ex -2&0&0\\0&0&\hskip-1ex -2&0\\0&0&0&\hskip-1ex -4\end{array}\right]\,.
\end{align}
In the case of the Kimura model $Q^{K3ST}$, a diagonalizing matrix $H$ (different from $X$ above) can likewise be chosen. This is the so-called $4\times 4$ \emph{Hadamard} matrix\footnote{$H$ can in turn be regarded as a tensor product $H = h\otimes h$ where (relabelling rows and columns as $i,j=0,1$), $h^i{}_j  = \frac{1}{\sqrt{2}}(-1)^{ij}$ is the corresponding Hadamard matrix for the $K=2$, binary case. The intimate connections with the group ${\mathbb Z}_2\cong {\mathfrak S}_2$ and the discrete Fourier transform, will be developed in more detail in \S\S \ref{subsec:TreesNetworks}, \ref{subsubsec:HadamardFourier} below, exemplified for the three character $K=3$ case.}
,
\begin{align}
\label{eq:Hadamard4x4Def}
%H =  &\, \textstyle{\frac 12}\left[\begin{array}{cccc} 1 &1&1&1\\ 1&\oo &1&\oo\\1& 1&\oo&\oo\\1& \oo&\oo&1\end{array}\right]\, = H^{-1}.
H =  &\, \textstyle{\frac 12}\left[\begin{array}{rrrr} 1 &1&1&1\\ 1&-1 &1&-1\\1& 1&-1&-1\\1& -1&-1&1\end{array}\right]\, = H^{-1}.
\end{align}
We record for completeness the form of the generators in the transformed basis  $\widehat{L}= H^{-1}{L}H$ \cite{bashford2004BF}:
\[
\widehat{L}_{(12)(34)}= \!\left[ \begin{array}{rrrr}
					 1 & 0 & 0 & 0 \\
					0 & \hskip-1ex -1  & 0 & 0\\
					0 & 0 & 1 & 0 \\
					0 & 0 & 0 & \hskip-1ex -1 \end{array}\right],\quad
\widehat{L}_{(13)(24)}=\left[ \begin{array}{rrrr}
					1 & 0 & 0 & 0 \\
					0 & 1  & 0 & 0\\
					0 & 0 & \hskip-1ex -1 & 0 \\
					0 & 0 & 0 & \hskip-1ex -1 \end{array}\right] \, \quad
\widehat{L}_{(14)(23)}  = \left[ \begin{array}{rrrr}
					1 & 0 & 0 & 0 \\
					0 & \hskip-1ex -1  & 0 & 0\\
					0 & 0 & -1 & 0 \\
					0 & 0 & 0 & 1 \end{array}\right]. 
\]

As is obvious from the diagonalized forms, or from explicit calculation in the standard basis, the Lie algebras ${\mathfrak L}^{K3ST}$
and ${\mathfrak L}^{T\!N93ef}$ are \emph{abelian}, and since (as noted above from their form in the affine basis) they are elements of the homogeneous $gl(3)$ subalgebra, it can be concluded that these models\footnote{In the Kimura case, the tracelessness of the generators means that they are also a Cartan subalgebra of the $sl(4)$ algebra affiliated to the $4\times 4$ matrices.}  correspond to \emph{distinct choices of Cartan (maximal abelian) subalgebra of $gl(3)$}. 

Model (3.3b) of the Lie-Markov hierarchy, $Q^{\widetilde{K3ST}}$\,, resembles  $Q^{{K3ST}}$
in having generators related to single permutations, its Lie algebra in this case being associated with the elements of one of the cyclic 
${\mathbb Z}_4$ subgroups of ${\mathfrak S}_4$, through
\[
Q^{\widetilde{K3ST}} = \alpha L_{(12)(34)} + \beta L_{(1324)} + \gamma L_{(1423)}\,,
\]
making it an instance of a group-based model.
While this rate matrix is no longer diagonalizable by a real orthogonal transformation, it is clear that the generators (up to shifts by the unit matrix) will have eigenvalues associated with complex fourth roots of unity, and the connection with a Cartan subalgebra is not apparent. However, in appendix \S \ref{subsec:CyclicBasis}, it is shown that for \emph{any} cyclic 
subgroup\footnote{Including the `principal' one generated by the element $(123\cdots K)$\,.} of ${\mathfrak S}_K$, there is a natural (unitary) transformation $S$ taking the standard $gl(K)$ Cartan  
generators to the permutation-adapted generators. Thus in these cases also, the fact that the model is abelian, can again be traced to
the choice of a certain Cartan subalgebra.

Consider however the remaining three-dimensional model $Q^{(3.4)}$, or its progenitor $Q^{(4.4b)}$. As well as 
\begin{align}
\label{eq:44aCommon}
L_{(12)(34)}\equiv L_{(12)}+L_{(34)} =:J_+ \,,\qquad 
L_{(1324)}+L_{(1423)} =: K\,,
\end{align}
encountered above, we have the additional generators
\begin{align}
\label{eq:44aExtras}
%\quad J_- = \left[ \begin{array}{cccc}\oo &1&0&0\\1&\oo&0&0\\ 0&0&1&\oo\\0&0&\oo&1 \end{array}\right]\,,
%\quad R = \left[ \begin{array}{cccc}1&1&1&1\\1&1&1&1\\ \oo&\oo&\oo&\oo\\ \oo&\oo&\oo&\oo \end{array}\right]\,.
\quad J_- = \left[ \begin{array}{rrrr}\hskip-1ex -1 &1&0&0\\1&\hskip-1ex -1&0&0\\ 0&0&1&\hskip-1ex -1\\0&0&\hskip-1ex -1&1 \end{array}\right]\,,
\quad R = \left[ \begin{array}{rrrr}1&1&1&1\\1&1&1&1\\ \hskip-1ex -1&\hskip-1ex -1&\hskip-1ex -1&\hskip-1ex -1\\ \hskip-1ex -1&\hskip-1ex -1&\hskip-1ex -1&\hskip-1ex -1 \end{array}\right]\,.
\end{align}
We find
${\mathfrak L}^{(3.4)} =\langle J_+, K, R\rangle$\, and ${\mathfrak L}^{(4.4b)}=  \langle J_-, J_+, K, R \rangle$\,,
with ${[}K,R{]}=  -4R$ the \emph{only} non-zero commutator bracket. For both of these (non-symmetric) models, the Lie algebra is certainly \emph{non-abelian}. With these examples, and indeed for the remaining Lie-Markov rate models, we have arrived at the generic situation of dealing with generators that are not (simultaneously) diagonalizable\footnote{In \S \ref{subsubsec:MarkovInvariantsLM}
we return to the transformation of the additional rate generators needed for $Q^{(3.4)}$, $Q^{(4.4b)}$ in the affine basis corresponding to matrix $X$(equation (\ref{eq:33c44aXdef}) above), where it will be seen that a useful $2\!\times \!2$ block structure (and corresponding decomposition of state space) arises.}.

Returning to the abelian case, the role of the Hadamard matrix was recognised early on by Kimura \cite{kimura1981estimation}, and formalized by \cite{evans1993} (see also \cite{hendy1994}) in terms of the group characters of the invariance subgroup of the model, the ${\mathbb Z}_2\! \times\! {\mathbb Z}_2$ Klein 4-group. In fact, in this case the derived transform methods can be applied across the entire phylogenetic tree \cite{szekely1993}, so that the `Hadamard coordinates' give immediate access to model parameters. Indeed, under the discrete Fourier transform\footnote{The Hadamard, and in general discrete Fourier, bases are instances of the ``affine'' basis choice.}, extant tree edges are encoded precisely by the nonzero coordinates, with their values directly giving the edge length parameters \cite{szekely1993} (generalizations will be treated in  \S\S \ref{subsec:TreesNetworks}, \ref{subsubsec:HadamardFourier} below).
%%%%%%%%%%%%

	%\input{secs/TreesNetworks.tex}
	%\subsection{Phylogenetic trees and networks}
	%%\label{subsec:TreesNetworks}
\subsection{Phylogenetic trees and networks}
\label{subsec:TreesNetworks}

%%%%%%%%%%%%%%%%%%%%%%%%%%
%%%%%%%%%%%%
\subsubsection{Coproducts and phylogenetic bialgebras.}
\label{subsubsec:Coproducts}
As we have emphasized, in contrast to the analysis of phylogenetic models and their substitution matrices via discrete group properties, 
our present perspective is on continuous-time models and their affiliated Lie algebras. We now turn to a
formalism which also effects the identical model parameter recovery across trees as in the Hadamard case, in a way complementary to the discrete group approach, and which also extends to other models and adds considerably to the algebraic context of modelling on trees and networks.

Recall the tensorial construction of the `general Markov model' on trees, presented in \S \ref{subsec:ModellingGMM} above (see figure
\ref{fig:6LeafTree} and equation (\ref{eq:AbstractTree}) above). As explained, a key object is the linear `splitting operator' 
$\delta: V \rightarrow V\otimes V$ which symbolizes the abstract event of speciation in the stochastic models. Given the algebraic context, it is reasonable to ask for an operator or operators, related to components of the rate matrix, which intertwines the action of $\delta$ -- that is, whose matrix elements after splitting, recovers the corresponding action prior to splitting. In terms of the selected basis 
of ${\mathfrak L}_{GM}$, we require
\begin{align}
\label{eq:IntertwiningDef}
\nabla\big(L_{ij}\big)\cdot \delta(e_k)= &\, \delta(L_{ij} e_k)\,.
\end{align}
As noted already, $L_{ij}e_k = \delta_{jk}(e_i -e_j)$, and a solution is easily verified to be
\begin{align}
\label{eq:CoproductNablaDef}
\nabla\big(L_{ij}\big) = &\, L_{ij} \otimes {\mathbb I} + {\mathbb I}\otimes L_{ij} + L_{ij} \otimes L_{ij}\,,
\end{align}
up to any operator which vanishes on $e_k \otimes e_k$\,, and is otherwise undetermined on $e_k\otimes e_\ell$\,, $k \ne\ell$\,.
In order to complete the definition of $\delta$ as a coproduct, we further require that \emph{it extends by linearity} to a mapping
 $\nabla: gl_1(K) \rightarrow gl_1(K)\otimes gl_1(K)$\,, defined with respect to the selected basis.

In terms of the algebraic construction of phylogenetic tensors presented in \S \ref{subsec:ModellingGMM}, the existence of $\nabla$
allows the formal possibility of ``pulling back'' the splitting operations to the root. In compensation, there occur iterated coproducts, extending the action of substitutions on edges closer to the root over several positions in tensor slots belonging to the generated sub-trees. Given that $\delta$ is associative, the entire ``tree topology'' is thus encoded by these iterated coproducts, all acting on an initial, ``star tensor'', of the form
\[
%\delta^{(L\!-\!1)}\big(\pi\big)
{\delta^{\underline{L\!-\!1}}\,\pi} = \sum_i \pi^i e_i \otimes e_i \otimes \cdots \otimes e_i\,.
\]

This structure will be explored in \S \ref{subsec:TreesNetworks} below, in the context of the binary general Markov model, where it will prove instructive in identifying obstructions to extending phylogenetic modelling from trees to networks. 
Here, we continue with the implications of the $\nabla$ coproduct, and the algebraic setting. As noted, the definition 
(\ref{eq:CoproductNablaDef}) fulfils the intertwining condition (\ref{eq:IntertwiningDef}), but leaves arbitrary the effect on off-diagonal basis elements $e_k\otimes e_\ell$\,, $k\ne \ell$\,. We here denote this as $\nabla^L$, and note the 
following variants $\nabla^K$, $\nabla^F$, which also satisfy (\ref{eq:IntertwiningDef}), but differ from (\ref{eq:CoproductNablaDef}):

%\noindent
\begin{quotation}
\noindent
\textbf{Forms of $\nabla$:}
\begin{description}
\item[$L_{ij}$ basis:] \mbox{}\\
We have 
\begin{align}
\nabla^L\big(L_{ij}\big) :=&\, L_{ij} \otimes {\mathbb I} + {\mathbb I}\otimes L_{ij} + L_{ij} \otimes L_{ij}\,; 
\\
\label{eq:LijCoproduct}
\nabla^L\big(L_{ij}\big)\cdot\big(e_k\otimes e_\ell\big) = &\,\delta_{jk}\big(e_i \otimes e_i - e_j \otimes e_j \big)\,; \nonumber
\end{align}
\item[Permutation basis:] \mbox{}\\
Let $L_\sigma = K_\sigma - \unit = \sum_i L_{i \sigma^{-1}i}$\,. We have 
\begin{align}
\nabla^K\big(L_\sigma\big) :=&\, K_{\sigma} \otimes  K_{\sigma} - {\mathbb I}\otimes {\mathbb I}\,;\\
\label{eq:LsigmaCoproduct}
\nabla^K\big(L_\sigma\big)\cdot e_k \otimes e_\ell  =&\, e_{\sigma k} \otimes  e_{\sigma \ell}- e_{k}\otimes e_{\ell}\,; 
\nonumber
\end{align}
\item[Equal input basis:] \mbox{}\\
Let $F_i = \sum_j L_{ij} := R_i -{\mathbb I}$\,. We have 
\begin{align}
\nabla^F\big(R_i \big) :=&\, R_i  \otimes  R_i  - {\mathbb I}\otimes {\mathbb I}\,\\
\label{eq:RiCoproduct}
\mbox{and } \qquad
\nabla^F\big(F_i \big) =&\, F_i  \otimes  F_i  - {\mathbb I}\otimes {\mathbb I}\,,\\
\label{eq:FiCoproduct}
\nabla^F\big(F_i \big)\cdot e_k \otimes e_\ell  =&\, e_i  \otimes  e_i  - e_k \otimes e_\ell \,.\nonumber
\end{align}
\end{description} 
\mbox{}\\[-1cm]
\mbox{}\hfill $\Box$ %\\[-.5cm]
\end{quotation}

Given two basis elements $L,L'$ of $gl_1(K)$\,, comparison of $\nabla^L\big([L,L']\big)$ and $[\nabla^L(L),\nabla^L(L')]$
reveals equality is tantamount to imposing a set of quadratic constraints on the structure constants, which turn out only to be valid for $K=2$\,. However, the more general\footnote{The superscripts ${\cdot}^K$\,, ${\cdot}^L$\, should not be confused with 
the whole numbers $K$\,, $L$ which are the number of states and number of leaves; the labels are chosen
to coincide with the notation for the generators in the different cases.} $\nabla^K$\,, $\nabla^F$ definitions are latent in the $K=2$ case, in the sense that the in symmetric instance of the latter, the rate matrix is of course obtainable from the permutation operator $K_{(12)}$\,, and furthermore, the general Markov model $GM_2$ is identical to an equal-input model. 
It is a further step to verify the fact $\nabla^K$\,, $\nabla^F$ provide \emph{homomorphisms} -- not only of the Lie algebra, but of the underlying associative matrix algebra. In the case of the permutation matrices, this is in turn related to the permutation group algebra. Recall the multiplication rules $K_\sigma \cdot K_\rho = K_{\rho \sigma}$\, and $R_i \cdot R_j = R_j$\, We have \\[-.6cm]

%\noindent
\begin{quotation}
\noindent
\textbf{Homomorphism property of $\nabla$:}
\begin{description}
\item[Permutation basis.] \mbox{}\\[-.8cm]
\begin{align*}
L_\sigma \cdot L_\rho = &\, L_{\rho \sigma} -L_{\rho}-L_{\sigma}+\unit\,;\\
\nabla^K\big(L_\sigma\big) \cdot \nabla^K\big(L_\rho\big) =&\, \nabla^K\big(L_\sigma \cdot L_\rho\big) =
 L_{\rho \sigma}\otimes  L_{\rho \sigma}-L_{\rho}\otimes  L_{\rho}-L_{\sigma}\otimes L_{\sigma}+\unit\otimes \unit\,;\\
[\nabla^K\big(L_\sigma\big) , \nabla^K\big(L_\rho\big)] =&\,   L_{\rho \sigma}\otimes  L_{\rho \sigma}- L_{\sigma\rho }\otimes  \
L_{\sigma\rho } = \nabla^K\big([L_\sigma , L_\rho]\big) \,.
\end{align*}
\item[Equal-input model.]  \mbox{}\\[-.8cm]
\begin{align*}
F_i\cdot F_j = &\, -F_j\,,\\
\nabla^F\big(F_i \big)\cdot \nabla^F\big(F_j \big) =&\, -F_j  \otimes  F_j + {\mathbb I}\otimes {\mathbb I} = -\nabla^F\big(F_j \big)\,,\\
{[} \nabla^F\big(F_i \big), \nabla^F\big(F_j \big){]} =&\, F_i  \otimes  F_i-F_j  \otimes  F_j = \nabla^F\big({[}F_i,F_j{]}\big)\,.
\end{align*}
\end{description} 
\mbox{}\hfill $\Box$ \\[-.6cm]
\end{quotation}
\mbox{}\\[-.6cm]
\noindent
In summary, we have seen that there is deep algebraic structure bound up with the tensorial construction of the theoretical; phylogenetic branching models. From the Lie algebraic point of view, the emergence of the original coproduct $\nabla^L$ with its additional `quadratic' cross term, is unusual, in that the usual coproduct\footnote{While primitive elements $p$ of a Hopf algebra have coproduct
$\Delta(p) =p\otimes \unit + \unit \otimes p$\,, grouplike elements $g$ have coproduct $\Delta(g) =g\otimes g$\,. The structure of 
$\nabla$ is heuristically explicable if it is borne in mind that the exponentials of rate generators (Markov matrices, $M$ ) can generically be expressed as $\unit + \lambda Q$ for some scalar $\lambda$ and rate generator $Q$ (not necessarily $\ln M$\,). See \cite{johnson1985}.}
 is the minimal (or primitive) choice $\Delta(L) =
L\otimes \unit + \unit \otimes L$. We emphasize that $\nabla$ is also a linear operator, but whose cross terms are defined with reference to the form of the generators in the standard, distinguished basis. It is unknown to what extent the homomorphism property
extends to other phlyogenetic Lie-Markov models, for example those within the ${\mathfrak S}_2\! \wr\! {\mathfrak S}_2$ class 
(see figure \ref{fig:LieMarkovModelsFlowChart}). The above establishes that the property holds for any such model based on permutation matrices, and also, as one example going beyond this, that it also holds for models of Felsenstein type, or models with Lie algebras which are regular subalgebras of the Lie algebra of the Felsenstein model. We call the Lie-Markov models for whose Lie algebras the homomorphism holds, \emph{phylogenetic bialgebras}.

%%%%%%%%%%%%

%%%%%%%%%%%%
%%%%%%%%%%%%

%%%%%%%%%%%%%%%%%%%%%%%%%%%%%%%%%%%%%%%%%%
%%%%%%%%%%%%%%%%%%%%%%%%%%%%%%%%%%%%%%%%%%%%%%%%%%%%%%%%%%%%%%%%%%%%%%%%%%%%%%%%%%%%%
\subsubsection{Phylogenetic tensors and the Star Lemma}
\label{subsubsec:StarLemma}
\mbox{}\\
Our focus so far has been on the general theoretical setting of phylogenetic models, augmented by intensive examination of the structure and symmetry properties of Markov transition matrices for character mutation processes applicable to evolution on individual branches, including diagonalisation or at least simplification, via various changes of basis; in the previous 
section, the intertwining properties of rate generators with the edge splitting operation have been investigated. In this section, we investigate the phylogenetic tensor as a whole, and determine how to extend 
such manipulations and similarity transformations to act across the whole underlying tree. 
For phylogenetic coalgebras this leads to a presentation of the phylogenetic tensor as some `tree' operator acting on the 
phylogenetic ``star tree'' tensor for all leaves emanating from the root.
As we shall see, for certain cases there also  exists an ``inversion'', amounting to a mapping from the observed data -- taken as a sample of the expected pattern frequencies representing the phylogenetic tensor in question -- back to the basic data, of the choice of tree, and its specified spectrum of edge lengths (and possible Markov model parameters). In accord with our approach in this review however, we merely establish the 
admissibility of such inversions, and do not enter into statistical judgements about their reliability.

For this section we revert to the consideration of model classes with a generic number of characters $K$\,.
Recall the fundamental property of Markov rate models which are affiliated with phylogenetic bialgebras, namely that the coproduct of generators defined through intertwining with the comultiplication $\delta$ as $\delta(L_a \pi) = (\nabla L_a) \delta \pi$, also
satisfies the homomorphism condition $\nabla(L_aL_b) = \nabla L_a  \nabla L_b$, and hence for rate matrices
$Q = \sum_a \alpha_a L_a$, $Q = \sum_a \alpha'_a L_a$,we have
\[
\nabla\big(e^Q e^Q{}'\big) = \big(e^{\nabla Q} e^{\nabla Q'}\big)\,.
\]
In what follows, we require notation which reflects the injection of operators across different subspaces (different parts of the multi-way tensor space). 
For the comultiplication itself define $\delta^{\underline{1}}\equiv \delta$, and note that, by associativity,
\[
\delta^{\underline{2}}:= ({\sf Id}\otimes \delta) \scirc \delta^{\underline{1}} \equiv (\delta \otimes {\sf Id})\otimes \delta^{\underline{1}} \,,
\]
which extends iteratively\footnote{In the notation of  \ref{subsec:ModellingGMM}, $\delta^{\underline{n}}:= 
\delta^{(1)}_{(n\!-\!1)}\scirc \delta^{(1)}_{(n\!-\!2)}\scirc \cdots \scirc \delta^{(1)}_{(2)}\scirc \delta^{(1)}_{(1)}$\,.} to 
\[
\delta^{\underline{n}}: V \rightarrow \otimes^n V\,, \qquad \delta^{\underline{n}} := \big( \delta \otimes (\otimes^{n\!-\!1} {\sf Id})\big)\scirc \delta^{\underline{n\!-\!1}}\,.
\]

We denote each edge $e$ by a multi-index string comprising the labels of its descendant subtree\footnote{For brevity, in the form of a 
tuple, without separating commas.}. Now for $|L|$ leaves\footnote{Where confusion does not arise, it is convenient to regard $L$ as both the leaf set, as well as its cardinality.}, define $\underline{L} = \{1,2,3,\cdots, L\}$ and let\footnote{For a planted tree, the string $(1234\cdots L)$ would be the ancestral root edge.} $e\subseteq {[}L{]}$. 
Define\footnote{In the following \scalebox{.95}{$\subseteq$} means \raisebox{-.3ex}{$\stackrel{\scalebox{.6}{$\ne \phi$}}{\hskip-1ex\scalebox{.95}{$\subseteq$}}$} unless otherwise indicated.}
\begin{align}
\label{eq:InclExclCoproduct}
L_a^{[e]}= &\, \sum_{e'\subseteq e} L_a^{e'} \qquad \mbox{where}\qquad 
L_a^{e'} = \prod_{i\in e'} L_a^{(i)}\, \qquad \\
 \mbox{with}\qquad L_a^{(i)} =&\, 
\underbrace{{\mathbb I} \otimes \cdots \otimes{\mathbb I}}_{\text{$i\!-\!1$}} \otimes L_a \otimes 
\underbrace{{\mathbb I} \otimes \cdots \otimes{\mathbb I}}_{\text{$L\!-\!i\!-\!1$}}\,, \nonumber
\end{align}
wherein the operator $L_a$ is inserted at position $i$ with $i\!-\!1$  leading, and $L\!-\!i\!-\!1$ trailing, products with ${\mathbb I}$, respectively. Using this notation, we have for example
\[
\nabla L_a = L_a \otimes L_a + L_a \otimes {\mathbb I} + {\mathbb I}\otimes L_a \equiv L_a^{[(12)]}\,,
\]
and the above enumerative set notation of course reflects coassociativity of $\nabla$\,.
The following properties can be established by induction \cite{sumner2010}:%\\

\begin{quotation}
\textbf{Lemma: iterated coproducts:}
\begin{align*}
\delta^{\underline{n\!-\!1}}\cdot L_{a}=&\, {L}_{a}^{[{\underline{n}}]}\cdot \delta^{\underline{n\!-\!1}}\,, \\
\delta^{\underline{n\!-\!1}}\cdot \exp\left[\left.\sum\right._a \alpha_a L_a\right]=&\, \left[\exp
\left.\sum\right._a \alpha_a L_a^{[{\underline{n}}]}\right]\cdot \delta^{\underline{n\!-\!1}}\,.
\end{align*}
\mbox{}\\[-.7cm]
\mbox{}\hfill $\Box$ %\\[-.5cm]
\end{quotation}
%\mbox{}\\

\begin{figure}[tbp]
%  \centering  
%  $\psmatrix[colsep=.3cm,rowsep=.4cm,mnode=circle]
%  && && &\rho\\
%  && && && 234 \\
%  && && && &34 \\
%  1 && &&  2 && 3 && 4 
%  \ncline{1,6}{4,1}
%  \ncline{1,6}{2,7}
%	\ncline{2,7}{4,5}
%	\ncline{2,7}{3,8}
%	\ncline{3,8}{4,9}
%	\ncline{3,8}{4,7}
%  \endpsmatrix
%  $
\centering
 \scalebox{.8}{
\begin{tikzpicture} 
%1. \draw [help lines] (0,0) grid (7,8); 
3. \shade[fill, top color=gray!20, bottom color=gray!0] (-.3,0) rectangle (7.3,2);
1. \draw[thick] (-.3,0) -- (7.3,0); 
%
%3. \shade[fill, top color=gray!20, bottom color=gray!0] (-.3,1) rectangle (7.3,2);
%1. \draw[thick] (-.3,1) -- (7.3,1); 

%
3. \shade[fill, top color=gray!20, bottom color=gray!0] (-.3,2) rectangle (7.3,4);
1. \draw[thick] (-.3,2) -- (7.3,2); 
3. \shade[fill, top color=gray!20, bottom color=gray!0] (-.3,4) rectangle (7.3,7);
1. \draw[thick] (-.3,4) -- (7.3,4); 
3. \shade[fill, top color=gray!20, bottom color=gray!0] (-.3,7) rectangle (7.3,8);
1. \draw[thick] (-.3,7) -- (7.3,7); 
%
%3. \shade[fill, top color=gray!20, bottom color=gray!0] (-.3,0) rectangle (7.3,1);
%1. \draw[thick] (-.3,8) -- (7.3,8); 
%
2. \draw[ultra thick] (0,0) --  (3.5,7) -- (7,0);
4. \draw[fill, color=white] (0,0) circle (.1);
4. \draw (0,0) circle (.1);
1. \node[below] at (0,-.1) {\scalebox{1.1}{$1$}};
4. \draw[fill, color=white] (7,0) circle (.1);
4. \draw (7,0) circle (.1);
1. \node[below] at (7,-.1) {\scalebox{1.1}{$4$}};
2. \draw[ultra thick] (6,2) --  (5,0); 
4. \draw[fill, color=white] (6,2) circle (.1);
4. \draw (6,2) circle (.1);
1. \node[right] at (6,2.3) {\scalebox{1.1}{$34$}};
4. \draw[fill, color=white] (5,0) circle (.1);
4. \draw (5,0) circle (.1);
1. \node[below] at (5,-.1) {\scalebox{1.1}{$3$}};
%2. \draw[ultra thick] (5.5,1) --  (6,0);
%4. \draw[fill, color=white] (6,0) circle (.1);
%4. \draw (6,0) circle (.1);
%4. \draw[fill, color=white] (5.5,1) circle (.1);
%4. \draw (5.5,1) circle (.1);
4. \draw[ultra thick] (5,4) -- (3,0);
4. \draw[fill, color=white] (5,4) circle (.1);
4. \draw (5,4) circle (.1);
1. \node[right] at (5,4.3) {\scalebox{1.1}{$234$}};
4. \draw[fill, color=white] (3,0) circle (.1);
4. \draw (3,0) circle (.1);
1. \node[below] at (3,-.1) {\scalebox{1.1}{$2$}};
4. \draw[fill, color=white] (3.5,7) circle (.1);
4. \draw (3.5,7) circle (.1);
1. \node[above] at (3.5,7.1) {\scalebox{1.1}{$1234$}};
\end{tikzpicture}}
  \caption{A rooted tree on four leaves.}
  \label{fig:tree2}
\end{figure}
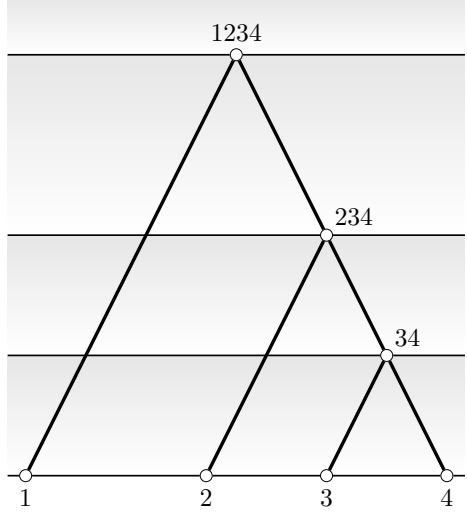

\noindent
In order to motivate the reformulation of the phylogenetic tensor, we examine a specific example. Consider the tree presented in figure~\ref{fig:tree2}, which would be coded as $(r_1,r_2,r_3)=(1,2,3)$ in the notation of \S \ref{subsec:ModellingGMM}. Given a root distribution $\pi$, a rate matrix $Q=\sum_a \alpha_a L_{a}$, and edge weights $\tau_1,\tau_2,\tau_3,\tau_{34}$ and $\tau_{234}$, the phylogenetic tensor corresponding to this tree is
\[
P=e^{\tau_1Q}\otimes e^{\tau_2Q}\otimes e^{\tau_3Q}\otimes e^{\tau_4Q}\scirc {\sf Id} \otimes {\sf Id} \otimes  \delta \scirc {\mathbb I} \otimes {\mathbb I} \otimes e^{\tau_{34}Q}\scirc {\sf Id}\otimes \delta \scirc {\sf Id}\otimes e^{\tau_{234}Q}\scirc (\delta   \pi)\,.\nonumber
\]
Now using the lemma to pull back the Markov transition matrices through successive applications of $\delta$ (at the expense of the iterative coproducts of generators),
\begin{align*}
{\sf Id}\otimes {\sf Id} \otimes \delta\scirc {\mathbb I} \otimes  {\mathbb I} \otimes e^{\tau_{34}Q}=&\, {\mathbb I} \otimes  {\mathbb I} \otimes \left(\delta\scirc e^{\tau_{34}Q}\right)= {\mathbb I} \otimes  {\mathbb I} \otimes 
e^{\tau_{34}Q^{[(34)]}}\scirc {\sf Id}\otimes {\sf Id} \otimes \delta\,;\\
{\sf Id}\otimes \delta^2\scirc {\mathbb I}\otimes e^{\tau_{234}Q}=&\,{\mathbb I}\otimes 
e^{\tau_{234}Q^{[(234)]}}
\scirc 
{\sf Id} \otimes \delta^2\,.
\end{align*}
Thus finally
\[
P=e^{\tau_1Q}\otimes e^{\tau_2Q}\otimes e^{\tau_3Q}\otimes e^{\tau_4Q}\scirc {\mathbb I}\otimes {\mathbb I}\otimes 
e^{\tau_{34}Q^{[(34)]}}\scirc 
{\mathbb I}\otimes 
e^{\tau_{234}Q^{[(234)]}}\scirc {\delta^3\pi}\,.
\]
This example leads to a very pleasant canonical way of writing the phylogenetic tree tensor, as follows\footnote{Here we consider $L$ to be the leaf set with $|L|$ leaves.}:
%\\[.3cm]
\begin{quotation}
\noindent
\textbf{Theorem: Star Lemma for phylogenetic trees:}\\
Consider a rooted tree $\mathcal{T}$ with $|L|$ leaves, and edges $E=\{e_1,e_2,\ldots , e_{2L\!-\!2}\}$ labelled by subsets 
$e \subset \underline{L}$ constituting a valid \emph{split system} for the tree (see \S \ref{subsubsec:Convergence} below).
Given a root distribution $\pi$\,, a rate generator $Q=\sum_a \alpha_a L_a$\,, and edge weights $\{\tau_{e_1},\tau_{e_2},\ldots, \tau_{e_{2L+2}}\}$, the phylogenetic tensor $P$ (or joint distribution at the leaves) is
\begin{equation}
\label{eq:StarFormL}
P=\exp\left[{\mathscr Q}^{[1]}\right]\cdot\exp\left[{\mathscr Q}^{[2]}\right]\cdot \ldots \cdot \exp\left[{\mathscr Q}^{[L\!-\!1]}\right]\cdot  
{\delta^{\underline{L\!-\!1}} \pi}\,,
\end{equation}
where
\begin{equation}
\label{eq:StarFormQi}
{\mathscr Q}^{[\ell]} = \left.\sum\right._{|e| =\ell} \tau_e Q^{[e]}\,, \quad Q^{[e]} := \sum \tau_a L_a^{[e]}\,,
\end{equation}
and  
\[
({\delta^{\underline{L\!-\!1}}\,\pi})^{i_1 i_2 \cdots i_L} :=\left\{ \begin{array}{rl} {\pi}^i, & i_1=i_2 = \cdots =i_L =i\,;\\
0\,,&\mbox{otherwise.} \end{array} \right. 
\]
%\mbox{}\\
\mbox{}\hfill $\Box$\\

\noindent
\textbf{Corollary: Star Lemma in the permutation basis:}\\
From the form of the coproduct of rate generators given earlier (see equ. (\ref{eq:LsigmaCoproduct})), where the model parametric
dependence expresses the rate generator as a sum of permutations, $Q=\sum_\sigma \tau_\sigma \big(K_\sigma-{\mathbb I}\big)$\,, we have 
\begin{equation}
\label{eq:SpecialStarPermsFormP}
P=e^{-\lambda}\exp\left[{\mathscr K}^{[1]}\right]\cdot\exp\left[{\mathscr K}^{[2]}\right]\cdot \ldots \cdot \exp\left[{\mathscr K}^{[L\!-\!1]}\right]\cdot 
{\delta^{\underline{L\!-\!1}}\,\pi}\,,
\end{equation}
where 
\begin{equation}
\label{eq:SpecialStarPermsFormQi}
{\mathscr K}^{[\ell]} = \left.\sum\right._{|e| =\ell} \tau_{\sigma,e} K_\sigma^{e}\,, \quad \mbox{with}\quad K^{e} := \prod_{i\in e} K^{(i)}\,,
\end{equation}
%where in the notation of Eq. (\ref{eq:InclExclCoproduct}),
%\[
%K^{e} = \prod_{i\in e} K^{(i)}\,,
%\]
and $\lambda := \sum_e \tau_e$ is the total edge weight (see equation (\ref{eq:InclExclCoproduct})).
\mbox{}\\
\mbox{}\hfill $\Box$\\
\end{quotation}

The fact that the contributing edge rate generators $Q^{(e)}$ can be ordered by edge cardinality in this way is a rather striking circumstance, but guaranteed by the topology of the tree -- the ${\mathscr Q}^{[\ell]}$ necessarily refer to distinct tensor parts (or ``multi-flattenings") of the overall space; whereas $Q^{(e')}$ for some $e' \supset e$ will definitely entail iterated coproducts of rate generators which overlap with those of $Q^{(e)}$, and hence must appear in a different exponential, ``closer'' to the progenitor star tree
represented by the tensor $\delta^{\underline{L\!-\!1}}\pi$\, because of the larger number of pull-back operations\footnote{This theorem is equivalent to the well-known `spider lemma' in the context of graphical structures and Frobenius algebras \cite{Coecke:Duncan:2008,fauser:2012a}; for further comments see \S\S \ref{subsubsec:HadamardFourier},\ref{subsubsec:Convergence} below and the concluding remarks.}. For parallel formulations we refer to the `multi-taxon process' presentation of \cite{Bryant2009}, and also \cite{Klaere:Liebscher:2012}\,.

In the next sections we further exploit the the Star Lemma reformulation of phylogenetic trees, firstly for a more restricted class of group-like models (\S \ref{subsubsec:HadamardFourier} below, making use of the corollary), and then to display some preliminary results regarding possible generalizations
of Markov models from trees to networks, or divergence-convergence models, when a joining operation, complementary to the splitting operation $\delta$, is introduced.

%%%%%%%%%%%%%%%%%%%%%%%%%%%%%%%%%%%%%%%%%%
%%%%%%%%%%%%%%%%%%%%%%%%%%%%%%%%%%%%%%%%%%%%%%%%%%%%%%%%%%%%%%%%%%%%%%%%%%%%%%%%%%%%%
%%%%%%%%%%%%%%%%
\subsubsection{Hadamard and Fourier transforms across trees}
\label{subsubsec:HadamardFourier}
\mbox{}\\
Here we examine the class of (abelian) group-based models -- those Lie-Markov models associated with abelian permutation subgroups, leaving rate parameters unaltered -- and which moreover form phylogenetic coalgebras. While the present discussion is framed for a generic number of characters $K$\, we restrict the analysis to the exemplifying cases $K=2, 3,4$. The binary symmetric, $K=2$ case, has been intensively studied using the method of the so-called Hadamard transformation, under the aegis of `Fourier analysis on evolutionary trees'
(for a review see \cite{Szekeley:Erdos:Steel1992}\,), and with its extension to the Kimura models (\S \ref{subsubsec:LieMarkovClass}) for $K=4$, provides a paradigmatic model, exemplary case study and starting point for much of the broader analysis of systematics and symmetry underlying the whole class of Lie-Markov models and their diagonalizations that we have presented (\S\S \ref{subsec:ModellingGMM}\,,\ref{subsec:LieMarkov} above).
 
For the binary case we refer the reader to the original literature and reviews \cite{hendy1993,hendy1993c,hendy1994,hendy2008,Szekeley:Erdos:Steel1992,Szekeley:Erdos:Steel1993}\,. Our present analysis relies on exploiting the alternative formulation of the phylogenetic coproduct $\nabla$\,, which applies in these cases of permutation-based actions, noted already in \S \ref{subsubsec:Coproducts} above, which we here briefly recall\footnote{Specifically, we use the operator $\nabla^K$ in forming iterated coproducts.}.
This is exemplified following \cite{sumner:jarvis:holland:2014} with the inversion of the group-based phylogenetic model with $G=\mathbb{Z}_3$ (rather than the binary case $\mathbb{Z}_2$\,); for the extension to the general case of $\mathbb{Z}_r$\, as well as the 4-state case $\mathbb{Z}_2\times \mathbb{Z}_2$, we cite \cite{sumner:jarvis:holland:2014}\,.

In the spirit of \S\S \ref{subsubsec:Catalogue}\,, \ref{subsubsec:Diagonalization}\,, we therefore consider the rate matrix for $K=3$ characters,
\beqn
Q^{{\mathbb Z}_3}&=\left[\begin{array}{ccc}
-(\alpha+\beta) & \beta & \alpha \\
\alpha & -(\alpha+\beta) & \beta \\
\beta & \alpha & -(\alpha+\beta) \\
\end{array}\right]
\equiv -(\alpha+\beta){\mathbb I}+\alpha K_1+\beta K_2,\nonumber
\eqn
where 
\beqn
K_1=\left[\begin{array}{ccc}
0 & 0 & 1 \\
1 & 0 & 0 \\
0 & 1 & 0 \\
\end{array}\right] \equiv K_{(123)},\qquad
K_2=\left[\begin{array}{ccc}
0 & 1 & 0 \\
0 & 0 & 1 \\
1 & 0 & 0 \\
\end{array}\right]\equiv K_{(132)}\,.\nonumber
\eqn
This model $Q^{{\mathbb Z}_3}$ (which we refer to as $Q$ in the following), as a circulant matrix, is a natural extension of the $2\times 2$ `general Markov model'\,, 
\begin{equation}
\label{eq:GM2Defn}
Q^{GM_2}=\left[\begin{array}{rr}
-{\beta} & \alpha \\
\beta & -{\alpha}
\end{array}\right]\,,
\end{equation}
but in common with its symmetric specialization $Q^{{\mathbb Z}_2}$ (with $\alpha = \beta$), admits a decomposition as a linear combination of permutation matrices, with $K_1$, $K_2$ above representing the permutations $(123)$ and $(123)^2= (132)$ under the regular representation, respectively. The matrix which diagonalizes such $Q$,
%We take $\mathbb{Z}_3=\{0,1,2\}_{+\text{ (mod 3)}}\cong\lra{\sigma|\sigma^3=\epsilon}$ and, following the usual conventions in the binary $\mathbb{Z}_2$ case, we label tensor components with indices $i,j=0,1,2$ and allow multiplication $\times$ by extending $\mathbb{Z}_3$ to the ring $\mathbb{F}_3=\{0,1,2\}_{+,\times \text{ (mod 3)}}$. In this case the general rate matrix is given by
%
%
%We define $\omega=e^{{2\pi i}/{3}}$, and present the character table of $\mathbb{Z}_3$ is given in Table~\ref{tab:Z3chartab}.
%The decomposition of the regular representation is $\rho_{\text{reg}}=id\oplus\omega\oplus\omega^2$, and the columns of the character table give the projection operators onto the (one-dimensional) irreducible subspaces:
%\beqn
%\Theta_{id}:&=\fra{1}{3}\left[\epsilon+\sigma+\sigma^2\right]\\\nonumber
%\Theta_{\omega}:&=\fra{1}{3}\left[\epsilon+\omega\sigma+\omega^2\sigma^2\right]\\\nonumber
%\Theta_{\omega^2}:&=\fra{1}{3}\left[\epsilon+\omega^2\sigma+\omega\sigma^2\right]\\\nonumber
%\eqn
\beqn
F=\frac{1}{\sqrt{3}}\left[\begin{array}{ccc}
1 & 1 & 1 \\
1 & \omega & \omega^2 \\
1 & \omega^2 & \omega \\
\end{array}\right]\,, \qquad\mbox{where} \quad \omega = e^{2\pi i/3}\,, \nonumber
\eqn
bears the same relationship with the cyclic group $\mathbb{Z}_3$, as the two-state Hadamard matrix $h$ and its four-state generalization $H$ (see (\ref{eq:Hadamard4x4Def})), have with the groups 
$\mathbb{Z}_2$ and $\mathbb{Z}_2\times \mathbb{Z}_2$\,, respectively (being character tables, up to scaling, and of course implementing the appropriate discrete Fourier transform). Specifically, we have
%%\begin{table}[t]
%%\centering
%%\begin{tabular}{|c|ccc|}
%%\hline
%% & $\texttt{id}$ & $\omega$ & $\omega^2$  \\
%%\hline
%%$[e]$ & 1 & 1 & 1  \\
%%%\hline
%%$[\sigma]$ & 1 & $\omega$ & $\omega^2$ \\
%%$[\sigma^2]$ & 1 & $\omega^2$ & $\omega$ \\
%%\hline
%%\end{tabular}
%%\caption{The character table of $\mathbb{Z}_3$.}
%%\label{tab:Z3chartab}
%%\end{table}
%
%
%
%The similarity transformation by $F$ thus effects a discrete Fourier transform, giving
\beqn
\widehat{Q}=FQF^{-1}=
\left[\begin{array}{ccc}
0 & 0 & 0 \\
0 & \alpha\omega+\beta\omega^2 & 0 \\
0 & 0 & \alpha\omega^2+\beta\omega \\
\end{array}\right],\nonumber
\eqn
or, equivalently,
\beqn
\widehat{K}_1=FK_1 F^{-1}=
\left[\begin{array}{ccc}
1 & 0 & 0 \\
0 & \omega & 0 \\
0 & 0 & \omega^2 \\
\end{array}\right],\qquad
\widehat{K}_2=FK_2 F^{-1}=
\left[\begin{array}{ccc}
1 & 0 & 0 \\
0 & \omega^2 & 0 \\
0 & 0 & \omega \\
\end{array}\right].\nonumber
\eqn
We recall the results stated in the Star Lemma corollary (see (\ref{eq:SpecialStarPermsFormP}),
(\ref{eq:SpecialStarPermsFormQi}) above) that for group-based models, a generic phylogenetic tensor can be expressed as\footnote{In the following \scalebox{.95}{$\subseteq$} means \raisebox{-.3ex}{$\stackrel{\scalebox{.6}{$\ne \phi$}}{\hskip-1ex\scalebox{.95}{$\subseteq$}}$} unless otherwise indicated.}
\beqn
P=e^{-\lambda}\exp\left(\sum_{\emptyset \neq e\subseteq \underline{L\!-\!1}} \left(\alpha_{e}K^{e}_1+\beta_{e}K^{e}_2\right)\right)\cdot 
\delta^{\underline{L\!-\!1}}\cdot \pi,\quad 
\mbox{where} \quad \lambda=\sum_{\emptyset \neq e\subseteq \underline{L\!-\!1}}\left(\alpha_e+\beta_e\right)\,.\nonumber
\eqn
For the present case, this crucial rearrangement
turns out to be compatible with the diagonalization extended to the phylogenetic tensor in its entirety, in such a way that a linear inversion is possible (effectively, a discrete Fourier transform across the whole tree), which maps the phylogenetic tensor $P$ directly on to the set of 
extant edges, and the corresponding edge length parameter spectrum. The diagonal algebraic form of the phylogenetic tensor $\widehat{P}$ thus obtained in the Fourier transform basis, is best summarized in terms of some combinatorial definitions, whose extension to the general $K$-state case should be clear.

Rather than multi-index notation, we adopt an equivalent \emph{ordered tripartition} component tensor labelling convention:
that is, we replace the string $i_1 i_2 \cdots i_L$\, $i_\ell = 0,1,2$ with the string $\underline{u}=\underline{u}_0\sep\underline{u}_1\sep\underline{u}_2$ where
$\underline{u}_a \subseteq  \underline{L}$\,, with $\bigcup_a \underline{u}_a =  \underline{L}$\,, indicating the position of the occurrence of each ternary digit $0,1,2$ in the string; that is,
where $j\in \underline{u}_a$ \,, $a=0,1,2$\,, if and only if $i_j =a$\,. For example with $L=5$, we have
\beqn
00000&\equiv \{1,2,3,4,5\}\sep\emptyset \sep\emptyset, \nonumber\\
00120&\equiv \{1,2,5\}\sep \{3\}\sep \{4\},\\
01122&\equiv \{1\}\sep \{2,3\}\sep \{4,5\}.
\eqn
Let $e \subseteq \underline{L}$ be an edge. We define the following combinatorial attributes associated to 
such $\underline{u}\big(i_1i_2 \cdots i_L\big)$:
%\begin{align*}
%|\underline{u}| =&\, \sum_{a=0}^2 |\underline{u}_a|\equiv L\,, &\, &\, \mbox{(cardinality)}\,;\\
%||\underline{u}|| =&\, \sum_{a=0}^2 a|\underline{u}_a| \equiv 0|\underline{u}|_0 + |\underline{u}|_1+ 2|\underline{u}|_2\,,
%&\, &\, \mbox{(mod 3 weight)}\,;\\
%||e \cap \underline{u}|| =&\, \sum_{a=0}^2 a|e \cap \underline{u}|_a \equiv 0|e\cap \underline{u}|_0 + |e \cap \underline{u}|_1+ 
%2|e \cap \underline{u}|_2\,;\\
%||\underline{u} \cap \underline{v}|| =&\, \sum_{a,b=0}^2 ab |u_a\cap v_b|\,.
%\end{align*}
\begin{align*}
|\underline{u}| :=&\, \sum_{a=0}^2 |\underline{u}_a|\equiv L\,, &\, 
||\underline{u}|| :=&\, \sum_{a=0}^2 a|\underline{u}_a| \equiv 0|\underline{u}_0| + |\underline{u}_1|+ 2|\underline{u}_2| \mod 3\,; \\
 ||\underline{u} \cap \underline{v}|| :=&\, \sum_{a,b=0}^2 ab |u_a\cap v_b|\mod 3\,, &\,
||e \cap \underline{u}|| :=&\, \sum_{a=0}^2 a|e \cap \underline{u}|_a \equiv 0|e\cap \underline{u}_0| + |e \cap \underline{u}_1|+ 
2|e \cap \underline{u}_2|\mod 3\,.
\end{align*}
Here of $|\underline{u}|$ is of course the cardinality, $||\underline{u}||$ is the tripartition weight $\!\!\mod 3$, while 
the full overlap $||\underline{u} \cap \underline{v}||$\, has $||e \cap \underline{u}||$ as a special case.
With the stationary distribution $\pi = \fra 13 (1,1,1)^\top$ we record
\[
\delta^{\underline{L\!-\!1}}\pi = \delta_{||\underline{u}||, 0}\,,\qquad
\big[ F^{\underline{L}} \big]^{\underline{u}}{}_{\underline{v}}= \omega^{||\underline{u}\cap \underline{v}||}/{3}^{\frac 12L}\,,
\quad \mbox{as well as} \quad
\big[\widehat{K_a}{}^{e}\big]^{\underline{u}}{}_{\underline{v}}=
\delta_{\underline{u}\underline{v}}\omega^{a||e\cap\underline{u}||}\,
\]
for $a=0,1,2$ with $K_0 :={\mathbb I}$\,. We can now state the main outcome of the Fourier basis transform \cite{sumner:jarvis:holland:2014}:
%\beqn
%\big[\widehat{K_a}{}^{e}\big]^{\underline{u}}{}_{\underline{v}}=
%\delta_{\underline{u}\underline{v}}\omega^{a||e\cap\underline{u}||}\,, \nonumber
%\eqn
%for $a=1,2$ (as well as $a=0$ if we define $K_0 :={\mathbb I}$)\,. We also record
%\[
%\big[ F^{\underline{L}} \big]^{\underline{u}}{}_{\underline{v}}= \omega^{||\underline{u}\cap \underline{v}||}/{3}^{\frac 12L}\,;
%\]
%and finally 
\beqn
\widehat{P}{\,}^{\underline{u}}%{}_{\underline{v}}
&=e^{-\lambda}\big[\exp\sum_{\emptyset \neq e \subseteq \underline{L\!-\!1}} \alpha_e\widehat{K_1}{}^{e}+\beta_e\widehat{K_2}{}^e \big]
^{\underline{u}}{}_{\underline{u}}%\cdot\delta_{\underline{u}\underline{u}}
\cdot \delta_{||\underline{u}||,0}\,. \nonumber
\eqn
Not only is the operator, which acts to construct the pylogenetic tensor acting on $\delta^{\underline{L\!-\!1}}\pi$\, via the Star Lemma, diagonal in this basis, but also, because of the stucture of $\delta^{\underline{L\!-\!1}}\pi$\,,
the components of the phylogenetic tensor must vanish \emph{unless} the tripartition satisfies $||\underline{u}||=0 \mod 3$\,.

In turn, enumeration of such tripartitions $\underline{u}$ of $\underline{L}$ can be arranged simply by indexing by a choice of tripartition, say $u$, of $\underline{L\!-\!1}$\,, and augmenting the appropriate component\footnote{The fact that the phylogenetic tensor thus effectively depends on a choice of (at most) $2^{L\!-\!1}$ tripartitions is consistent with the fact that the edges $e$ of the model tree
are indeed thus labelled. Effectively, the zeroes of $\widehat{P}$ amount to a set of linear constraints on the components of $P$ in the natural basis arising from the very specific nature of the construcrtion of such tensors under the rules of \S \ref{subsec:ModellingGMM}\,, specialized here to the $Q^{{\mathbb Z}_3}$ model.}
 $\underline{u}_a$\,, $a=0,1,2$\,. 
To process the nonzero components of $\widehat{P}$\,, we therefore consider $u\subseteq \underline{L\!-\!1}$, nominally for the string
$i_1\i_2\cdots i_{L-1}$ say, and introduce $i_L$ as $i_L = 3 \!-\!||u|| \mod 3$\,.
%
%This guarantees that the $\underline{L}$-tripartition for the full string so obtained, $\underline{u}\big(i_1i_2\cdots i_L\big)$\,, will trivially satisfy
%the required condition $||\underline{u}|| = ||u|| + 3 -||u||  \equiv 0 \quad \mbox{(mod 3)}$\,. Given a 
%progenitor $\underline{L\!-\!1}$-tripartition  $u=u_0\sep u_1\sep u_2$\,, 
We denote the augmented
$\underline{L}$-tripartition so obtained as $\underline{u}^+$\,.
Introducing the $3^{L\!-\!1}$-component vectors,
\begin{align}
\label{eq:1stpartZ3}
\mathcal{P}_u:=&\, \widehat{P}{\,}^{\underline{u}^+}\,,\qquad\eta_{u}
:=\big[\left.\sum\right._{\emptyset \neq e\subseteq\underline{L}}\alpha_e\widehat{K_1}{}^{e}+\beta_e\widehat{K_2}{}^e\big]^{\underline{u}^+}{}_{\underline{u}^+}\,,\nonumber\\
\mbox{we have}\qquad 
\mathcal{P}_u=&\, e^{-\lambda}\exp\left(\eta_u\right),\qquad \eta_u=\ln \mathcal{P}_u +\lambda\,.
%\label{eq:1stpartZ3} 
\end{align}
%%%%%%%%%%%%%%%%%%%%NEW ENDING
In order to complete the full inversion for this model, we would like to use (\ref{eq:1stpartZ3}) to recover the rate parameters $\alpha_e,\beta_e$ for all possible edges $\emptyset \neq e\subseteq \underline{L}$\,. This amounts to solving the conditions
\[
\ln {\mathcal P}_u = - \lambda+ \left.\sum\right._e \big( \big[{\mathcal F}_1\big]^u{}_e \alpha_e +  \big[{\mathcal F}_2\big]^u{}_e \beta_e \big)
\]
for $\alpha_e$\,, $\beta_e$\,, where the coefficients ${\mathcal F}_a$ are the $3^{L\!-\!1}\times 2^{L}$ matrices defined by the matrix elements of $\widehat{K_a}$\,, $a=1,2$\, in (\ref{eq:1stpartZ3}).   Remarkably, these can also be recovered as matrix elements of particular tripartition components of the Fourier transform itself, which moreover admit formal $2^{L}\times 3^{L\!-\!1}$ left inverses $\overline{\mathcal F}_a$\,. With the help of the complement $e^c := \underline{L}\backslash e$, we have (as a consequence of $F F^{-1}= {\mathbb I}$ and coproducts thereof) we have
%\begin{align*}
%\big[{\mathcal F}_1\big]^e{}_u:=&\,\big[F^{{[}L-1{]}}\big]^{{e}^c\sep e\sep \emptyset}{}_{u_0\sep u_1\sep u_2}=\omega^{|e \cap u_1|+2|e \cap u_2|}=\big[\widehat{K_1}^{e}\big]^{\underline{u}^+}{}_{\underline{u}^+}\,;\nonumber \\
%\big[{\mathcal F}_2\big]^e{}_u:=&\,\big[F^{{[}L-1{]}}\big]^{{e}^c\sep \emptyset\sep e }{}_{u_0\sep u_1\sep u_2}=\omega^{2|e \cap u_1|+|e \cap u_2|}=\big[\widehat{K_2}^{e}\big]^{\underline{u}^+}{}_{\underline{u}^+}\,;\nonumber
%\end{align*}
\begin{align*}
\big[{\mathcal F}_1\big]^u{}_e:=&\,\big[F^{\underline{L\!-\!1}}\big]^{{e}^c\sep e\sep \emptyset}{}_{u_0\sep u_1\sep u_2}\,; \qquad 
\big[{\mathcal F}_2\big]^u{}_e:=\big[F^{\underline{L\!-\!1}}\big]^{{e}^c\sep \emptyset\sep e }{}_{u_0\sep u_1\sep u_2}\,;\\
\big[\overline{\mathcal F}_1\big]^e{}_u:=&\,\big[{F^{-1}}^{\underline{L\!-\!1}}\big]^{u_0\sep u_1\sep u_2}{}_{{e}^c\sep e\sep \emptyset}\,; \qquad 
\big[\overline{\mathcal F}_2\big]^e{}_u:=\big[{F^{-1}}^{\underline{L\!-\!1}}\big]^{u_0\sep u_1\sep u_2}{}_{{e}^c\sep \emptyset \sep e}\,,
\end{align*}
so that the full inversion for this model is
\beqn
\label{eq:2ndpartZ3}
{\alpha}_e=
%\left[ \big[\overline{\mathcal F}_1\big]^e{}_u +
%\frac{\big[\overline{\mathcal F}_1\big]^e \big[\overline{\mathcal F}_1\big]_u}{1-\langle \overline{\overline{\mathcal F}}\rangle} \right] \eta_u
%\,;
%\qquad {\beta}_e=\left.\sum\right._u \big[{\mathcal F}_2^{-1}\big]^e{}_u {\eta}_u\,.
\left.\sum\right._u \left[ \big[\overline{\mathcal F}_1\big]^e{}_u +
\frac{\big[\overline{\overline{\mathcal F}}_1]^e{}_u}{1-\langle{\overline{\mathcal F}}\rangle} \right] \eta_u
\,;
\qquad {\beta}_e=\left.\sum\right._u \left[ \big[\overline{\mathcal F}_2\big]^e{}_u +
\frac{\big[\overline{\overline{\mathcal F}}_2]^e{}_u}{1-\langle{\overline{\mathcal F}}\rangle} \right] \eta_u\,,
\eqn
where  
\begin{align*}
\lambda = &\, \left.\sum\right._{a,u} \frac{\big[{\overline{\mathcal F}}_a]_u}{1-\langle{\overline{\mathcal F}}\rangle} \eta_u\,,
\quad \big[{\overline{\mathcal F}}_a]_u  = \left.\sum\right._{e} \big[{\overline{\mathcal F}}_a\big]^e{}_u\,;\\
\big[\overline{\overline{\mathcal F}}_a]^e{}_u=&\, \left.\sum\right._{f,v}\big[\overline{\mathcal F}_a\big]^e{}_v \big[\overline{\mathcal F}\big]^f{}_u\,,\qquad
\langle{\overline{\mathcal F}}\rangle = \left.\sum\right._{a,e,u}\big[\overline{\mathcal F}_a\big]^e{}_u\,,
\end{align*}
Together with (\ref{eq:1stpartZ3}), these equations give a one to one map between pattern probabilities and edge weights for the group-based model ${\mathfrak Q}^{\mathbb{Z}_3}$.

As mentioned above, we have developed this case as a template for our tensorial approach to model diagonalization across the tree based on the Star Lemma rearrangement property. More generally, the $K$-state case with cyclic group ${\mathbb Z}_K$ can be 
treated analogously, by appropriate minor adjustment of the notation, as a natural generalization of this and the ${\mathbb Z}_2$ case. Without use of the Star Lemma, the application of the standard Hadamard transformation in the binary case entrains inclusion-exclusion
counting methods over tree path sets, and hence is more intimately related to the geometry of the tree than our direct method. On the other hand, the use of multi-partition labelling remains a natural and powerful tool: \emph{bi}partitions in the binary ${\mathbb Z}_2$ case;
\emph{bi-bi}partitions (\emph{quadri}partitions, that is, ordered pairs of bipartitions) in the ${\mathbb Z}_2\times {\mathbb Z}_2$ case (which applies to the inversion of the Kimura models for example); \emph{tri}partitions for ${\mathbb Z}_3$\, treated here; and \emph{tetra}partitions\footnote{As noted in \S 
\ref{subsubsec:Diagonalization}, model (3.3b) of the Lie Markov hierarchy, with rate matrix $Q^{\widetilde{K3ST}}$\,, is a 3-parameter group-based model with cyclic symmetry ${\mathbb Z}_4$ and admits such a discrete Fourier inversion.} 
for ${\mathbb Z}_4$. 
%%%%%%%%%%%%%%%%%%%%%%%%%%%%%%%%%%%%%%%%%%
%%%%%%%%%%%%%%%%%%%%%%%%%%%%%%%%%%%%%%%%%%%%%%%%%%%%%%%%%%%%%%%%%%%%%%%%%%%%%%%%%%%%%
\subsubsection{Modelling phylogenetic convergence}
\label{subsubsec:Convergence}
\mbox{}\\
We now take up a further implication of the Star Lemma rearrangement of phylogenetic tensors, which as we shall see, bears on the significant  question of how to extend phylogenetic models beyond trees, and towards networks. The construction of parametric stochastic tensors
based on networks, as opposed to trees, for phylogenetic modelling, is currently an open question. Our analysis leads instead to variant class of `divergence-convergence' models, which nonetheless have some network-like features.

In this section we restrict attention to the binary general Markov model (equation (\ref{eq:GM2Defn})), although the initial discussion applies generically.
The rate generator (\ref{eq:GM2Defn}) can be written 
\[
Q^{GM_2} = \alpha L_1 + \beta L_2\,, 
\]
with $L_1=L_{12}$ and $L_2:= L_{21}$ the basis of stochastic generators (see \S \ref{subsec:ModellingGMM}). As in the previous section on the Fourier inversion, we shall in fact specialize to the \emph{symmetric} case $Q^{{\mathbb Z}_2}$ with $\beta =\alpha$\,; in view of the 
scaling by edge lengths, we can take $\beta=1=\alpha$ without loss of generality. As we have seen, in this case we have
$L_1+L_2 = K -{\mathbb I}$, where $K\equiv K_{(12)}$ is the permutation matrix which interchanges characters in the natural basis. For the present discussion, in view of the Star Lemma and its corollary, we refer to these as the ``$L$'' and ``$K$'' forms, respectively.

Consider again the detailed statement of the Star Lemma \S \ref{subsubsec:StarLemma}, equation (\ref{eq:StarFormL})\,. The phylogenetic tensor is constructed as a product of a string of exponentials 
\beqn
\exp\left[{\mathscr Q}^{[\ell]}\right]:=\exp\left[\left.\sum\right._{|e|=\ell}\tau_{e}{Q}^{[e]}\right]=\left.\prod\right._{e,|e|=\ell}
\exp\left[\tau_e{Q}^{[e]}\right]\,,\qquad 
{Q}^{[e]} = \sum_{e'\subseteq e} \alpha_e L_1^{(e')} +\beta_eL_2^{(e')}\,,
\eqn
pertaining to the edge decomposition of the tree (with $\alpha_e=1=\beta_e$ in the symmetric case), with $2L\!-\!2$ assigned lengths 
$\tau_{e_1}\, \tau_{e_2}\,\cdots, \tau_{e_{2L\!-\!2}}$\,.

There is an obvious extension \cite{sumner2010,Bryant2009} of the above tree-based phylogenetic tensor to a general stochastic tensor, where the sum of exponential contributions extends over an arbitrary collection of subsets $\emptyset\subset A \subset {[}L{]}$\,, not only those subsets $e$ present as tree edges. In principle such a list (conventionally written with $\underline{L}$-complements as $\{ A|A^c \}$\, technically a \emph{split system} for 
$\underline{L}$\,, could include edges which can only be displayed on subtrees of a graphical network diagram (see for example \cite{bryant2005}). Of course, as such a network loses directedness, the resulting tensor, while still stochastic, no longer corresponds to a single evolutionary branching process.

These difficulties notwithstanding, let us pursue the technical issues arising from this suggested generalization, in relation to the Star Lemma and its corollary (which applies for the symmetric model). As shown in \cite{sumner2010}, for the edges chosen from a tree, if 
%$|e'|<|e|$ then either $e'\subset e$ or 
$e'\cap e=\emptyset$\,, in these cases $L_1^{(e')}\, ,L_2^{(e')} $ commute with
$L_1^{(e)}\, ,L_2^{(e)} $\, (as evident from the Star Lemma construction)\,, while each set has the same algebra as the basic $L_1,L_2$\,. 
Moreover, we have that both
$L_1^{(e)} + L_2^{(e)}$ and $K^e - {\mathbb I}\otimes {\mathbb I}\otimes\cdots {\mathbb I}$ have the same action on the image of 
the splitting operator $\delta^{\underline{L\!-\!1}}$\, -- ensuring of course that (in the symmetric case) both the ``$L$'' presentation
and the ``$K$'' presentation construct the same phylogenetic tensor, consistent with the Star Lemma corollary (see also 
\cite{jarvis2001}).

For \emph{arbitrary} split systems, however, this equivalence of presentations is no longer true, and the resolution
of the difference leads to the divergence-convergence model class for the ``$L$'' presentation, as we now show in the context of a concrete example.

\begin{figure}[tbp]
%\label{fig:3TaxonConv}
\centering
%\begin{tabular}{cc}
%(a)\hspace{2em}
%$
%\psmatrix[colsep=.3cm,rowsep=.7cm,mnode=circle,arrowscale=2]
%&&&\rho\\
%[mnode=dot,dotscale=.00001]&&[mnode=dot,dotscale=1]&&&&&&[mnode=dot,dotscale=.00001]\\
%[mnode=dot,dotscale=.00001]\\
%&1&&2&&&&3
%\ncline{1,4}{2,3}
%\ncline{1,4}{4,8}
%\ncline{2,3}{4,2}
%\ncline{2,3}{4,4}
%\psset{linestyle=dashed}\ncline{2,1}{2,9}
%\endpsmatrix
%$ 
%
%&
%(b)\hspace{2em} 
%$
%\psmatrix[colsep=.3cm,rowsep=.7cm,mnode=circle,arrowscale=2]
%&&&&\rho\\
%[mnode=dot,dotscale=.00001]&&&[mnode=dot,dotscale=1]&&&[mnode=dot,dotscale=1]&[mnode=dot,dotscale=.00001]\\
%[mnode=dot,dotscale=.00001]&&&&[mnode=dot,dotscale=1]&[mnode=dot,dotscale=1]&&[mnode=dot,dotscale=.00001]\\
%&1&&2&&&3
%\ncline{1,5}{2,4}
%\ncline{1,5}{2,7}
%\ncline{2,4}{4,2}
%\ncarc[arcangle=40]{2,4}{3,5}
%\ncarc[arcangle=-40]{2,7}{3,6}
%\ncline{3,5}{4,4}
%\ncline{3,6}{4,7}
%\psset{linestyle=dashed}\ncline{2,1}{2,8}
%\psset{linestyle=dashed}\ncline{3,1}{3,8}
%\endpsmatrix
%$ 
%\\
%\end{tabular}
\scalebox{.8}{\begin{tikzpicture} 
3. \shade[fill, top color=gray!25, bottom color=gray!0] (-.3,0) rectangle (5.3,3);
3. \shade[fill, top color=gray!25, bottom color=gray!0] (-.3,3) rectangle (5.3,5);
1. \draw[thick] (-.3,3) -- (5.3,3); 
3. \shade[fill, top color=gray!20, bottom color=gray!0] (-.3,5) rectangle (5.3,5.6);%
4. \draw[ultra thick] (1.5,3) -- (2.5,5);
4. \draw[ultra thick] (4.5,0) -- (2.5,5);
4. \draw[fill, color=white] (2.5,5) circle (.1);
4. \draw (2.5,5) circle (.1);
4. \draw[ultra thick] (0.5,0) -- (1.5,3);
4. \draw[ultra thick] (2.5,0) -- (1.5,3);
4. \draw[fill, color=white] (1.5,3) circle (.1);
4. \draw (1.5,3) circle (.1);
4. \draw[fill, color=white] (0.5,0) circle (.1);
4. \draw (0.5,0) circle (.1);
4. \draw[fill, color=white] (2.5,0) circle (.1);
4. \draw (2.5,0) circle (.1);
4. \draw[fill, color=white] (4.5,0) circle (.1);
4. \draw (4.5,0) circle (.1);
1. \node[below] at (0.5,-.1) {$1$};
1. \node[below] at (2.5,-.1) {$2$};
1. \node[below] at (4.5,-.1) {$3$};
1. \node[above] at (2.5,5.1) {$\pi$};
1. \node[left] at (-.3,.1) {\scalebox{1.1}{$(a)$}};
%\draw [help lines] (0,0) grid (5,6); 
\end{tikzpicture}}
\mbox{}\hskip6ex
%
%%%%%%%%%%%%%%%%%%%%%%%%%
%%%%%%%%%%%%%%%%%%%%%%%%% RHS of bendy graph
\scalebox{.8}{\begin{tikzpicture} 
3. \shade[fill, top color=gray!20, bottom color=gray!0] (-.3,0) rectangle (5.3,1.5);
%3. \shade[fill, top color=gray!25, bottom color=gray!0] (-.3,1.5) rectangle (5.3,3);
3. \shade[fill, top color=gray!25, bottom color=gray!0] (-.3,1.5) rectangle (5.3,3);
\path [fill, color=gray!30] (1,3) -- (4,3) to [out=210,in=90] (3,1.5) -- (2,1.5) to [out=90, in= 330] (1,3);
3. \shade[fill, top color=gray!25, bottom color=gray!0] (-.3,3) rectangle (5.3,5);
1. \draw[thick,dashed] (-.3,3) -- (5.3,3); 
3. \shade[fill, top color=gray!20, bottom color=gray!0] (-.3,5) rectangle (5.3,5.6);%
1. \draw[thick,dashed] (-.3,1.5) -- (5.3,1.5); 
4. \draw[ultra thick] (1,3) -- (2.5,5);
4. \draw[ultra thick] (4,3) -- (2.5,5);
4. \draw[ultra thick] (0,0) -- (1,3);
4. \draw (2,1.5) circle (.1);
4. \draw[fill, color=white] (0,0) circle (.1);
4. \draw (0,0) circle (.1);
\draw[ultra thick] (1.5,0) -- (2,1.5);
\draw[ultra thick,color=gray!100] (3,1.5) to [out=90,in=210] (4,3);
4. \draw[ultra thick,color=gray!100] (2,1.5) to [out=90,in=330] (1,3);
4. \draw[fill, color=gray!90] (4,3) circle (.1);
4. \draw (4,3) circle (.1);
4. \draw[fill, color=white] (1.5,0) circle (.1);
4. \draw (1.5,0) circle (.1);
4. \draw[fill, color=gray!90] (3,1.5) circle (.1);
4. \draw (3,1.5) circle (.1);
4. \draw[fill, color=gray!90] (2,1.5) circle (.1);
4. \draw (2,1.5) circle (.1);
4. \draw[fill, color=gray!90] (1,3) circle (.1);
4. \draw (1,3) circle (.1);
4. \draw[fill, color=white] (2.5,5) circle (.1);
4. \draw (2.5,5) circle (.1);
\draw[ultra thick] (3,1.5) -- (4,0);
\draw[fill, color=white] (4,0) circle (.1);
\draw (4,0) circle (.1);
4. \draw[fill, color=gray!90] (3,1.5) circle (.1);
4. \draw (3,1.5) circle (.1);
%4. \draw[fill, color=white] (5,0) circle (.1);
%4. \draw (5,0) circle (.1);
1. \node[below] at (0,-.1) {$1$};
1. \node[below] at (1.5,-.1) {$2$};
1. \node[below] at (4,-.1) {$3$};
1. \node[above] at (2.5,5.1) {$\pi$};
1. \node[left] at (-.3,.1) {\scalebox{1.1}{$(b)$}};
%\draw [help lines] (0,0) grid (5,6); 
\end{tikzpicture}}
%\label{fig:3TaxonConvTree}
\caption{\label{fig:3TaxonConv} A three taxon tree (a) is modified by the introduction of the additional split ${23}|{1}$ in (b).}
\end{figure}
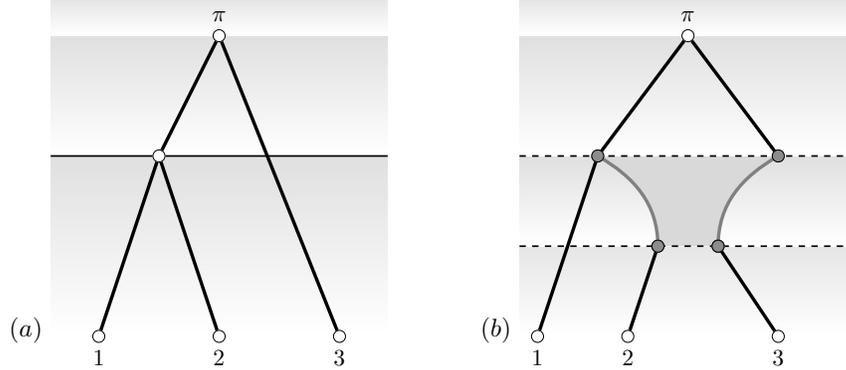

Consider the three taxon phylogenetic tree given in figure \ref{fig:3TaxonConv}(a) with edges $1, 2, 3$ and $12$.
For the binary symmetric model, we get an identical probability distribution if we use either the ``${L}$''-presentation 
or the ``$K$''-presentation\,, respectively, where
\begin{align*}
P=&\,\exp\left[\tau_{1}{ Q}^{[1]}+\tau_{2}{ Q}^{[2]}+\tau_{3}{ Q}^{[3]}\right]\cdot\exp
\left[\tau_{12}{ Q}^{[12]}\right]\cdot {\delta^{\underline{2}} \pi}\,,\\
P=&\,e^{-\lambda}\exp\left[\tau_{1}K^{(1)}+\tau_{2}K^{(2)}+\tau_{3}K^{(3)}+\tau_{12}K^{(12)}\right] \cdot {\delta^{\underline{2}} \pi}\,,
\qquad \mbox{with} \quad \lambda=\tau_{1}+\tau_{2}+\tau_{3}+\tau_{12}\,.
\end{align*}

We wish to introduce an additional parameter $\tau_{23}$ associated an ``imaginary'' split ${23}|{1}$ to these probability distributions.  
Consistently with the design given in figure \ref{fig:3TaxonConv}(b), the evolutionary history is broken up into three epochs: I: divergence of taxon 3 away from 1 and 2; II: concurrent evolution of taxa 2 and 3, with independent divergence of taxon 3; and III: independent divergence of all taxa.
To model this situation we must introduce the additional edge to each representation. 
For simplicity, we set a molecular clock on the model such that $\tau_{2}=\tau_{1}$ and $\tau_{3}=\tau_{{12}}+\tau_{{1}}$, and we introduce a scaling parameter $\theta\in\left[0,1\right]$ to control the length of the second epoch as a proportion of the third epoch by setting $\tau_{{23}}=\theta\tau_{1}$ .
As all operators in the $K$-representation commute, the only choice available in this case is to take
\beqn
P'_K=e^{-\lambda}\exp\left[\tau_1K^{(1)}+(1\!-\!\theta)\tau_1K^{(2)}+\left(\tau_{12}+(1\!-\!\theta)\tau_1\right)K^{(3)}+\tau_{12}K^{(12)}+\theta\tau_{1} K^{(23)}\right]\cdot {\delta^{\underline{2}} \pi}\,.\nonumber
\eqn
This is exactly consistent with the generalizations given in \cite{bryant2005,Bryant2009}.

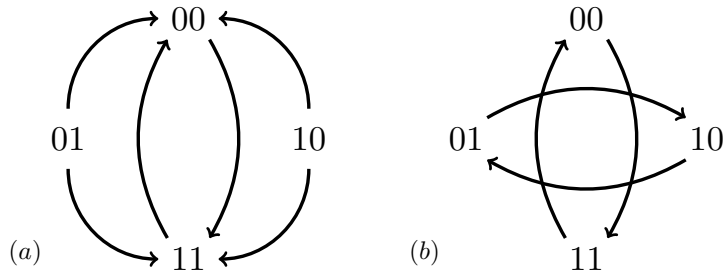
\begin{figure}
%\label{fig:ratediag}
\centering
%\begin{tabular}{cccc}
%(a)\hspace{2em}$
%\psmatrix[colsep=.3cm,rowsep=.3cm]
%&\Ket{x00}&\\
%\Ket{x01}&&\Ket{x10}\\
%&\Ket{x11}&
%\ncarc[arcangle=30]{<-}{3,2}{2,1}
%\ncarc[arcangle=30]{->}{2,1}{1,2}
%\ncarc[arcangle=-30]{<-}{3,2}{1,2}
%\ncarc[arcangle=-30]{<-}{1,2}{3,2}
%\ncarc[arcangle=-30]{<-}{3,2}{2,3}
%\ncarc[arcangle=30]{<-}{1,2}{2,3}
%\endpsmatrix
%$
%&\hspace{4em}&
%(b)\hspace{2em}$
%\psmatrix[colsep=.9cm,rowsep=.9cm]
%\Ket{x00}&\Ket{x01}\\
%\Ket{x11}&\Ket{x10}
%\ncarc[arcangle=30]{->}{1,1}{2,1}
%\ncarc[arcangle=30]{->}{2,1}{1,1}
%\ncarc[arcangle=30]{->}{1,2}{2,2}
%\ncarc[arcangle=30]{->}{2,2}{1,2}
%\endpsmatrix
%$
%\end{tabular}
\scalebox{.8}{\begin{tikzpicture} 
%\draw [help lines] (0,0) grid (4,4); 
\node at (0,2) {\scalebox{1.5}{$01$}};
\node at (2,4) {\scalebox{1.5}{$00$}};
\node at (4,2) {\scalebox{1.5}{$10$}};
\node at (2,0) {\scalebox{1.5}{$11$}};
\draw[ultra thick, ->] (0,2.5) to [out=90,in=180] (1.5,4);
\draw[ultra thick, ->] (0,1.5) to [out=270,in=180] (1.5,0);
\draw[ultra thick, ->] (4,2.5) to [out=90,in=0] (2.5,4);
\draw[ultra thick, ->] (4,1.5) to [out=270,in=0] (2.5,0);
\draw[ultra thick, ->] (1.65,.35) to [out=120,in=240] (1.65,3.65);
\draw[ultra thick, ->] (2.35,3.65) to [out=300,in=60] (2.35,.35);
1. \node[left] at (-.3,.1) {\scalebox{1.1}{$(a)$}};
\end{tikzpicture}
\hskip6ex
\begin{tikzpicture} 
%\draw [help lines] (0,0) grid (4,4); 
\node at (0,2) {\scalebox{1.5}{$01$}};
\node at (2,4) {\scalebox{1.5}{$00$}};
\node at (4,2) {\scalebox{1.5}{$10$}};
\node at (2,0) {\scalebox{1.5}{$11$}};
%\draw[ultra thick, ->] (0,2.5) to [out=90,in=180] (1.5,4);
%\draw[ultra thick, ->] (0,1.5) to [out=270,in=180] (1.5,0);
%\draw[ultra thick, ->] (4,2.5) to [out=90,in=0] (2.5,4);
%\draw[ultra thick, ->] (4,1.5) to [out=270,in=0] (2.5,0);
\draw[ultra thick, ->] (1.65,.35) to [out=120,in=240] (1.65,3.65);
\draw[ultra thick, ->] (2.35,3.65) to [out=300,in=60] (2.35,.35);
\draw[ultra thick, ->] (.35,2.35) to [out=30,in=150] (3.65,2.35);
\draw[ultra thick, ->] (3.65,1.65) to [out=210,in=330] (.35,1.65);
1. \node[left] at (-.3,.1) {\scalebox{1.1}{$(b)$}};
\end{tikzpicture}}
\caption{\label{fig:ratediag}Transitions undergone by states across an alignment if an operator for an incompatible 
split is introduced in the construction. In the three leaf case, the action on tensor components representing taxa 2 and 3 are given, under the operators (a) ${Q}^{{[23]}}$, and (b) $K^{{(23)}}-
{\mathbb I}\otimes {\mathbb I}\otimes {\mathbb I}$.}
\end{figure}

For the ${L}$-representation however we do \emph{not} have commutativity of the operators ${ Q}^{[2]},{ Q}^{[3]},{ Q}^{{[12]}}$ with the new operator ${ Q}^{{[23]}}$. Using the diagram and its three epochs as a guide, we take
\beqn
P'_{{L}}=\exp\Big[(1\!\!-\!\!\theta)\tau_{1}\left({ Q}^{[1]}\!+\!{ Q}^{[2]}\!+\!{ Q}^{[3]}\right)\Big]\cdot\exp\Big[\theta\tau_{1}\left({ Q}^{[1]}\!+\!{ Q}^{[23]}\right)\Big]\cdot\exp\Big[\tau_{12}\left({ Q}^{[3]}\!+\!{ Q}^{[12]}\right)\Big] \cdot {\delta^{\underline{2}} \pi}\,.\nonumber
\eqn
The $K$\!-\!representation in epoch form reads 
\beqn
P'_K=e^{\!-\!\lambda}\exp\left[(1\!\!-\!\!\theta)\tau_{1}\left(K^{{(1)}}\!+\!K^{{(2)}}\!+\!K^{{(3)}}\right)\right]\cdot&\exp\left[\theta\tau_{1}\left(K^{{(1)}}\!+\!K^{{(23)}}\right)\right]
\cdot\exp\left[\tau_{{12}}\left(K^{{(3)}}\!+\!K^{{(12)}}\right)\right]\cdot {\delta^{\underline{2}}\pi}\,.\nonumber
\eqn
Now consider the state of the probability distribution at the beginning of epoch II. 
As we are dealing with the binary symmetric model, it is clear that the probability of the any state ${ijk}$ is invariant to permutation of the states $0$ and $1$.
Also, the structure of the tree up to the start of epoch II implies that any state of the form ${ijk}$ where $i\neq j$ is of probability zero.
Thus at the start of epoch II the distribution $P$ is of the form
%\footnote{In index bipartition notation, condensed 
%from $u_0:u_1$ to simply $u_1$\,, thus $\emptyset \leftrightarrow 000, [3] \leftrightarrow 123$\, and so on.}
%\begin{align*}
%P^\emptyset = &\, P^{\{123\}}= \textstyle{\frac 12}(1-q)\,,\qquad &\, P^{\{3\}}= &\, P^{\{12\}}= \textstyle{\frac 12}(1-q)\,.
%\end{align*}
\begin{align*}
P^{000} = &\, P^{111}= \textstyle{\frac 12}(1-q)\,,\qquad  P^{001}=  P^{110}= \textstyle{\frac 12}(1-q)\,,
\end{align*}
for some parameter $0\le q < \textstyle{\frac 12}$ in a continuous-time model.
Using the definitions we have
\begin{align*}
{Q}^{[(23)]}=&\, L_1^{[(23)]} + L_2^{[(23)]}=
{\mathbb I}\otimes\left(L_1\otimes L_1+L_1\otimes {\mathbb I}+{\mathbb I}\otimes L_1+L_2\otimes L_2+L_2\otimes {\mathbb I}+{\mathbb I}\otimes L_2\right)\,,\nonumber\\
\mbox{versus}\quad 
K^{(23)}=&\,{\mathbb I}\otimes K\otimes K={Q}^{[(23)]}+{\mathbb I}\otimes\left(L_1\otimes L_2+L_2\otimes L_1\right)\,.\nonumber
\end{align*}
It follows that transition rates between the four existing states in the two cases are given by the two graphs in figure \ref{fig:ratediag}, where all transition rates are equal.
The crucial thing to note is that ${Q}^{[(23)]}$ ``corrects'' patterns that are inconsistent with the split ${23}|{1}$, whereas 
$K^{(23)}$ simply \emph{permutes} these two states. 
The ``${L}$'' presentation thus introduces a natural notion of the ``coming together'' of taxa. 
In fact it is easy to see directly from the diagram that in the limit of extension of the edge $\tau_{{23}}$ to infinity, that the probability distribution will converge to 
\begin{align*}
P^{000} = &\, P^{111}= \textstyle{\frac 12}-\textstyle{\frac 14}q\,,\qquad  P^{001}=  P^{110}= \textstyle{\frac 14}q\,,
\end{align*}
%
%
%
%\beqn
%P&=\left(\left(1-\fra{1}{2}q\right)\fra{1}{2}+\fra{1}{4}q\right)\left(\Ket{000}+\Ket{111}\right)+\fra{1}{2}q\fra{1}{2}\left(\Ket{011}+\Ket{100}\right)\\
%&=\left(1-\fra{1}{2}q\right)\fra{1}{2}\left(\Ket{000}+\Ket{111}\right)+\fra{1}{2}q\fra{1}{2}\left(\Ket{011}+\Ket{100}\right),\nonumber
%\eqn
which is consistent with a probability distribution where taxon 1 has diverged from 2 and 3, but there has been zero divergence of taxa 2 and 3 themselves. This behaviour motivates the graphical representation given in figure \ref{fig:3TaxonConv}~(b).
The ``$K$"-presentation cannot achieve this type of convergence, with its limiting state being
%\beqn
%(1-q)\fra{1}{4}\left(\Ket{000}+\Ket{011}+\Ket{111}+\Ket{100}\right)+q\fra{1}{4}\left(\Ket{001}+\Ket{010}+\Ket{110}+\Ket{101}\right).\nonumber
%\eqn
\begin{align*}
P^{000} =  P^{011}=   P^{011}=  P^{100}= \textstyle{\frac 14}-\textstyle{\frac 14}q\,,\qquad  
P^{000} =  P^{010}=   P^{110}=  P^{101}= \textstyle{\frac 14}-\textstyle{\frac 14}q\,.
\end{align*}

In \cite{mitchell2018distinguishing} the possibility of such phylogenetic convergence has been developed, from the point of view of model consistency and identifiability. In some circumstances, with three or more taxa, it turns out that identical phylogenetic tensors derived from standard models can be constructed, with the same number of parameters, from alternative trees which also display convergence events. In other cases, indeed, the divergence-convergence constructions are in direct competition of standard tree models, and in principle map on to non-overlapping phylogenetic tensors. Returning to our general introduction where the limits of phylogenetic modelling were identified, and given the abundance of evidence that evolution is not strictly treelike, divergence-convergence models of the type treated here serve as a salutory reminder of the necessity for careful analysis of the structural form of phylogenetic models.

	%%%%%%%%%%%%%%%%%%%%%%%%%%%%%%%%%%%%%%%%%%%%%
	%\input{secs/Entanglement.tex}
	%\section{Entanglement in phylogenetic data}
	%\label{sec:Entanglement}
%%%%%%%%%%%%%%%%%%%%%%%%%%%%%%%%%%%%%%%%%%%%
%%%%%%%%%%%%%%%%%%%%%%%%%%%%%%%%%%%%%%%%%%%%%%%%%%%%%%%%%%%%%%%%%%%%%%%%%%%%%%%%%%%%%%%%%%%%%%%%%%%%%%%%%%%%%%%%%%%%%%%%%%
\section{Entanglement and Markov invariants in phylogenetic data}
\label{sec:Entanglement}
%%%%%%%%%%%%%%%%%%%%%%%%%%%%%%%%%%%%%%%%%%%%%%%%%%%%%
\subsection{Entanglement measures: quantum versus stochastic setting}
\label{sec:EntanglementQvsS}

In this review we have been at pains to expose structural aspects of molecular phylogenetic modelling at the theoretical level, with a view to improved understanding of limitations on the adequacy of the standard models, model choice and consistency, and ultimately, in 
facilitating parameter recovery. Some of these themes are brought together in the present section. We return to the analysis 
of the phylogenetic tensors and underlying models, and their behaviour under stochastic transformations. We identify key
quantities which deal with these transformations as `nuisance' parameters, and whose properties thus give direct information on the underlying tree. This entails the identification of certain \emph{polynomial invariants} which characterize the algebraic properties 
known as \emph{entanglement} of the tensors and their marginalisations. There are remarkable parallels between 
this setting, involving what we term the `Markov invariants' in phylogenetic models, and the setting of composite quantum systems
and their affiliated local unitary invariants and entanglement measures.

We should note at the outset that other polynomial invariants, of a somewhat different nature from the Markov invariants, have long been known 
and applied in phylogenetics. These are the `phylogenetic invariants' of Cavender and Felsenstein \cite{CavenderFelsenstein1987}  and Lake \cite{lake1987}\,, which can access model
inference and parameter refinement via the methods of algebraic geometry (see for example \cite{allman2003, allman2008}, for an overview and recent progress see \cite{eriksson2008}, \cite{CasanellasFernandezSanchez2016}). In the literature  the term `phylogenetic invariant', or perhaps better, `phylogenetic identity' \cite{draisma2008}, is used to refer to any polynomial which vanishes on all distributions arising from a subset of phylogenetic tree topologies (understood as leaf-labelled trees). If the subset is proper, the phylogenetic invariant is referred to as `tree informative'. In our terminology, however, the `Markov invariants' are invariants in the sense of classical invariant theory \cite{olver2003}\,, that is, invariants under a group action, in this case (as discussed in the foregoing sections) the action of the Markov group of (complex) stochastic transformations. In \cite{sumner2017developing}, we explored the interrelationship between phylogenetic identities and Markov invariants in the special case of two state models and quartet trees, but in general, the two types of polynomial are distinct.

We now turn to our theme of Markov invariants and entanglement. Recall the basic arena of theoretical phylogenetics as outlined in \S\S \ref{sec:IntroOverview}, \ref{sec:ModellingIntro}. The ambient model space $V \cong {\mathbb C}^K$ acts as a receptacle for the convex space of (real) probability tensors, and provides through its tensor products $V \otimes V \otimes \cdots \otimes V \cong \otimes^L V$, the raw material for probability arrays populated by alignment sampling pattern frequencies. These in turn are subject to independent  transformations by stochastic substitution matrices on each part. According to equation (\ref{eq:PfromClipP}), a general $L$-way phylogenetic tensor modelling a tree has the form 
\begin{align}
\label{eq:PfromClipP2}
{P}_{\mathcal T}=&\, M_1 \otimes M_2 \otimes \cdots \otimes M_L \cdot \overline{P}_{\mathcal T}\,.
\end{align}
Evidently, ${P}_{\mathcal T}$ is obtained from $\widehat{P}_{\mathcal T}$ by an element of the extended group
$\times^L GL_1(K) \cong GL_1(K) \times GL_1(K)\times \cdots \times GL_1(K)$\,; more generally, we consider how
${P}_{\mathcal T}$ changes under arbitrary local stochastic transformations, taken independently on each component:
\begin{align}
\label{eq:P'fromP}
{P}_{\mathcal T}\rightarrow &\,{P}_{\mathcal T}' = M_1 \otimes M_2 \otimes \cdots \otimes M_L \cdot {P}_{\mathcal T}\,.
\end{align}

This then is the mathematical setting of \emph{classical invariant theory}, and there is an exact parallel in the description of composite quantum systems -- the ambient space $V$ describes single system pure states (e.g. the ubiquitous `qubit'), or mixed states (for density operators one takes instead $W \cong V \otimes V^*$), and for multipartite systems the state space is the appropriate tensor product.
Pure states or their mixed state counterparts are subject to time evolution (by the local unitary group, e.g. $U(2)$ for qubits), or general quantum evolution incorporating measurement, and other invertible operations (for example $SL(2,{\mathbb C})$ in the qubit case).
Quantum information protocols for manipulating quantum states rely on entanglement, and these in turn are characterized by the polynomial invariants which are local entanglement measures, as we now describe for the phylogenetics case.
In fact, the analogy between the two fields is sufficiently close that in selected instances, precisely the same invariants are involved!

%%%%%%%%%%%%%%%%%%
\subsection{Markov invariants for the general Markov model.}
\label{subsubsec:MarkovInvariantsGMM}

As a first example we look at the famous \emph{tangle} quantity, quartic in three qubit wavefunctions whose values are known to distinguish between the different classes of entanged states in tripartite systems \cite{DurVidalCirac2000thq2e}. The phylogenetics equivalent \cite{sumner2006} is for alignments $P^{ijk}$ of binary traits on three species, $L=3$\, and coding these as $\{1,2\}$\,, the tangle is the degree four homogeneous polynomial
\begin{align*}
\tau(P) =&\,
(P^{111})^2(P^{222})^2 + (P^{112})^2(P^{221})^2 + (P^{121})^2(P^{212})^2 + (P^{211})^2(P^{122})^2 \\
&\,\qquad +4P^{111}P^{122}P^{212}P^{221} + 4P^{112}P^{121}P^{211}P^{222}  \\
&\,\qquad -2P^{111}P^{112}P^{221}P^{222}-2P^{111}P^{121}P^{212}P^{222}-2P^{111}P^{122}P^{211}P^{222} \\
&\, \qquad -2P^{112}P^{121}P^{212}P^{221} -2P^{112}P^{122}P^{221}P^{211}-2P^{12 1}P^{122}P^{212}P^{211} 
\end{align*}
known in mathematics as the Cayley hyperdeterminant function. This has the special property of being invariant 
under transformations of the type Eq (\ref{eq:P'fromP}), up to scaling by the product of determinants\footnote{
This property holds for arbitrary (nonsingular) complex transformations (elements of $\times^3 GL(2)$), not just under stochastic
substitutions (see text). Indeed, the square root $\sqrt{|\tau(P)|}$ of the complex modulus of $\tau$ has the strong property of being an \emph{entanglement monotone} \cite{VerStraeteEtAl2002fq9e} for the local adjoint action under $SL(2,{\mathbb C})$ on the associated density operator.},  in particular
\[
\tau(P) = \mbox{Det}(M_1) \mbox{Det}(M_2) \mbox{Det}(M_3) \tau(\overline{P})\,.
\]
For the binary continuous time general Markov model, we have from (\ref{eq:GM2Defn}) $M^{GM_2}=\exp Q^{GM_2}$ where 
\[
Q^{GM_2} = \left(\begin{array}{rr} -\beta & \alpha \\ \beta & -\alpha \end{array}\right)\,.
\]
Thus $-\log \mbox{Det}(M) = \alpha + \beta$\,. Recall that in phylogenetic models based on trees, the
root location is not identifiable (see \S\S \ref{sec:IntroOverview},\ref{subsec:ModellingGMM}), so that in the present case we can assume the tree is a 
star with three leaves. Clearly, $\tau\big(\overline{P}\big)\equiv (\pi^1){}^2(\pi^2){}^2$, the only contribution being from the first term combining the nonzero components $\overline{P}{}^{111}=\pi^1$ and $\overline{P}{}^{222}=\pi^2$\,.
Incorporating the determinants on each edge (the traces of the rate generators $Q_1$, $Q_2$, $Q_3$\,) the conclusion from the above analysis \cite{sumner2006} is that there is a convenient measure given directly from the tangle invariant evaluated on the data,
\[
-\log \big(\tau(P)\big) + 2 \log(\pi^1\pi^2) \equiv (\alpha_1 +\beta_1 )+  (\alpha_2 +\beta_2 )+ (\alpha_3 +\beta_3 )
\]
which is a theoretical indicator of total evolutionary change undergone by any triplet of taxa, whose three-way pattern frequency tensor may come as a marginalization of a larger alignment of many taxa.

It turns out that in this binary case for triplets of taxa incorporating the tangle, this measure which we might call the 
`$\log \mbox{HDet}$'\,, is a rather specific generalization of a well-known quantity known as the `$\log \mbox{Det}$'\,\cite{barry1987,lake1994,lockhart1994}. This latter is available for any number $K$ of traits, but restricted to two taxa. The determinant function of the two-way phylogenetic tensor is of course a degree $K$ polynomial, and by an analysis paralleling the above discussion, and up to assumptions about the probability distribution at the root, it can be seen to yield a measure of `length', or `distance', between pairs of taxa.  Again, in the case of bipartite quantum systems, the analogous quantity (termed the \emph{concurrence}) provides a local unitary invariant, and can be used as an entanglement measure\footnote{In our case the determinant polynomial is a $GL(K) \times GL(K)$ invariant, and hence has simple transformation properties under both the unitary subgroup $U(K) \times U(K)$ and the stochastic subgroup $GL_1(K) \times GL_1(K)$\,; the hyperdeterminant is of course a degree 4 relative invariant for
the action of $GL(2)\times GL(2)\times GL(2)$ and hence of the stochastic subgroup.}.

The `$\log \mbox{Det}$' and variations on it have been used for many years in phylogenetics \cite{barry1987,lake1994,lockhart1994}, as the basis of so-called `distance methods' in inferring evolutionary trees (as discussed in the introduction, \S \ref{sec:IntroOverview} above). Our demonstration of the utility of the binary tangle leads to the question of characterizing and classifying all such quantities, for differing numbers of traits and leaves. 
%The mathematical task is to characterize the invariant ring, that is, polynomials in the coordinates of the relevant representation of the transformation group (in this case, phylogenetic tensors under sochastic evolution, as in equation (\ref{eq:P'fromP}) above). 

The $\log \mbox{Det}$ and the $\log \mbox{HDet}$ examples are invariants under the general linear group of arbitrary invertible matrices. It turns out that there is a rich spectrum of 
corresponding \emph{Markov invariants}, specialized to the stochastic transformation case; that is, polynomials in the coordinates of the relevant representation of the transformation group (in this case, phylogenetic tensors) invariant up to scaling, under stochastic evolution, as in equation (\ref{eq:P'fromP}) above  \cite{sumner2008}.
Thus there are binary-, three- and four-state \emph{stangle} invariants for triplets of taxa (\underline{s}tochastic t\underline{angle}s), generalizations of the binary tangle \cite{sumner2006}, and with similar properties to it. In the binary case, we have for example in the affine basis the simple cubic expression
\[
S\!T(P)=-2 P^{100}P^{010}P^{001}+(P^{100}P^{011}+P^{010}P^{101}+P^{001}P^{110}-P^{111}P^{000})P^{000}\,.
\] 
The rules for handling group representation operations via their characters, and theorems allowing invariants such as the stangles (if not the full invariant ring) to be enumerated at each degree, are developed in detail in appendix \S \ref{subsec:GroupCharacters} below. To conclude the discussion, we describe some further examples.

For quartets of taxa, and for DNA models ($L=K=4$\,), there is a symmetrical set of three degree five Markov invariants dubbed the `\emph{squangles} (\underline{s}tochastic \underline{qu}artet t\underline{angles})\cite{sumner2008}, and by analyzing their behaviour under leaf permutations of the quartet isotropy group \cite{sumner2009}, it is possible to provide a robust way of resolving quartets under the general Markov model (without any further special assumptions about the types of rate matrices in the model) \cite{holland2013low}. 
The squangles are degree 5 polynomials in the components of the $4^4 = 256$-element array, and given their combinatorial origins, it is perhaps not surprising to find that they each have over 50,000 terms in an affine basis (their expansions in the natural basis are not known). Once defined however, there is no computational difficulty with
evaluations. These methods are useful because of the known result that correctly specifying all quartets arising from a tree, is sufficient to reconstruct the full tree (see for example \cite{holland2013low,strimmer1996,bryant2001}). Moreover, the analysis here is quite independent of `distance' considerations, showing that Markov invariants in their most general form, provide flexible new tools for phylogenetics. 

In recent work it has been possible to complete the mathematical task of characterizing the Markov invariant ring for the 2 state general Markov model, and triplets of taxa \cite{sumner:jarvis:inprep}, but the explicit methods used are not feasible for higher $L$ or $K$\,. It is likely though, that the Markov invariants of lowest degree are the most practical from a statistical point of view; for DNA models ($K=4$) the quintic quartet squangles are clearly of great importance, given that the determinant (for two taxa)  is already of degree 4.

%%%%%%%%%%%%%%%%%%
\subsection{Markov invariants for selected Lie-Markov models.}
\label{subsubsec:MarkovInvariantsLM}

We have developed Markov invariants with the general Markov model in mind, but where data supports a particular model class of the Lie-Markov type, there is of course a larger ring of invariants arising from the restriction of the stochastic transformations to the subgroup in question, and it is a question of practical and theoretical importance to enumerate them. While we have not attempted to implement a systematic enumeration of Markov invariants for all of the 35 models in the Lie-Markov hierarchy, the following case studies demonstrate some of the salient considerations. 

Consider for example the three abelian models discussed in \S\S \ref{subsec:LieMarkov}, \ref{subsubsec:HadamardFourier} above (namely models (3.3a), (3.3b) and (3.3c) in the Lie-Markov hierarchy (Fig. \ref{fig:LieMarkovModelsFlowChart} above, and \cite{woodhams:fernandez-sanchez:sumner:2015a}), all of which were able to be related via similarity transformations to choices of Cartan subalgebras (of $gl(4)$ or $sl(4)$ within the Lie algebra $gl_1(4)$ of the general Markov model). From the point of view of invariants, implementing the similarity transform on a general $K$-way phylogenetic tensor array, simply diagonalizes the $3K$ Cartan generators. The components thus provide the relevant weight decomposition and are by definition, one-dimensional representations, and hence \emph{linear} Markov invariants, for all of these models. This statement provides the representation-theory equivalent of the discrete Fourier transform given by `Hadamard coordinates' for the Kimura models (as presented in \cite{bashford2004BF}), but the method is more general, in that it also encompasses the two remaining abelian models in the Lie-Markov scheme\footnote{The intertwining property of the rate generators with respect to the splitting operator, and the ability to lift the inversion to the whole tree, requires further structure, discussed in detail in \S \ref{subsubsec:HadamardFourier} above.}.

The smallest non-abelian Lie-Markov algebras, models (3.4) and (4.4b) have been discussed in connection with rate matrix diagonalization in \S \ref{subsubsec:Diagonalization}. Recall that both models have essentially three generators in common with model (3.3b), and hence can be simplified via the similarity transformation via matrix $X$ induced by the same change of basis, equation (\ref{eq:33c44aXdef}). We have with the additional generators $J_-:= L_{(12)} -L_{(34)}$, $R:= L_{(1324)} -L_{(1423)}$\,,
\begin{align*}
\widehat{J}_+ = &\, \left[\begin{array}{rrrr} 0&0&0&0\\0&\hskip-1ex -2&0&0\\0&0&\hskip-1ex -2&0\\0&0&0&0\end{array}\right]\,,
\quad \, \widehat{J}_- = \left[\begin{array}{rrrr} 0&0&0&0\\0&\hskip-1ex -2&0&0\\0&0&2&0\\0&0&0&0\end{array}\right]\,,\quad \,
\widehat{K} =  \left[\begin{array}{rrrr} 0&0&0&0\\0&\hskip-1ex -2&0&0\\0&0&\hskip-1ex -2&0\\0&0&0&\hskip-1ex -4\end{array}\right]\,,\quad \,
\widehat{R} =\left[\begin{array}{rrrr} 0&0&0&0\\0&0&0&0\\0&0&0&0\\\hskip-1ex -4&0&0&0\end{array}\right]\,.
\end{align*}
On the subspace corresponding to rows and columns $\who, \whf$ in this basis, the generators $\widehat{K} , \widehat{R}$ are  identical up to scaling with those of the binary general Markov model in the affine basis\footnote{Up to scaling, coincident with the $K=2$ generators $L_{12}\pm L_{21}$, after a Hadamard transformation.}, while the remaining subspace (rows and columns $\wht,\whth$) reduces to
a direct sum of distinct weights of the abelian generators $J_{\pm}$ together with $K$. 
This implies that ${\mathfrak L}^{(4.4b)}$ is isomorphic to the direct sum of the binary general Markov model, and a two-dimensional abelian subalgebra (omitting $J_-$ for ${\mathfrak L}^{(3.4)}$).

A case of intermediate complexity is that of the strand symmetric model SSM, model (6.6) (see figure \ref{fig:LieMarkovModelsFlowChart}). With the base ordering $A,G, C,T$ that we have been adopting, the rate matrix is
\[
Q^{SSM} = \left[\begin{array}{cccc} \!-\!\lambda \!-\!\mu\!-\!\nu &\alpha & \beta & \nu \\ \lambda &\!-\!\alpha\!-\!\beta\!-\!\gamma& \gamma & \mu \\ 
\mu & \gamma & \!-\!\alpha\!-\!\beta\!-\!\gamma & \lambda \\ \nu & \beta & \alpha & \!-\!\lambda \!-\!\mu\!-\!\nu\end{array}\right]\,,
\]
which shows the symmetry with respect to canonical Watson-Crick pairs (hence \emph{strand} symmetry)\footnote{This is model 
(6.6)$_{\scalebox{.9}{\texttt{WS}}}$, but the algebraic analysis applies to the variants (6.6)$_{\scalebox{.9}{\texttt{RY}}}$ and 
(6.6)$_{\scalebox{.9}{\texttt{MK}}}$, whose rate matrix patterns would reflect the appropriately modified dihedral symmetry on the other pairs.}; for example the rate equalities $A \leftarrow G  = T \leftarrow C= \alpha$\, and
$G \leftarrow A  = C \leftarrow T= \lambda$\,. 

The 6 dimensional Lie algebra ${\mathfrak L}^{SSM}$ can be given \cite{jarvis:sumner:2016missm} a standard Levi-type decomposition as the sum of the four dimensional semisimple Lie algebra $gl(2) \cong gl(1) +sl(2)$, together with the Lie algebra $gl_1(2)$ of the binary general Markov model. A suitable transformation confirms that the four nucleotide state space in the natural basis reduces to a sum of two doublets -- a two-dimensional copy of the binary general Markov model, as in models (3.4), (4.4b), but in contrast to these, with the second doublet the defining representation of the nonabelian part $sl(2)$.

Appendix \S \ref{subsec:GroupCharacters} gives technical details of the group character manipulations required for enumeration and evaluation of low degree Markov invariants for the strand symmetric model, and also the submodels (3.4), (4.4a). In particular we give a count of  linearly independent quadratic invariants for any number of taxa (see equation (\ref{eq:quadraticLMsubmodels}); for the strand symmetric model these number  $5, 13, 41, \cdots$ for $L=2,3,4, \cdots$\,. The very existence of invariants at low degree is in stark contrast to the more restricted selection of degrees and numbers of taxa for general Markov model invariants at various numbers of characters \cite{sumner2008,sumner2006aBF}, reflecting the fact that these models have very specific structure. As we have noted, the \emph{abelian} 
Lie-Markov models have many \emph{linear} invariants, associated with the coordinate transformation to the appropriate weight basis; here, the invariants at quadratic (and cubic) degree indicate that these \emph{nonabelian} models are the next simplest in complexity as far as their paramerizations are concerned. Indeed, for the strand symmetric model Markov invariants, we have shown \cite{jarvis:sumner:2016missm} that the usual monolithic `distance' measures (associated with the $\log \mbox{Det}$, as we have described), become more ramified, in that they allow extraction of total evolutionary change of both intra- and inter-pairing type (for the canonical Crick-Watson or strong-weak pairing, this would be within and across base pairs $CG|AT$).

	%%%%%%%%%%%%%%%%%%%%%%%%%%%%%%%%%%%%%%%%%%%%%
	%\input{secs/Conclusions.tex}
	%\label{sec:Conclusions}
%%%%%%%%%%%%%%%%%%%%%%%%%%%%%%%%%%%%%%%%%%%%
%%%%%%%%%%%%%%%%%%%%%%%%%%%%%%%%%%%%%%%%%%%%%%%%%%%%%%%%%%%%%%%%%%%%%%%%%%%%%%%%%%%%%%%%%%%%%%%%%%%%%%%%%%%%%%%%%%%%%%%%%%
\section{Conclusions}
\label{sec:Conclusions}

In this review we have been at pains to present and introduce the subject of theoretical phylogenetics in a language amenable to a primarily physically or mathematically trained audience, whose expertise is not necessarily in biology. In our discussion, beyond our brief introductory survey, we have not attempted to cover `observational' aspects regarding the body of biological data supporting the evolutionary theories leading to ancestral reconstruction via phylogenetic trees; neither have we provided case studies of the `success' of the phylogenetic models under study, nor the details of statistical analyses which underly inference in these parametric stochastic models.

It should be pointed out that there is considerable controversy in the literature as to the appropriateness of trees as an encapsulation of evolutionary change in all situations. The ubiquity of interaction networks in other contexts such as metabolic, immune, energetic, and other major systems of biological organisation, certainly suggest that
inheritance and speciation are likely to be similarly complex. However, notwithstanding these critiques, 
the original brief of phylogenetics -- the use of quantitative, inter-species comparison data (in the modern context, molecular sequence data) to infer the evolutionary ancestry of species in the form of a binary tree -- remains central. Such is their importance that there is a great variety of theoretical constructs and formalisms in which they emerge, and we close with a brief survey of a selection of the literature in this direction.

Most directly, it will have been noticed that the graph-based construction of phylogenetic tensors decribed above 
(\S \ref{subsec:ModellingGMM}) has much in common with the description of other physical branching processes such as scattering, or radioactive decay.
There is a formal identification between writing down phylogenetic tensors and applying a set of `Feynman rules' for 
a type of reaction-diffusion process described in second-quantized language 
\cite{jarvis2001,jarvis2005} as in other statistical physics models of this sort \cite{Doi:1976,Peliti:1985}. This theoretical direction is in turn affiliated with a general paradigm of `stochastic mechanics' \cite{BaezBiamonte2014}, purporting to handle master equations and associated Dirichlet operators as 
fundamental theoretical tools for the investigation of conservation laws and symmetry principles, in the same way as is known for Hamiltonian dynamics.

One of the themes of our work has been to exploit eclectically the manifest similarities in the (mathematical) formalisms between different specializations in `physics' and `biology', and bringing to bear the tradition of mathematical physics in adopting appropriate levels of abstraction to applications at hand. A fascinating case in point concerns the subtleties of a  choice of selected basis. We have emphasized, for example, the `natural' (or `biological') basis for stochastic models, and the `affine' (or `computational') basis class which renders transition matrices upper diagonal, with one component (corresponding to the probability mass) necessarily preserved; there are many instances of such basis transformations, including to Hadamard or discrete Fourier (including complex) bases, which are convenient for the analysis of specific phylogenetic models. In elementary quantum physics, on the other hand, unitarily equivalent basis choices are well understood, and are central to many issues of measurement and prediction. Nonetheless, in abstract, categorical formulations, a `measurement basis' is accorded a separate
significance, and is regarded as a foundational axiomatic datum equated to the presence of Frobenius algebraic structures \cite{abramsky2004categorical,Coecke:Duncan:2008}. Indeed, the basic `splitting operator' $\delta$\,, in our multilinear tensor formulation in \S \ref{subsec:ModellingGMM}, $\delta (e_i) = \left.\sum\right._i e_i \otimes e_i $\,,
is precisely the Frobenius comultiplication of \cite{Coecke:Duncan:2008} (see also \cite{fauser:2012a}). In that case, there is the accompanying multiplication,
$\mu(e_i \otimes e_j) = e_i\delta_{i,j}$\,.  In the strict phylogenetic context, the introduction of such a move into the graphical 
rules would lead to directed acyclic graphs, rather than trees, and probability violation. However, we have seen in 
\S \ref{subsubsec:Convergence} above that there is a role for such `phylogenetic convergence' viewed asymptotically, in a large-time limit. This feature arises through the intimate relationship between
the comultiplication $\delta$ and the phylogenetic coproduct $\nabla$ (see 
(\ref{eq:CoproductNablaDef}), \S \ref{subsubsec:Coproducts} above), $\nabla(L_{ij}) =L_{ij} \otimes \unit + \unit \otimes L_{ij}  +  L_{ij} \otimes L_{ij}$\,, and the emergence of special classes of substitution models, which admit various types of direct inversion methods
which act on the model as a whole (irrespective of parameter choice) and across the entire tree (as opposed to simplification on each edge alone).

Remarkably, the rich mathematical structure that we have presented here within the context of molecular phylogenetics, in fact also has older connections, in the complementary area of inheritance and population genetics. Starting with the efforts in the 'forties of some statisticians to encapsulate the algebraic structure of Mendelian genetics, what is known in the literature as `genetic coalgebras', has developed into an area of active study (see for example \cite{tian2004coalgebraic} and references therein). It is hard to escape the speculation that future progress on the intellectual journey introduced in our opening remarks, may indeed engender a further refinement in our understanding of evolution in all its ramifications. Increasing mathematical sophistication 
\cite{manon2009algebra,baez2017operads} will be an inherent part of this progress.

The quantitative models in molecular phylogenetics that we have described,
are fundamentally underpinned by foundations of bioinformatics, and the central dogma of information coding, storage and transmission in biological systems. Whether these also truly have any foundation in `quantum biology', is an open question, well beyond our remit here\footnote{See for example \cite{Bashford:Jarvis:2008} and articles in the same collection, and also \cite{ellinas:jarvis:2011}.}. The interested reader 
is referred to relevant literature (see for example the inspiring reviews of \cite{carbone2001mathematical, Kauffman2002} and references therein). It is appropriate to end with
Kauffman \cite{Kauffman2002}, as an encapsulation of these broader and deeper questions: \\
%\vfill
\begin{quotation}
``We have been trained to think of physics as the foundation of biology, but it is possible to realize that indeed biology can also be regarded as a foundation for thought, language, mathematics and even physics." \\
\mbox{}\hfill Louis Kauffman, \emph{Biologic} \cite{Kauffman2002}
\end{quotation}
\vfill

\noindent
\textbf{Acknowledgements}\\
The authors wish to record their appreciation to many colleagues and coauthors for discussion, encouragement, constructive criticism and collaboration for joint work which has contributing to this review. These include Jim Bashford, Ioannis Tsochantzis, Mike Steel and Demosthenes Ellinas, as well as current members of our group, and students both undergraduate and postgraduate. PDJ is grateful for support from the Alexander von Humboldt Foundation, and the Australian-American Fulbright Foundation, for part of this work. We also acknowledge the Australian Research Council for discovery research grants supporting this work, and the University of Tasmania for funding several collaborative visits under the University of Tasmania Visiting Fellowship scheme.
 
\pagebreak 
%\vfill
\begin{appendix}
\section{Appendix}
\renewcommand{\theequation}{A--\arabic{equation}}
	%%%%%%%%%%%%%%%%%%%%%%%%%%%%%%%%%%%%%%%%%%%%%%%%%%%%%%%%%%%%%%%
	%\input{secs/CyclicBasis.tex}
	%\section{Group character manipulations for Markov models in phylogenetics}
	%\label{subsec:CyclicBasis}
\subsection{Cyclic basis for $gl(K)$ and group-based models.}
\setcounter{equation}{0}
\label{subsec:CyclicBasis}
Suppose $\sigma$ is cyclic generating ${\mathbb Z}_K$. Then there is a set of fundamental generators corresponding to 
$K_\sigma$, $K_{\sigma^2}$, etc., namely as usual $Q = K - \unit$, where the $K$'s are sums of elementary matrices,
\[
E = K_\sigma = \sum E_{i, \sigma i}
\]
and powers, $E^2 = K_{\sigma^2}$\,, etc. This whole set can be obtained as a similarity transformation of the standard basis by introducing the concept of the \emph{index} $\underline{i}$ of an element,
$i := \sigma^{\uli}1$, $i, \underline{i}=1,2,\cdots, K$ along with a primitive $K$'th root of unity:
\begin{align*}
A_{pq} := &\, \frac{1}{K} \sum_{k,\ell}\omega^{p \ulk} \ovw^{q\ull} E_{k\ell}\,;\\
S_{rs} =&\, \frac{1}{\sqrt{K}} \omega^{s\ulr}\,,\\
%\mbox{claim}\qquad S S^\dagger =&\, \unit\,,\qquad \mbox{and}\qquad A_{pq} = S E_{pq}S^\dagger\,;\\
{[}E, A_{pq}{]} =&\, (\omega^p - \omega^q)A_{pq}
\end{align*}
where we claim $S S^\dagger = \unit$ and that $A_{pq} = S E_{pq}S^\dagger$\,; thus the $A_{pq}$ and elementary matrices $E_{k\ell}$ are related by a similarity transformation; and hence the $A_{pq}$ have identical commutation relations to $E_{k\ell}$. In particular we can choose the Cartan basis for the $A$'s, to be the map of the standard Cartan basis for the $E$'s. In fact, we wish to take the 
Cartan basis to be the powers $E$, $E^2$, $\cdots$ $E^{K-1}$, as these clearly commute and have standard spectrum. Note that these $E^m$ generators have constant relative index $1,2,\cdots, K-1$ between $k$ and $\ell$ in $E_{k\ell}$ in their unifom sums, so they ought to be constructible from projections of the $A_{ii}$ (which themselves can be regarded as discrete Fourier transform combinations of uniform sums with fixed relative projection). So the required definition of $E^m$'s in terms of $A_{ii}$ is the inverse discrete Fourier transform\footnote{The whole exercise can be rectified by working with a conjugation $\gamma$ which converts the cyclic generator into the standard rising form
$(123\cdots K)$, say $\sigma = \gamma^{-1} (123\cdots K) \gamma$. The method developed shows directly that the rate generators are indeed Cartan elements within a specific basis $gl(K)$.}.

Returning to the transformation discussion, this shows that a basis transformation of any cyclic generator underlying a model is related to a Cartan basis choice, and so (in principle) if the usual $\delta$ splitting intertwining property holds, there is an appropriate inversion on the tree.

	%%%%%%%%%%%%%%%%%%%%%%%%%%%%%%%%%%%%%%%%%%%%%%%%%%%%%%%%%%%%%%%
	%\input{secs/BCHapproximants.tex}
	%\section{Group character manipulations for Markov models in phylogenetics}
	%\label{subsec:BCHapproximants}
\subsection{Multiplicative closure and BCH approximants.}
\setcounter{equation}{0}
\label{subsec:BCHapproximants}
%\mbox{}\\
Here we reiterate the setting of our discussion from \S \ref{subsubsec:ModelClosure} above, which provides important underpinning for 
our emphasis on Lie algebras and group representations in analyzing phylogenetic modelling. Following \cite{Sumner_2017mult},
and the general treatment of model types in \S\S \ref{subsubsec:Catalogue}, \ref{subsubsec:LieMarkovClass}, the most general type of phylogenetic rate model can be considered simply to be some set ${\mathfrak R}^+ := {\mathfrak R}\cap {\mathfrak L}^+ $ of stochastic rate generators. The characterization of the set ${\mathfrak R}$ is that the parameters form the real solution space
of a set of homogeneous polynomial constraints. We further note that such matrices can be arbitrarily scaled without
leaving ${\mathfrak R}$ -- so that in particular, the stochastic rate matrices of the model admit scaling by nonnegative real numbers. 
Then, the demand of multiplicative closure for the model is simply that products of the affiliated Markov substitution matrices, that is, the matrix exponentials, should also be compatible with the required parametrization -- that is, at the level of the corresponding matrix logarithms (where these are defined, in the standard way), and any scalings thereof, $\ln\big(e^{Q_1}e^{Q_2}\big) \in {\mathfrak R}$\,.
%\[
%{\mathbb R} \ln\big(e^{Q_1}e^{Q_2}\big) \in {\mathfrak R}\,.
%\]

The characterization of $\ln\big(e^{X}e^{Y}\big)$ is a difficult problem in matrix analysis, and is intimately related to the structure and convergence of the famous Baker Campbell Hausdorff (BCH) series of terms with nested commutators \cite{Baker1902,Campbell1897,Hausdorff1906} essentially belonging to the free Lie algebra of $X$ and $Y$. For the current application, it is sufficient to take careful consideration of relevant convergence results, as reviewed, for example, by \cite{BlanesCasas2004}. The following discussion is adapted from this work.

The most fruitful avenue for the derivation of estimates of BCH series convergence is via the Magnus expansion (essentially a form of interaction representation) for an equivalent operator equation,
\[
\frac{dU(t)}{dt} = A(t)U(t)\,, \qquad U(0) = {\mathbb I}\,.
\]
The iterative solution $U(t)= \exp \Omega(t)$\,, $\Omega = \sum_{n=1}^\infty \Omega_n(t)$ recovers the BCH expansion for the value $t = 2$\, if the driving term $A(t)$ is the piecewise continuous operator
\[
A(t) = \left\{ \begin{array}{rl}Y\,,& 0\le t \le 1\,,\\ X\,,& 1< t \le 2 \end{array}\right.\,.
\]
Moreover, if an approximation $U_1(t) = \exp Z(t)$ is taken (for example by truncating the BCH series to a finite number of terms), then it is possible to derive a limit on the norm $\nrm U_1(t)-U(t) \nrm = \nrm U_1(t)\big(I-U_1(t)^{-1}U(t)\big)\nrm$\, by in turn approximating the solution of the induced equation
\[
\frac{dU_1(t)}{dt} = A_1(t)U_1(t)\,, \qquad U_1(0) = {\mathbb I}\,.
\]
(and that for the inverse $\big(U_1(t)\big)^{-1}$).
The relevant truncations $Z(t)$ are those leading for $t=2$ to $\Omega_1(2) = X + Y$\, (the first term of the BCH series), as well as $\Omega_2(2) = X + Y + \textstyle{\frac 12}{[}X,Y{]}$\, (the first two terms of the BCH series, here with the commutator bracket present). We have the estimates: 
\begin{quotation}
\noindent
\textbf{Lemma: BCH approximants}\footnote{The norm is taken to be submultiplicative, $\nrm XY\nrm \le \nrm X\nrm \nrm Y \nrm$\,.}:
\\[-.6cm]
%\begin{align*}
%\lim_{\nrm Q_1\nrm, \nrm Q_2\nrm \rightarrow 0} \nrm \exp Q_1 \exp Q_2 - \exp\big(Q_1 + Q_2) \nrm  \le 
\begin{center}
%\fbox{
\mbox{}\hskip 4ex\parbox{16cm}{
\textbf{First bound}:
\[
\nrm e^X e^Y - e^{X+Y}\nrm \le e^{K_0}\left(K_1 e^{K_1}\right)\,,
\quad K_0 = \nrm X\nrm +\nrm Y\nrm\,, \quad K_1 := \frac{\big( 1-  {e^{2K_0}}{(1-2K_0)}\big)}{4K_0^2}\nrm{[}X,Y{]}\nrm
\]
}%}
\\
\parbox{16cm}{
\textbf{Second bound}:
\begin{align*}
\!\nrm e^X e^Y - e^{X+Y+ \frac 12{[}X,Y{]}}\nrm \le \, e^{\overline{K}_0}\left(\overline{K}_1 e^{\overline{K}_1}\right)\,,&
\qquad \overline{K}_0 = \nrm X\nrm\! +\!\nrm Y\nrm\!+ \textstyle{\frac 12}\nrm{[}X,Y{]}\nrm\,,\\
\overline{K}_1 :=\,  \frac{\big( 1- \overline{K}_0+ e^{2\overline{K}_0}(3\overline{K}_0-1)\big)}{4\overline{K}_0^2}\nrm{[}X,Y{]}\nrm&
 +{\textstyle{\frac 14}}\frac{\big( 1-  {e^{2\overline{K}_0}}{(1-2\overline{K}_0)}\big)}{4\overline{K}_0^2}\nrm{[}X,{[}X,Y{]}{]}\nrm
\end{align*}
}%}
\end{center}

\noindent
\textbf{Proof}:
The first bound is proven in \cite{BlanesCasas2004}\,. Our derivation of the second bound follows the same method. (We believe this result to be new). \mbox{} \hfill $\Box$\\
\end{quotation}
\mbox{}\\[-.5cm]
\noindent
To complete our multiplicative closure result, let
$\Pi(Q_1,Q_2) = \exp Q_1 \exp Q_2$\,, and $\Omega_1(Q_1,Q_2) = Q_1+Q_2$,
$\Omega_2(Q_1,Q_2) = Q_1+Q_2 +\textstyle{\frac 12}{[}Q_1,Q_2{]}$\,. We assume $\log\big(\Pi(Q_1,Q_2)\big) \in {\mathfrak R}$\, and by pointwise convergence under the BCH approximant bounds\footnote{The coefficients of the commutator bracket terms in ${K}_1$ and 
$\overline{K}_1$\,, which are functions of $K_0$ and $\overline{K}_0$ respectively,
have smooth behaviour for small arguments.} both $\Omega_1(Q_1,Q_2)\rightarrow \log\big(\Pi(Q_1,Q_2)\big)$ and $\Omega_2(Q_1,Q_2) \rightarrow \log\big(\Pi(Q_1,Q_2)\big)$ in the limit $\nrm Q_1\nrm$,$\nrm  Q_2\nrm\rightarrow 0$\,.  Invoking smoothness, we \emph{assume} that, for the cases at hand, for small enough $\nrm Q_1\nrm$,$\nrm Q_2\nrm >0$, there are neighbourhoods $N_1$, $N_2$ of 
$\log\big(\Pi(Q_1,Q_2)\big)$ (within the solution space of the parametric constraints defining the subset ${\mathfrak R} \subseteq {\mathbb R}^{K(K\!-\!1)}$) which contain $\Omega_1(Q_1,Q_2)$ and $\Omega_2(Q_1,Q_2)$ respectively. These results obtain for \emph{arbitrary}
$Q_1$, $Q_2$ with sufficiently small nonzero norm. By scaling, the first assumption implies that  
${\mathfrak R}$ is closed under conical combinations; thus the homogeneous constraints defining ${\mathfrak R}$ can \emph{only be linear}, whence it is a vector space. The second assumption, taken with the first, in turn implies that the commutator bracket ${[}Q_1,Q_2{]}$ of elements of small enough norm, also belongs to ${\mathfrak R}$. Hence, again by scaling, ${\mathfrak R}$ is in fact a Lie algebra (and hence forms a Lie-Markov model, under our definition).  \hfill $\Box$

	%%%%%%%%%%%%%%%%%%%%%%%%%%%%%%%%%%%%%%%%%%%%%%%%%%%%%%%%%%%%%%%
	%\input{secs/GroupCharacters.tex}
	%\section{Group character manipulations for Markov models in phylogenetics}
	%\label{sec:GroupCharacters}
\subsection{Group character manipulations for Markov invariants in phylogenetics \\
and local unitary invariants in composite quantum systems.}
\setcounter{equation}{0}
\label{subsec:GroupCharacters}

%%%%%%%%%%%%
%%%%%%%%%%%%%%%%%%%%%%%%%%%%%%%%%%%%%%%%%%%%%%%%%%%%%%%%%%%%%
In this section we provide technical details of group character notation and methods for handling finite dimensional representations of the relevant matrix Lie groups needed for the study of Markov invariants in the phylogenetics context (as discussed in \S\S
\ref{subsec:LieMarkov}, \ref{subsubsec:HadamardFourier}, \ref{sec:Entanglement} above), including both the general Markov model, as well as models within the Lie-Markov hierarchy. This text is adapted from \cite{jarvis2014adventures}, and also from several other related technical discussions (see for example \cite{jarvis:sumner:2016missm}), which have provided similar background information\footnote{This appendix extends \cite{jarvis2014adventures} in providing some details of the enumeration of Markov invariants for the strand symmetric model \cite{jarvis:sumner:2016missm}, and in the extension to the low-dimensional (non-abelian) cases
(3.4), (4.4a).}. In line with our objectives for this review article of presenting the ideas of molecular phylogenetics to a wide audience, we have retained (from \cite{jarvis2014adventures}) the group theoretical explorations of entanglement in quantum systems, in order to underline the close connections between both areas, and to provide a familiar context for readers more accustomed to the physics setting\footnote{For a recent overview of the topic of quantum entanglement, see \cite{Eltschka:Siewert:2014}; see below for further 
remarks on mixed state systems.} 

For the following subsection only, the notations adopted in the main text (state space dimension $K$, $L$-way tensors from phylogenetics alignments) have been modified (state space dimension $D$, matrix Lie algebras indexed by $D$, for example $gl(D)$; $K$-way phylogenetic probability distributions, or $K$-partite quantum states or density operators; polynomial invariants of degree $n$).

The mathematical setting for both the study of entanglement measures for composite quantum systems, and of analogous quantities for the setting of phylogenetics, is that there is a model space
$V$ which is a $K$-fold tensor product, $V \cong {\mathbb C}^D\otimes {\mathbb C}^D\otimes \cdots 
\otimes  {\mathbb C}^D$. In the case of quantum mechanics the components of $V$ in some standard basis describe the state; for example in Dirac notation a pure state is a ket $|\Psi \rangle \in V$ of the form
$
|\Psi \rangle = \sum_{0}^{D\!-\!1} \Psi_{i_1 i_2 \cdots i_K} |i_1,i_2, \cdots, i_K \rangle
$
in the case of qu$D$its (see below for mixed states). In the phylogenetic case we simply have a $K$-way frequency array
${\{} P^{i_1 i_2 \cdots i_K} {\}}$ sampling the probability of a specific pattern, say ${i_1 i_2 \cdots i_K}$, where each $i_k \in {\{} A,C,G,T {\}}$ for nucleotide data, at a particular site 
in a simultaneous alignment of a given homologous sequence across all $K$ of the species under consideration. 

We focus attention on the linear action of the appropriate matrix group $G = G_1 \times G_2 \times \cdots \times G_K$ on $V$.
In the quantum qu$D$it case each local group $G_k$ is a copy of $U(D)$, but given the irreducibility of the fundamental representation, for polynomial representations the analysis can be done using the character theory of the complex group. This group is too large for the phylogenetic case, where the pattern frequency array $P$ evolves as $P \rightarrow P' := g \cdot P$, namely
\[
P' = M_1 \otimes M_2 \otimes \cdots \otimes M_K \cdot P
\]
where each $M_k$ belongs to the Markov group $GL_1(D,{\mathbb C})$ (the group of nonsingular complex unit row-sum $D \!\times\!D$ matrices). 

We compute the terms in the Molien \cite{molien1897invarianten} series  $h(z) = \sum_0^\infty h_n z^n$ for ${\mathbb C}{[} V{]}^G$ degree-by-degree using combinatorial methods based on classical character theory for $GL(D)$, adapted slightly for the stochastic case $GL_1(D)$, which we now describe. All evaluations are carried out using the group representation package ${}^\copyright \!$\texttt{Schur} \cite{SCHUR}. 

In terms of class parameters
(eigenvalues) $x_1,x_2,\cdots, x_D$ for a nonsingular matrix $M \in GL(D)$, the defining representation, the character is simply $Tr(M) =  x_1+ x_2+ \cdots + x_D$; the contragredient has character
$Tr((M^T)^{-1}) =  x_1{}^{-1}+ x_2{}^{-1}+ \cdots + x_D{}^{-1}$. Irreducible polynomial and rational characters of $GL(D)$ are given in terms of the celebrated Schur functions \cite{Weyl1939,littlewood1940} denoted $s_\lambda(x)$, where $\lambda = (\lambda_1,\lambda_2,\cdots,\lambda_D)$, $\lambda_1 \ge \lambda_2 \ge \cdots \ge \lambda_D$, is an integer partition of at most $D$ nonzero parts. 
$\ell(\lambda)$, the length of the partition, is the index of the last nonzero entry (thus $\ell(\lambda)=D$ if $\lambda_D >0$). $|\lambda|$, the weight of the partition, is the sum $|\lambda|=\lambda_1+\lambda_2 + \cdots + \lambda_D$, and we write $\lambda \vdash |\lambda|$. For brevity we write the Schur function simply as $\{\lambda \}$ where the class parameters are understood. Thus the space $V$ as a representation of $G$ as a $K$-fold Cartesian product is endowed with the corresponding product of $K$ characters of the above defining representation of each local group, $\chi= \{1\} \cdot \{1\}\cdot \, \cdots \,  \cdot \{1\}$ in the quantum mechanical pure state and stochastic cases, and 
$\chi = (\{1\} \{\overline{1}\}) \! \cdot \! (\{1\} \{\overline{1}\})\!\cdot \, \cdots \, \cdot\!(\{1\} \{\overline{1}\})$ in the quantum mechanical mixed state case (where $\{1\}$ is the character of the defining representation, and $\{\overline{1}\}$ that of its contragredient), as appropriate for the transformation properties in the density matrix description.
The space of polynomials of degree $n$ in $\Psi$ or $P$, ${\mathbb C}{[}V{]}_n$, is a natural object of interest and by a standard result is isomorphic to the $n$-fold symmetrised tensor product $V \vee V \vee \cdots \vee V$, a specific case of a Schur functor: ${\mathbb S}_{\{n\}}(V)$. Its character is determined by the corresponding Schur function \emph{plethysm}, 
$\chi \underline{\otimes} \{n\}$, and the task at hand is to enumerate the one-dimensional representations occurring therein.

Before giving the relevant results it is necessary to note two further rules for combining Schur functions. The \emph{outer} Schur function product, is simply the pointwise product of Schur functions, arising from the character of a tensor product of two representations. Of importance here is the \emph{inner} Schur function product $\ast$ defined via the Frobenius mapping between Schur functions and irreducible characters of the symmetric group. We provide here only the definitions sufficient to state the required counting theorems in technical detail. For a more comprehensive, Hopf-algebraic setting for symmetric functions and characters of classical (and some non-classical) groups see \cite{FauserJarvis2003hl,FauserJarvisKing2006nbr}. 

Concretely, we introduce structure constants for inner products in the Schur function basis as follows:
\[
\{\lambda \} \ast \{ \mu \} = \sum_\nu g^\nu_{\lambda,\mu} \{\nu \}.
\]
For partitions $\lambda$, $\mu$ of equal weight\footnote{If 
$|\lambda| \ne |\mu|$ then $\{\lambda \} \ast \{ \mu \}=0$.}, $|\lambda| = |\mu|= n$, say, this expresses the reduction of a tensor product
of two representations of the symmetric group ${\mathfrak S}_n$ labelled by partitions $\lambda$, $\mu$. By associativity, we can extend the definition of the structure constants to $K$-fold inner products,
\[
\{\tau_1 \} \ast \{\tau_2 \} \ast \cdots \ast \{\tau_K \} = \sum_\nu g^\nu_{\tau_1, \tau_2, \cdots, \tau_K}
\{\nu \}.
\]
\mbox{}\\[-.3cm] 
\noindent
\textbf{Theorem: Counting invariants:}
\begin{description}
\item[(a) Quantum pure states:]\mbox{}\\
Let $D$ divide $n$, $n = rD$, and let $\tau$ be the partition $(r^D)$ (that is, with Ferrers diagram a rectangular array of $r$ columns of length $D$). Then 
\[
h_n = g^{(n)}_{\tau,\tau,\cdots,\tau}\quad \mbox{($K$-fold inner product)}.
\] 
If $D$ does not divide $n$, then $h_n =0$.
\item[(b) Quantum mixed states:]\mbox{}\\ We have\footnote{See \cite{KingWelshJarvis2007, Jarvis2014m2qt} for the case of the mixed two qubit and two qutrit systems, respectively.}
\[
h_n = \sum_{|\tau|= n,\ell(\tau) \le D^2}
\left( \sum_{|\sigma|= n, \ell(\sigma) \le D} g^{\tau}_{\sigma,\sigma}\right)^{\!\!\!\!2}.
\]
\item[(c) Phylogenetic $K$-way pattern frequencies, general Markov model:]\mbox{}\\
%Let $n = rD+s$, $s \ge 0$. Then
We have
\[
h_n = g^{(n)}_{\tau_1,\tau_2,\cdots,\tau_K}\quad \mbox{($K$-fold inner product)},
\] 
for each $\tau_k$ of the form $(r_k+s_k, r_k^{(D\!-\!1)})$ such that $n = r_kD+s_k$, $s_k \ge 0$.
\item[(d) Phylogenetic $K$-way pattern frequencies, doubly stochastic model:]\mbox{}\\
%Let $n = r(D\!-\!1)+s+t$, $0\le t \le r$, $s \ge 0$. Then
We have 
\[
h_n = g^{(n)}_{\tau_1,\tau_2,\cdots,\tau_K}\quad \mbox{($K$-fold inner product)},
\] 
for each $\tau_k$ of the form $(r_k+s_k, r_k^{(D\!-\!2)},t_k)$ such that 
$n = r_k(D\!-\!1)+s_k+t_k$, $0\le t_k \le r_k$, $s_k \ge 0$.
\end{description}
\mbox{}\hfill $\Box$\\

The enumeration and identification of entanglement invariants, in the case of quantum systems, and Markov invariants, in the phylogenetic context, is of practical importance in characterising general properties of the systems under study -- in the quantum case, because they are by definition impervious to local unitary operations, and form the raw material for constructing interesting entanglement measures; and in the phylogenetic case, because they tend to be independent of how the specific Markov change model is parametrized, but nonetheless they can give information about the underlying tree. 

An example of identifying invariants is the case of the three squangle quantities (see \S \ref{subsubsec:MarkovInvariantsGMM}). We find ${g^{(5)}}_{\tau\tau\tau\tau}=4$, where $\tau$ is the partition $(2,1^3)$ which is of course of dimension $4$ and irreducible in $GL(4)$, but indecomposable in $GL_1(4)$, as it contains a one-dimensional representation. One of the four linearly independent degree five candidates is discounted, because of algebraic dependence on lower degree invariants. Recourse to the appropiate quartet tree isotropy group under leaf permutations \cite{sumner2009} reveals that one of the remaining three is not tree informative. Further, the situation with respect to the final two objects is expressed symmetrically in terms of the \emph{three} squangle quantities $Q_1$, $Q_2$, $Q_3$, which satisfy $Q_1+Q_2+Q_3=0$, and their evaluation on the three possible unrooted trees. For tree 1, $12|34$ (whose internal edge nodes have leaves 1 and 2 together, and 3 and 4 together), we have on evaluation with stochastic parameters, $Q_1= 0$, and $-Q_3=Q_2 > 0$. This pattern recurs cyclically for the other two unrooted quartet trees: for tree 2, $13|24$, $Q_2=0$, whereas $-Q_1=Q_3 > 0$, and for tree 3, $14|23$, $Q_3=0$, and $-Q_2=Q_1>0$. As noted above, the (strict) inequalities entailed in the above evaluations are crucial for the validity of the least squares method for ranking quartet trees using squangles.

As mentioned in the main text (\S \ref{sec:Entanglement}), our approach to the analysis of Markov invariants has been mainly enumerative, except for the
binary case \cite{sumner:jarvis:inprep} in which the complete structure of the invariant ring has been determined.  As a variant on the  triplet \underline{s}tochastic \underline{tangle} (`stangle') invariants for $K=3$\,, we have evidence \cite{sumner2006aBF, sumner2008} at degree 8 for invariants with mixed scaling behaviour with determinantal products. We have
\[
g^{(8)}_{(51^3),(2^4),(2^4)} =1 \quad (\equiv g^{(8)}_{(2^4),(51^3),(2^4)}\equiv g^{(8)}_{(2^4),(2^4),(51^3)})\,.
\]
Thus there are three mixed weight stangle candidates, which distinguish between edges of their ancestral star tree; a suitable $\log$ measure would potentially enable reconstruction of each edge length separately.

Beyond the general Markov model, Markov invariants for different models and Lie-Markov subgroups have not been systematically studied (but for an example not in the Lie-Markov hierarchy see \cite{jarvis:sumner:2012miesm}). As illustrative of the group character manipulations involved in the subgroup counting process, we briefly review the potentially important model (6.6), the strand symmetric model $SSM$ \cite{jarvis:sumner:2016missm}, and discuss how the analysis extends further to sub-models such as the nonabelian models 
(3.4) and (4.4b)\footnote{Note that a subgroup invariant ring necessarily contains (is larger than) that for the group in question.}. 

To this end we require an extension \cite{jarvis:sumner:2016missm} of the above Counting invariants Theorem, which applies to any submodel of Lie-Markov type with symmetry group $G < GL_1(K)$. We regard the model space $V \cong C^D$ as the (decomposable) $G$-module corresponding to the linear representation generated by the natural basis of $D$ character traits (branching from the defining representation of $GL_1(D)$):\\

\noindent
\textbf{Theorem: Counting invariants -ctd:}
\begin{description}
\item[(e) Phylogenetic $K$-way pattern frequencies, Lie-Markov model:]\mbox{}\\
1. For each partition $\sigma \vdash n$, compute the number $f_\sigma$ of one-dimensional representations  occurring
in the decomposition of the plethysm $V\underline{\otimes} \sigma$. \\
2. The number of Markov invariants at degree $n$ is
\[
h_n = \sum_{\sigma_1,\sigma_2,\cdots, \sigma_K\,; \sigma_i \vdash n} g^{(n)}_{\sigma_1,\sigma_2,\cdots, \sigma_K}
f_{\sigma_1}f_{\sigma_2}\cdots f_{\sigma_K}
\]
where $g^{(n)}_{\sigma_1,\sigma_2,\cdots, \sigma_K}$ is the inner product multiplicity for the occurrence of the module $(n)$ in the tensor product ${\sigma_1\otimes \sigma_2 \otimes \cdots \otimes \sigma_K}$ of modules of ${\mathfrak S}_n$.\\
\noindent
\end{description}
\mbox{}\\[-.5cm]
\mbox{}\hfill $\Box$\\

For the examples we have in mind, $K=4$ and $n=2$, the plethysms $V\underline{\otimes} \{2\}$ and $V\underline{\otimes} \{1^2\}$
can be computed in a two-stage process starting from the reduction of $V$ into a direct sum $V\cong U+W$ of two two-dimensional modules -- the first, $U$, equivalent to the standard binary general Markov model, and the second, $W$, either an irreducible $gl(2)$ doublet, or (in the case of (3.4) and (4.4b)) a direct sum of two one-dimensional modules. We use the right distributivity law of plethysm over direct sum to infer
\[
(U+W)\underline{\otimes} \{2\} \cong U\underline{\otimes} \{2\} + U \otimes W + W \underline{\otimes} \{2\}\,;
\qquad 
(U+W)\underline{\otimes} \{1^2\} \cong U\underline{\otimes} \{1^2\} + U \otimes W + W \underline{\otimes} \{1^2\}\,.
\]
This means that for model $SSM$ (6.6)   $f_{(2)}= 1$\, (from $ U\underline{\otimes} \{2\}$), and $f_{(1^2)}= 2$\, (from both 
($U\underline{\otimes} \{1^2\}$ and $W\underline{\otimes} \{1^2\}$). By contrast, for model (4.4b), there are now additional singlets
from the remaining terms: $2+3$ for the $\underline{\otimes} \{2\}$ piece, and $2+1$ for the
$\underline{\otimes} \{1^2\}$ part giving $f_{(2)}= 6$\,, $f_{(1^2)}= 4$\,. Since only for symmetric function inner products
with even powers of $\{1^2\}$ do we find $\{1^2\}^{2\ell}*\{2\}^{K-2\ell}=\{2\}$, we have from the binomial theorem (with the combinatorial factor for orderings)
\begin{align}
\label{eq:quadraticLMsubmodels}
SSM:\qquad h_2 = & \, \sum_{\ell=0}^{\lfloor{K/2}\rfloor}\left(\begin{array}{c}K\\2\ell\end{array}\right)2^{2\ell} =
\textstyle{\frac 12}\big(3^K + (-1)^K \big)\,;\\
(4.4b):\qquad h_2 = & \, \sum_{\ell=0}^{\lfloor{K/2}\rfloor}\left(\begin{array}{c}K\\2\ell\end{array}\right)4^{2\ell}6^{K-2\ell}
 = \textstyle{\frac 12}\big(10^{K}+(-2)^K \big)\,.
 \end{align}

%\end{itemize}

	%%%%%%%%%%%%%%%%%%%%%%%%%%%%%%%%%%%%%%%%%%%%%%%%%%%%%%%%%%%%%%%
	%\end{appendix}
	%\pagebreak
	%\bibliographystyle{unsrt}
	%{\small
	%\bibliography{master_Jan17,mistress}
	%}
	
	%{\small
	%\begin{thebibliography}{10}
	%
	%%\bibitem{}
	%\end{thebibliography}
	%}
\end{appendix}
\pagebreak

\end{document}